\documentstyle[aps]{revtex}
\begin{document}
\title{$KN$ and $\overline{K}N$ Elastic Scattering in the Quark Potential Model}
\author{Hai-Jun Wang$^2$, Hui Yang$^2$ and Jun-Chen Su$^{1,*,2}$}
\address{1. Department of Physics, Harbin Institute of Technology, Harbin 150006,\\
People's Republic of China\\
2. Center for Theoretical Physics, College of Physics, Jilin University,\\
Changchun 130023, People's Republic of China}
\maketitle

\begin{abstract}
The $KN$ and $\overline{K}N$ low-energy elastic scattering is consistently
studied in the framework of the QCD-inspired quark potential model. The
model is composed of the t-channel one-gluon exchange potential, the
s-channel one-gluon exchange potential and the harmonic oscillator
confinement potential. By means of the resonating group method, nonlocal
effective interaction potentials for the $KN$ and $\overline{K}N$ systems
are derived and used to calculate the $KN$ and $\overline{K}N$ elastic
scattering phase shifts. By considering the effect of QCD renormalization,
the contribution of the color octet of the clusters ($q\overline{q}$) and ($%
qqq$) and the suppression of the spin-orbital coupling, the numerical
results are in fairly good agreement with the experimental data.


* Corresponding author. E-mail address: junchens@public.cc.jl.cn
\end{abstract}

\section{Introduction}

Inspired by the achievement in the study of hadron spectroscopy within the
framework of quark potential model [1-3], there have been continuous efforts
to apply the quark potential model and the resonating group method (RGM) [4]
to study nucleon-nucleon, nucleon-meson and meson-meson interactions and
scattering [5-20]. Among these efforts, the study of kaon-nucleon ($KN$)
interaction arose a particular interest in the past. Due to the high
penetrating power of $K^{+}$ meson, it is expected that the study of $K^{+}N$
interaction would provide more information for nuclear structures and
properties. Since in the $K^{+}N$ interaction, the one-pion exchange is
forbidden, the two-pion exchange is suppressed [10,11] and there is no
annihilation of valence quarks to appear, it is expected that the t-channel
one-gluon exchange potential (OGEP) plus the harmonic oscillator confinement
potential would give a reasonable description of the $K^{+}N$ interaction.
With this idea, the authors in Ref.[10] calculated the S-wave phase shifts
of $K^{+}N$ elastic scattering and found that the theoretical results are in
quite good agreement with the experimental data in the low-energy domain. It
is noted that in the calculation, the authors only took the coulomb,
spin-spin and Darwin terms in the t-channel OGEP without considering the
spin-orbital coupling and tensor force terms which contribute to the higher
partial wave scattering. Subsequently, to investigate P-wave phase shifts,
the authors in Ref.[11] introduced into their model the spin-orbital
coupling terms originating from the t-channel one gluon exchange and a
scalar exchange term describing the confinement interaction. As one knows,
there are two kinds of spin-orbit coupling terms in the t-channel OGEP: the
spin-symmetric term and the spin-antisymmetric one. Correspondingly, there
also exist such two terms in the confining potential generated from the
scalar exchange. The spin-symmetric terms in the two potentials have
opposite signs. Therefore, the effect of the spin-orbital coupling is
suppressed just as required in the study of meson and baryon spectra [2,21].
However, the spin-antisymmetric terms are of the same sign and hence would
produce large splitting which is in contradiction with the experiment [11].
Such terms are therefore dropped out in Ref.[11]. The calculation in
Ref.[11] showed that if the spin-orbit coupling terms are considered only,
except for the $I=0,J=\frac 12$ channel phase shift, the sign and magnitude
of the other channel phase shifts are well reproduced; when the other terms
in the t-channel OGEP are taken into account together, there appears a
serious problem that the $I=1$ channel phase shifts all become negative,
conflicting with the experimental data. Later, the S-wave phase shifts of $%
KN $ scattering are restudied in Ref.[15] by employing the Born order
diagrammatic technique. In the study, although only the spin-spin coupling
term in the t-channel OGEP is considered, the calculated result looks fine.
Subsequently, the Born approximation was applied to investigate the $KN$
scattering more extensively in Ref.[17]. In the investigation, apart from
the hyperfine term in the OGEP and the linear scalar confinement, the
spin-orbital coupling and spin-independent terms in the OGEP are taken into
account in the evaluation of P-wave and D-wave phase shifts. Nevertheless,
the magnitudes of \ most of the calculated phase shifts are smaller than the
experimental ones. Particularly, the sign of the theoretical $P_{13}$ wave
phase shift is opposite to the experimental data.\ Recently, the $KN$ phase
shifts are recalculated in Ref.[18] within the constituent quark model by
numerically solving the Hill-Wheeler equation, trying to give a consistent
description for the $KN$ interaction and the relevant meson and baryon
spectra. In the calculation, besides the linear confining potential, the
authors used only the Coulomb term and the spin-spin interaction term
multiplied by a phenomenological coefficient function of Gaussian type. The
calculated $I=0$ channel S-wave phase shift is pretty good in comparison
with the experiment; whereas for the $I=1$ channel S-wave phase shift, there
appears a big discrepancy between the theoretical result and the
experimental one. The authors also calculated higher angular momentum phase
shifts without including the spin-orbit coupling and tensor force terms in
the OGEP. Even though the results were considered to be quite reasonable,
the calculation is not complete theoretically because the spin-orbit
coupling and tensor force terms in the OGEP were not taken into account.

From the previous works mentioned above, it is clearly seen that a precise
understanding of the $KN$ interaction at quark-gluon level still calls for a
sophisticated quark potential model which can give a consistently good
description for not only the $KN$ interaction, but also the $\overline{K}N$
interaction which has never been investigated in the past. For this purpose,
it is necessary to incorporate new physical ingredients into the model as
suggested in Refs.[11,18]. In this paper, we attempt to investigate the $KN$
and $\overline{K}N$ interactions and their low-energy elastic phase shifts
in a consistent way within the quark potential model . The new features of
this investigation include: (1) The potential model is composed of the
t-channel OGEP [1] and the s-channel OGEP [22] as well as a phenomenological
confining potential. The t-channel OGEP is responsible for the $KN$
interaction, while for the $\overline{K}N$ interaction where the
annihilation and creation of a quark-antiquark ($q\overline{q}$) pair
appear, the s-channel OGEP is necessary to be considered as demonstrated in
our previous investigations of $\pi \pi $ and $K\overline{K\text{ }}$
interactions [19,20]. In these investigations, it was shown that the
s-channel OGEP plays a dominant role for the $\pi \pi $ $I=0$ channel S-wave
scattering and is necessary to be considered for the formation of $K%
\overline{K\text{ }}$ molecular states. As one knows, the two OGEPs are
derived from QCD in the nonrelativistic approximation of order $p^2/m^2$ and
contain spin-independent terms such as the Coulomb, velocity-dependent terms
and spin-dependent terms such as the spin-spin interaction, spin-orbital
coupling and tensor force terms. All these terms are taken into account in
our calculation as should be done in a theoretically consistent treatment.
(2) Inclusion of the QCD renormalization effect. It is well-known that the
OGEPs are derived from the tree diagram approximation of the S-matrix
elements or the irreducible interaction kernels in the Bethe-Salpeter (B-S)
equation. Obviously, to refine the potential model, the QCD renormalization
effect is necessary to be incorporated into the model. This can be done by
replacing the QCD coupling constant and quark masses in the OGEP with their
effective ones which were derived in our previous work in the one loop
approximation and a mass-dependent momentum space subtraction [23]. Our
calculation indicates that the inclusion of the QCD renormalization effect
gives an appreciable improvement on the theoretical phase shifts,
particularly, for the P-wave phase shifts. (3) The contribution from the
color octet of the three quark cluster ($qqq$) and the quark-antiquark
cluster ($q\overline{q}$) to the $KN$ and $\overline{K}N$ scattering is
considered because when the kaon (antikaon) and nucleon interact, the color
singlet states of the clusters ($qqq$) and ($q\overline{q}$) are possibly
polarized. This consideration has been justified in the recent studies of
meson production and decay phenomena [24,25]. In these studies, the color
octet of $q\overline{q}$ cluster plays an essential role in explaining the
experimental data. According to our calculation, the consideration of color
octet can also improve the theoretical results. (4) Nonlocal $KN$ and $%
\overline{K}N$ effective interaction potentials are derived from the
underlying interquark potentials by employing the RGM and used to evaluate
the phase shifts. We do not use the localized version of the potentials
because an inappropriate localization would damage the hermiticity of the
potentials and induce other unexpected errors. (5) The effect of
spin-orbital coupling in the t-channel OGEP is necessarily suppressed in the
present investigation in order to reproduce the P-wave phase shifts. This is
consistent with the requirement in the study of hadron spectroscopy. The
suppression may be achieved by a proper change of the coefficient function
of the spin-orbit term in the effective intercluster potentials which are
derived from the corresponding term in the OGEP. With the considerations
mentioned above, we obtain in this paper a series of theoretical $KN$
S-wave, P-wave and D-wave phase shifts which are in fairly good agreement
with the experimental data. In addition, a series of theoretical phase
shifts for the $\overline{K}N$ elastic scattering are predicted.

The rest of this paper is arranged as follows: Section 2 is used to describe
the quark potential model and show how to derive the $KN$ and $\overline{K}N$
effective interaction potentials. Section 3 serves to describe the
calculation of the $KN$ and $\overline{K}N$ scattering phase shifts. In the
last section, the calculated results are presented and discussions are made.
There are four appendices. In Appendix A, we show the construction of the
color-flavor-spin wave function for the systems under consideration. In
Appendix B, the effective $KN$ and $\overline{K}N$ interaction potentials
derived in position space are listed. In Appendix C, we briefly describe the
derivation of the phase shift formula used in our calculation. Appendix D is
used to make some explanations on the QCD renormalization.

\section{Quark Potential Model and Effective $KN$ and $\overline{K}N$
Potentials}

According to the quark model, the $KN$ ( $\overline{K}N$ ) system may be
treated as two quark clusters: the $K$-cluster $(q\overline{s})$ ( the $%
\overline{K}$-cluster $(\overline{q}s)$) and the $N$-cluster ($qqq)$ where $%
q=u$ or $d$. The effective $KN$ interaction potential may be extracted from
the following Schr\"odinger equation for the interacting $q^4\overline{s}$
system by the RGM 
\begin{equation}
(T+V)\Psi =E\Psi  \eqnum{1}
\end{equation}
where $E,T,V$ and $\Psi $ stand for the total energy, the kinetic energy,
the interaction potential and the wave function of the system respectively.
In the center of mass frame, 
\begin{equation}
T=\sum_i\frac{\stackrel{\rightharpoonup }{p}_i^2}{2m_i}-T_c  \eqnum{2}
\end{equation}
where $T_c$ represents the center of mass kinetic energy, 
\begin{equation}
V=\sum_{i<j=1}^5(V_{ij}^t+V_{ij}^s+V_{ij}^c)  \eqnum{3}
\end{equation}
here $V_{ij}^t$, $V_{ij}^s$ and $V_{ij}^c$ denote the t-channel OGEP, the
s-channel OGEP and the confining potentials respectively. They are
separately written in the following. The t-channel OGEP represented in the
momentum space is [26] 
\begin{equation}
\begin{tabular}{l}
$V_{ij}^t=\frac{4\pi \alpha _sC_{ij}^t}{(\stackrel{\rightharpoonup }{q}-%
\stackrel{\rightharpoonup }{k})^2}\{1-\frac{\stackrel{\rightharpoonup }{P}^2%
}{m_{ij}^2}-\frac{(m_i^2+m_j^2)}{8m_i^2m_j^2}(\stackrel{\rightharpoonup }{q}-%
\stackrel{\rightharpoonup }{k})^2+\frac{\left( m_i-m_j\right) }{2m_im_jm_{ij}%
}\stackrel{\rightharpoonup }{P}\cdot (\stackrel{\rightharpoonup }{q}+%
\stackrel{\rightharpoonup }{k})$ \\ 
$+\frac{(\stackrel{\rightharpoonup }{q}+\stackrel{\rightharpoonup }{k})^2}{%
4m_im_j}+\frac i{4m_{ij}}[\stackrel{\rightharpoonup }{P}\cdot (\stackrel{%
\rightharpoonup }{q}-\stackrel{\rightharpoonup }{k})\cdot (\frac{\stackrel{%
\rightharpoonup }{\sigma _i}}{mi}-\frac{\stackrel{\rightharpoonup }{\sigma _j%
}}{m_j})]-\frac{(\stackrel{\rightharpoonup }{q}-\stackrel{\rightharpoonup }{k%
})^2}{4m_im_j}\stackrel{\rightharpoonup }{\sigma _i}\cdot \stackrel{%
\rightharpoonup }{\sigma _j}$ \\ 
$+\frac i{4m_{ij}}(\stackrel{\rightharpoonup }{q}\times \stackrel{%
\rightharpoonup }{k})\cdot [(2+\frac{m_j}{m_i})\stackrel{\rightharpoonup }{%
\sigma _i}+(2+\frac{m_i}{m_j})\stackrel{\rightharpoonup }{\sigma _j}]+\frac{(%
\stackrel{\rightharpoonup }{q}-\stackrel{\rightharpoonup }{k})\cdot 
\stackrel{\rightharpoonup }{\sigma _i}(\stackrel{\rightharpoonup }{q}-%
\stackrel{\rightharpoonup }{k})\cdot \stackrel{\rightharpoonup }{\sigma _j}}{%
4m_im_j}\}$%
\end{tabular}
\eqnum{4}
\end{equation}
where $m_{ij}=m_i+m_j$, $\alpha _s$ is the QCD fine structure constant, $%
\stackrel{\rightharpoonup }{\sigma _i}$ are the spin Pauli matrices for i-th
particle, $C_{ij}^t$ is the t-channel color matrix defined as

\begin{equation}
C_{ij}^t=\{ 
\begin{tabular}{ll}
$\frac{\lambda _i^a}2\frac{\lambda _j^a}2(\frac{\lambda _i^{a*}}2\frac{%
\lambda _j^{a*}}2),$ & $\text{for }qq(\overline{q}\overline{q})$ \\ 
$-\frac{\lambda _i^a}2\frac{\lambda _j^{a^{*}}}2,$ & $\text{for }q\overline{q%
}$%
\end{tabular}
\eqnum{5}
\end{equation}
with $\lambda ^a$ being the Gell-Mann matrix, $\stackrel{\rightharpoonup }{P}%
,\stackrel{\rightharpoonup }{k}$ and $\stackrel{\rightharpoonup }{q}$ are
the total momentum, the initial state relative momentum, and the final state
relative momentum of the two interacting particles.

The s-channel OGEP is [20, 22] 
\begin{equation}
\begin{tabular}{l}
$V_{ij}^s=\frac{\pi \alpha _sF_{ij}^sC_{ij}^s}{2mm^{^{\prime }}}[(3+%
\stackrel{\rightharpoonup }{\sigma _i}\cdot \stackrel{\rightharpoonup }{%
\sigma _j})-\frac{5(m^2+m^{\prime 2})-4m^{\prime }}{8m^2m^{\prime 2}}%
\stackrel{\rightharpoonup }{P}^2-\frac{2\stackrel{\rightharpoonup }{k}^2}{m^2%
}-\frac{2\stackrel{\rightharpoonup }{q}^2}{m^{\prime 2}}$ \\ 
$-(\frac{(m^2+m^{\prime 2})}{8m^2m^{\prime 2}}+\frac{\stackrel{%
\rightharpoonup }{k}^2}{m^2}+\frac{\stackrel{\rightharpoonup }{q}^2}{%
m^{\prime 2}})\stackrel{\rightharpoonup }{\sigma _i}\cdot \stackrel{%
\rightharpoonup }{\sigma _j}+\frac i{4m^{\prime 2}}(\stackrel{%
\rightharpoonup }{P}\times \stackrel{\rightharpoonup }{q})\cdot (\stackrel{%
\rightharpoonup }{\sigma _i}-\stackrel{\rightharpoonup }{\sigma _j})$ \\ 
$-\frac i{4m^2}(\stackrel{\rightharpoonup }{P}\times \stackrel{%
\rightharpoonup }{k})\cdot (\stackrel{\rightharpoonup }{\sigma _i}-\stackrel{%
\rightharpoonup }{\sigma _j})-\frac{\left( m-m^{\prime }\right) ^2}{4m^2}%
\stackrel{\rightharpoonup }{P}\cdot \stackrel{\rightharpoonup }{\sigma _i}%
\stackrel{\rightharpoonup }{P}\cdot \stackrel{\rightharpoonup }{\sigma _j}$
\\ 
$+\frac 1{4m^2}(\stackrel{\rightharpoonup }{P}\cdot \stackrel{%
\rightharpoonup }{\sigma _i}\stackrel{\rightharpoonup }{k}\cdot \stackrel{%
\rightharpoonup }{\sigma _j}-\stackrel{\rightharpoonup }{k}\cdot \stackrel{%
\rightharpoonup }{\sigma _i}\stackrel{\rightharpoonup }{P}\cdot \stackrel{%
\rightharpoonup }{\sigma _j}+4\stackrel{\rightharpoonup }{k}\cdot \stackrel{%
\rightharpoonup }{\sigma _i}\stackrel{\rightharpoonup }{k}\cdot \stackrel{%
\rightharpoonup }{\sigma _j})$ \\ 
$+\frac 1{4m^{\prime 2}}(\stackrel{\rightharpoonup }{P}\cdot \stackrel{%
\rightharpoonup }{\sigma _i}\stackrel{\rightharpoonup }{q}\cdot \stackrel{%
\rightharpoonup }{\sigma _j}-\stackrel{\rightharpoonup }{q}\cdot \stackrel{%
\rightharpoonup }{\sigma _i}\stackrel{\rightharpoonup }{P}\cdot \stackrel{%
\rightharpoonup }{\sigma _j}+4\stackrel{\rightharpoonup }{q}\cdot \stackrel{%
\rightharpoonup }{\sigma _i}\stackrel{\rightharpoonup }{q}\cdot \stackrel{%
\rightharpoonup }{\sigma _j})]$%
\end{tabular}
\eqnum{6}
\end{equation}
where $m$ and $m^{\prime }$ denote the quark (antiquark) masses before and
after annihilation respectively, $C_{ij}^s$ and $F_{ij}^s$ are the s-channel
color and flavor matrices, defined by

\begin{equation}
C_{ij}^s=\frac 1{24}(\lambda _i^a-\lambda _j^{a^{*}})^2  \eqnum{7}
\end{equation}
and

\begin{equation}
F_{ij}^s=\frac 23-(\frac 12\stackrel{\rightharpoonup }{\tau _i}\cdot 
\stackrel{\rightharpoonup }{\tau _j}%
+V_i^{-}V_j^{+}+V_i^{+}V_j^{-}+U_i^{-}U_j^{+}+U_i^{+}U_j^{-}+\frac 32Y_iY_j)
\eqnum{8}
\end{equation}
here $\stackrel{\rightharpoonup }{\tau _i}$ are the isospin Pauli matrices
for i-th particle, $Y_i$ the hypercharge operators, $V_i^{+}$ and $V_i^{-}$( 
$U_i^{+}$ and $U_i^{-}$) represent the rising and lowering operators of the $%
V$-spin ($U$-spin) respectively.

The confining potential, as was done in Ref.[10,11], is taken to be the
harmonic oscillator one. In the momentum space it is represented as

\begin{equation}
V_{ij}^c=C_{ij}^t(2\pi )^3\mu _{ij}\omega ^2\nabla _k^2\delta ^3(\stackrel{%
\rightharpoonup }{q}-\stackrel{\rightharpoonup }{k})  \eqnum{9}
\end{equation}
where $\mu _{ij}$ is the reduced mass of the interaction particles and $%
\omega $ force-strength parameter.

Now let us construct the wave function of the $KN$ system from the wave
functions of clusters $(q\overline{s})$ and ($qqq)$. Since there are
identical particles between the two clusters, the basis function of the
system may be represented as 
\begin{equation}
\begin{tabular}{l}
$\Phi _{TMsm}(\stackrel{\rightharpoonup }{p_1},\stackrel{\rightharpoonup }{%
p_2},\stackrel{\rightharpoonup }{p_3},\stackrel{\rightharpoonup }{p_4},%
\stackrel{\rightharpoonup }{p_5};\stackrel{\rightharpoonup }{\rho })$ \\ 
$=\frac 1{\sqrt{4}}(1-P_{14}-P_{24}-P_{34})\Psi _{TM\frac 12m}(1,2,3,4,5)R(%
\stackrel{\rightharpoonup }{p_1},\stackrel{\rightharpoonup }{p_2},\stackrel{%
\rightharpoonup }{p_3},\stackrel{\rightharpoonup }{p_4},\stackrel{%
\rightharpoonup }{p_5};\stackrel{\rightharpoonup }{\rho })$%
\end{tabular}
\eqnum{10}
\end{equation}
where we number the three quarks in the $N$-cluster as $1,2,3$ and the quark
and antiquark in the $K$-cluster (or the antiquark and quark in the $%
\overline{K}$-cluster) as $4$ and $5$, $P_{j4}$ ($j=1,2,3$) symbolize the
interchange operators, $\Psi _{TM\frac 12m}(1,2,3,4,5)$ and $R(\stackrel{%
\rightharpoonup }{p_1},\stackrel{\rightharpoonup }{p_2},\stackrel{%
\rightharpoonup }{p_3},\stackrel{\rightharpoonup }{p_4},\stackrel{%
\rightharpoonup }{p_5};\stackrel{\rightharpoonup }{\rho })$ represent the
color-isospin-spin wave function and the position space wave function
respectively which are constructed from the color-isospin-spin wave
functions and the coordinate space wave functions of nucleon and kaon. For
the $\overline{K}N$ system, noticing that there is no identical particles
between the two clusters $(\overline{q}s)$ and $(qqq)$, the basis wave
function of the system may simply be written as 
\begin{equation}
\Phi _{TMsm}(\stackrel{\rightharpoonup }{p_1},\stackrel{\rightharpoonup }{p_2%
},\stackrel{\rightharpoonup }{p_3},\stackrel{\rightharpoonup }{p_4},%
\stackrel{\rightharpoonup }{p_5};\stackrel{\rightharpoonup }{\rho })=\Psi
_{TM\frac 12m}(1,2,3,4,5)R(\stackrel{\rightharpoonup }{p_1},\stackrel{%
\rightharpoonup }{p_2},\stackrel{\rightharpoonup }{p_3},\stackrel{%
\rightharpoonup }{p_4},\stackrel{\rightharpoonup }{p_5};\stackrel{%
\rightharpoonup }{\rho })  \eqnum{11}
\end{equation}
where $\Psi _{TM\frac 12m}(1,2,3,4,5)$ and $R(\stackrel{\rightharpoonup }{p_1%
},\stackrel{\rightharpoonup }{p_2},\stackrel{\rightharpoonup }{p_3},%
\stackrel{\rightharpoonup }{p_4},\stackrel{\rightharpoonup }{p_5};\stackrel{%
\rightharpoonup }{\rho })$ are the color-isospin-spin wave function and the
position space wave function constructed from the corresponding wave
functions of nucleon and antikaon.

Since the $KN$ system is treated as two clusters, when they interact, each
cluster may be in color singlet \underline{1} or in color octet \underline{8}
as indicated in Refs.[12,20]. Thus, the color-spin-isospin wave function $%
\Psi _{TM\frac 12m}(1,2,3,4,5)$ of the whole system may be given by the
color singlet part $\Psi _{TM\frac 12m}^{(1)}(1,2,3,4,5)$ or the color octet
part $\Psi _{TM\frac 12m}^{(2)}(1,2,3,4,5)$ formed by the color singlets or
color octets of the two clusters. In principle, we may test a general color
structure of system under consideration which is given by the following
linear combination 
\begin{equation}
\Psi _{TM\frac 12m}(1,2,3,4,5)=\alpha \Psi _{TM\frac 12m}^{(1)}(1,2,3,4,5)+%
\beta \Psi _{TM\frac 12m}^{(2)}(1,2,3,4,5)  \eqnum{12}
\end{equation}
where the coefficients $\alpha $ and $\beta $ are required to satisfy 
\begin{equation}
\left| \alpha \right| ^2+\left| \beta \right| ^2=1  \eqnum{13}
\end{equation}
The wave functions $\Psi _{TM\frac 12m}^{(1)}(1,2,3,4,5)$ and $\Psi _{TM%
\frac 12m}^{(2)}(1,2,3,4,5)$ are listed in Appendix A. They are constructed
by antisymmetry of the wave functions of identical particles in nucleon.

Because we limit our discussion to the interaction in the low-energy regime,
it is appropriate to write the position space basis function of the $KN$ or $%
\overline{K}N$ system in the form 
\begin{equation}
R(\stackrel{\rightharpoonup }{p_1},\stackrel{\rightharpoonup }{p_2},%
\stackrel{\rightharpoonup }{p_3},\stackrel{\rightharpoonup }{p_4},\stackrel{%
\rightharpoonup }{p_5};\stackrel{\rightharpoonup }{\rho })=\phi
_{os}^{\left( +\right) }(\stackrel{\rightharpoonup }{p_1},\stackrel{%
\rightharpoonup }{\rho })\phi _{os}^{\left( +\right) }(\stackrel{%
\rightharpoonup }{p_2},\stackrel{\rightharpoonup }{\rho })\phi _{os}^{\left(
+\right) }(\stackrel{\rightharpoonup }{p_3},\stackrel{\rightharpoonup }{\rho 
})\phi _{os}^{\left( -\right) }(\stackrel{\rightharpoonup }{p_4},\stackrel{%
\rightharpoonup }{\rho })\phi _{os}^{\left( -\right) }(\stackrel{%
\rightharpoonup }{p_5},\stackrel{\rightharpoonup }{\rho })  \eqnum{14}
\end{equation}
where $\phi _{os}^{\left( +\right) }(\stackrel{\rightharpoonup }{p_i},%
\stackrel{\rightharpoonup }{\rho })$ and $\phi _{os}^{\left( -\right) }(%
\stackrel{\rightharpoonup }{p_j},\stackrel{\rightharpoonup }{\rho })$ are
the lowest-lying harmonic oscillator states of the $N$-cluster and $K$%
-cluster given in the momentum space, 
\begin{equation}
\phi _{os}^{\left( \pm \right) }(\stackrel{\rightharpoonup }{p_i},\stackrel{%
\rightharpoonup }{\rho })=(2\sqrt{\pi }b_i)^{3/2}\exp (-\frac 12b_i^2%
\stackrel{\rightharpoonup }{p_i}^2\mp i\lambda _{\pm }\stackrel{%
\rightharpoonup }{p_i}\cdot \stackrel{\rightharpoonup }{\rho })  \eqnum{15}
\end{equation}
in which $\stackrel{\rightharpoonup }{\rho }$ is the vector representing the
separation between the centers of mass of the two clusters and parameters $%
\lambda _{\pm }$ are defined by

\begin{equation}
\lambda _{-}=\beta _1=\frac{3m_1}{4m_1+m_2},\lambda _{+}=\beta _2=\frac{%
m_1+m_2}{4m_1+m_2}  \eqnum{16}
\end{equation}
here $m_1$ denotes the mass of d and u quarks, $m_2$ the mass of strange
quark. The wave function in Eq.(14) can be represented through the cluster
coordinates in the form 
\begin{equation}
R(\stackrel{\rightharpoonup }{p_1},\stackrel{\rightharpoonup }{p_2},%
\stackrel{\rightharpoonup }{p_3},\stackrel{\rightharpoonup }{p_4},\stackrel{%
\rightharpoonup }{p_5};\stackrel{\rightharpoonup }{\rho })=X_K(\stackrel{%
\rightharpoonup }{q})X_N(\stackrel{\rightharpoonup }{k_{1,}}\stackrel{%
\rightharpoonup }{k_2})\Gamma (\stackrel{\rightharpoonup }{Q},\stackrel{%
\rightharpoonup }{\rho })Z_{CM}(\stackrel{\rightharpoonup }{P})  \eqnum{17}
\end{equation}
where $X_K(\stackrel{\rightharpoonup }{q})$ and $X_N(\stackrel{%
\rightharpoonup }{k_{1,}}\stackrel{\rightharpoonup }{k_2})$ are the internal
motion wave functions of the $K(\overline{K})$-cluster $(q\overline{s})$ ($(%
\overline{q}s)$) and the $N$-cluster $(qqq)$ with $\stackrel{\rightharpoonup 
}{q}$ and $\stackrel{\rightharpoonup }{k_1,}\stackrel{\rightharpoonup }{k_2}$
being the relative momenta in the clusters $(q\overline{s})$ ($(\overline{q}%
s)$) and ($qqq)$ respectively, $\Gamma (\stackrel{\rightharpoonup }{Q},%
\stackrel{\rightharpoonup }{\rho })$ is the wave function describing the
relative motion between the two clusters with $\stackrel{\rightharpoonup }{Q}
$ being the relative momentum of the two clusters and $Z_{CM}(\stackrel{%
\rightharpoonup }{P}$ $)$ the wave function for the center-of-mass motion of
the whole system in which $\stackrel{\rightharpoonup }{P}$ is the total
momentum of the system. According to the RGM, the wave function of the two
clusters may be represented in the form

\begin{equation}
\overline{\Psi }_{TMsm}=\int d^3\rho \Phi _{TM\frac 12m_s}(\stackrel{%
\rightharpoonup }{p_1},\stackrel{\rightharpoonup }{p_2},\stackrel{%
\rightharpoonup }{p_3},\stackrel{\rightharpoonup }{p_4},\stackrel{%
\rightharpoonup }{p_5};\stackrel{\rightharpoonup }{\rho })f(\overrightarrow{%
\rho })  \eqnum{18}
\end{equation}
where $\Phi _{TM\frac 12m_s}(\stackrel{\rightharpoonup }{p_1},\stackrel{%
\rightharpoonup }{p_2},\stackrel{\rightharpoonup }{p_3},\stackrel{%
\rightharpoonup }{p_4},\stackrel{\rightharpoonup }{p_5};\stackrel{%
\rightharpoonup }{\rho })$ is the basis function defined in Eqs.(10) and
(11) and $f(\stackrel{\rightharpoonup }{\rho })$ is the unknown function
describing the relative motion of the two clusters. On substituting the
above wave function in Eq.(1), according to the well-known procedure, one
may derive a resonating group equation satisfied by the function $f(%
\stackrel{\rightharpoonup }{\rho }).$ Then, by the following transformation 
\begin{equation}
f(\overrightarrow{\rho })=\int d^3R\Gamma (\stackrel{\rightharpoonup }{\rho }%
,\stackrel{\rightharpoonup }{R})\overline{\Psi }(\stackrel{\rightharpoonup }{%
R})  \eqnum{19}
\end{equation}
where 
\begin{equation}
\Gamma (\stackrel{\rightharpoonup }{\rho },\stackrel{\rightharpoonup }{R})=%
\frac 1{\sqrt{2}(2\pi )^3}(\frac{3\beta _2}{\pi b^2})^{3/4}\int d^3ke^{\frac 
1{6\text{ }\beta _2}b^2\stackrel{\rightharpoonup }{k}^2+i\stackrel{%
\rightharpoonup }{k}\cdot (\stackrel{\rightharpoonup }{\rho }-\stackrel{%
\rightharpoonup }{R})}  \eqnum{20}
\end{equation}
in which $b$ is the harmonic oscillator size parameter, the resonating group
equation will be transformed to the following Schr\"odinger equation
satisfied by the relative motion of the two clusters 
\begin{equation}
-\frac 1{2\mu }\nabla _{\overrightarrow{R}}^2\overline{\Psi }(\stackrel{%
\rightharpoonup }{R})+\int d^3R^{^{\prime }}V(\stackrel{\rightharpoonup }{R},%
\stackrel{\rightharpoonup }{R^{\prime }})\overline{\Psi }(\stackrel{%
\rightharpoonup }{R^{\prime }})=\varepsilon \overline{\Psi }(\stackrel{%
\rightharpoonup }{R})  \eqnum{21}
\end{equation}
where $\varepsilon ,\mu $ and $\Psi (\stackrel{\rightharpoonup }{R})$ are
respectively the energy of relative motion, the reduced mass and the
Schr\"odinger-type wave function for the two clusters and 
\begin{equation}
V(\stackrel{\rightharpoonup }{R},\stackrel{\rightharpoonup }{R^{\prime }}%
)=V^t(\stackrel{\rightharpoonup }{R},\stackrel{\rightharpoonup }{R^{\prime }}%
)+V^s(\stackrel{\rightharpoonup }{R},\stackrel{\rightharpoonup }{R^{\prime }}%
)+V^c(\stackrel{\rightharpoonup }{R},\stackrel{\rightharpoonup }{R^{\prime }}%
)  \eqnum{22}
\end{equation}
is the nonlocal $KN$ ( $\overline{K}N$ ) effective interaction potential in
which $V^t(\stackrel{\rightharpoonup }{R},\stackrel{\rightharpoonup }{%
R^{\prime }})$, $V^s(\stackrel{\rightharpoonup }{R},\stackrel{%
\rightharpoonup }{R^{\prime }})$ and $V^c(\stackrel{\rightharpoonup }{R},%
\stackrel{\rightharpoonup }{R^{\prime }})$ are generated by the t-channel
OGEP, the s-channel OGEP and the confining potential. The potential $V(%
\stackrel{\rightharpoonup }{R},\stackrel{\rightharpoonup }{R^{\prime }})$ in
the Schr\"odinger equation is connected with the potential $V(\stackrel{%
\rightharpoonup }{\rho },\stackrel{\rightharpoonup }{\rho ^{\prime }})$
appearing in the resonating group equation by the following formula. 
\begin{equation}
V(\stackrel{\rightharpoonup }{R},\stackrel{\rightharpoonup }{R^{\prime }}%
)=\int d^3\rho d^3\rho ^{^{\prime }}\Gamma (\stackrel{\rightharpoonup }{R},%
\stackrel{\rightharpoonup }{\rho })V(\stackrel{\rightharpoonup }{\rho },%
\stackrel{\rightharpoonup }{\rho ^{\prime }})\Gamma (\stackrel{%
\rightharpoonup }{\rho ^{\prime }},\stackrel{\rightharpoonup }{R^{\prime }})
\eqnum{23}
\end{equation}
where $V(\stackrel{\rightharpoonup }{\rho },\stackrel{\rightharpoonup }{\rho
^{\prime }})$ is described in Appendix B. To compute the elastic scattering
phase shifts, we need to calculate the transition matrix between initial and
final plane wave functions as follows 
\begin{equation}
T_{fi}(\stackrel{\rightharpoonup }{k},\stackrel{\rightharpoonup }{k^{\prime }%
})=\int d^3Rd^3R^{\prime }e^{-i\stackrel{\rightharpoonup }{k}\cdot \stackrel{%
\rightharpoonup }{R}}V(\stackrel{\rightharpoonup }{R},\stackrel{%
\rightharpoonup }{R^{\prime }})e^{i\stackrel{\rightharpoonup }{k^{\prime }}%
\cdot \stackrel{\rightharpoonup }{R^{\prime }}}  \eqnum{24}
\end{equation}
where $\stackrel{\rightharpoonup }{k^{\prime }}$ and $\stackrel{%
\rightharpoonup }{k}$ are the $KN$ ($\overline{K}N$) relative momenta for
the initial and final states respectively. Upon substituting Eq.(20) into
Eq.(23), it is easy to find 
\begin{equation}
T_{fi}(\stackrel{\rightharpoonup }{k},\stackrel{\rightharpoonup }{k^{\prime }%
})=\frac 12(\frac{3\beta _2}{\pi b^2})^{3/2}e^{\frac{b^2}{6\beta _2}(%
\stackrel{\rightharpoonup }{k}^2+\stackrel{\rightharpoonup }{k^{\prime }}%
^2)}\int d^3\rho d^3\rho ^{^{\prime }}e^{-i\stackrel{\rightharpoonup }{k}%
\cdot \stackrel{\rightharpoonup }{\rho }}V(\stackrel{\rightharpoonup }{\rho }%
,\stackrel{\rightharpoonup }{\rho ^{\prime }})e^{i\stackrel{\rightharpoonup 
}{k^{\prime }}\cdot \stackrel{\rightharpoonup }{\rho ^{\prime }}}  \eqnum{25}
\end{equation}
This expression shows that to calculate the transition matrix, we may
directly use the potential $V(\stackrel{\rightharpoonup }{\rho },\stackrel{%
\rightharpoonup }{\rho ^{\prime }})$ instead of the potential $V(\stackrel{%
\rightharpoonup }{R},\stackrel{\rightharpoonup }{R^{\prime }})$.

\section{Calculation of Phase Shifts}

The phase shifts of the $KN$ ($\overline{K}N$) low-energy elastic scattering
are calculated in the Born approximation. As argued in Appendix C and
demonstrated in the previous literature [14, 15], the Born approximation can
reasonably describe hadron elastic low energy scattering processes. In this
approximation and in the center-of-mass frame, the $l$-th partial wave phase
shift is expressed by the following formula whose derivation will be
sketched in Appendix C

\begin{equation}
\delta _l^{IJ}=-2MkT_{fi}^{IJl}(k)  \eqnum{26}
\end{equation}
where 
\begin{equation}
M(E)=\frac{E^4-(m_K^2-m_N^2)^2}{4E^3}  \eqnum{27}
\end{equation}
in which $E$ is the total energy of the $KN$ ($\overline{K}N$) system, $m_K$
and $m_N$ are the masses of kaon and nucleon, $k=\mid \stackrel{%
\rightharpoonup }{k}\mid =\mid \stackrel{\rightharpoonup }{k^{\prime }}\mid $
is the magnitude of relative momenta $\stackrel{\rightharpoonup }{k}$ and $%
\stackrel{\rightharpoonup }{k^{\prime }}$ in the case of elastic scattering
and $T_{fi}^{IJl}(k)$ with isospin $I$, total angular momentum $J$ and
orbital angular momentum $l$ is the transition amplitude. This amplitude can
generally be expressed as 
\begin{equation}
T_{fi}^{IJl}(k)=\sum\limits_{m,m^{\prime },\mu ,\mu ^{\prime }}C_{lm\frac 12%
m_s}^{JM}C_{lm^{^{\prime }}\frac 12m_s^{\prime }}^{JM}\int d\Omega (%
\stackrel{\wedge }{k})d\Omega (\stackrel{\wedge }{k^{\prime }})Y_{lm^{\prime
}}^{*}(\stackrel{\wedge }{k^{\prime }})Y_{lm}(\stackrel{\wedge }{k})T_{fi}^I(%
\stackrel{\rightharpoonup }{k},\stackrel{\rightharpoonup }{k^{\prime }}%
;m_s,m_s^{\prime })  \eqnum{28}
\end{equation}
where $C_{lm\frac 12m_s}^{JM}$are the Clebsch-Gordan coefficients, $Y_{l%
\text{ }m}(\stackrel{\wedge }{k})$ are the spherical harmonic functions and 
\begin{equation}
T_{fi}^I(\stackrel{\rightharpoonup }{k},\stackrel{\rightharpoonup }{%
k^{\prime }};m_s,m_s^{\prime })=\langle C;I,M;\frac 12,m_s\mid V_{f\text{ }%
i}(\stackrel{\rightharpoonup }{k},\stackrel{\rightharpoonup }{k^{\prime }}%
)\mid I,M_I;\frac 12,m_s^{\prime };C\rangle  \eqnum{29}
\end{equation}
are the matrix elements of the operator $V_{fi}(\stackrel{\rightharpoonup }{k%
},\stackrel{\rightharpoonup }{k^{\prime }})$ defined in Eq.(25) between the
color-spin-isospin wave functions $\mid I,M_I;\frac 12,m_s^{\prime
};C\rangle $ and $\mid I,M_I;\frac 12;m_s;C\rangle $ in which $I$, $M_I$ and 
$\frac 12$, $m_s$ are the isospin and spin quantum numbers of the $KN$ ($%
\overline{K}N$) system respectively and $C$ denotes the color singlet of the
whole system. These matrix elements can be easily calculated. The explicit
expressions of the quantities $T_{fi}^I(\stackrel{\rightharpoonup }{k},%
\stackrel{\rightharpoonup }{k^{\prime }};m_s,m_s^{\prime })$ and $T_{f\text{ 
}i}^{IJ\text{ }l}(k)$, we think, are unnecessary to be listed in this paper.
We only show here numerical results of the theoretical phase shifts in
Figs.1-6 using the conventional partial wave notation $L_{I2J}$. It is noted
here that the formula in Eq.(28) is general for evaluating the transition
amplitude, particularly, in the case that the spin-orbit coupling and tensor
force terms are present in the nonlocal effective potentials.

We would like here to discuss the problem of suppression of the effect of
spin-orbital coupling. It is a common recognition in the study of hadron
spectroscopy that the effect of the spin-orbital coupling term in the
t-channel OGEP ought to be suppressed by the corresponding term in the
confining potential [2]. In Ref.[21], one of the authors in this paper and
his coworkers proposed a $q\overline{q}$ confining potential which was
obtained from a general Lorentz structure of the confinement. In the
confining potential, there are various terms among which the spin-orbit
coupling term is of a sign opposite to the corresponding one in the OGEP. In
this paper, to avoid the complexity of such a confining potential, we
alternatively take an effective treatment to achieve the spin-orbital
suppression. Looking at the expression of the potential $V^t(\stackrel{%
\rightharpoonup }{\rho },\stackrel{\rightharpoonup }{\rho ^{\prime }})$
shown in Appendix B, one can see that there is a kind of factorial functions
to appear in some terms of the potential $V^t(\stackrel{\rightharpoonup }{%
\rho },\stackrel{\rightharpoonup }{\rho ^{\prime }})$ which are of the form 
\begin{equation}
g(x,\stackrel{\rightharpoonup }{\rho })=\int \frac{d^3r}{4\pi r}e^{-xr^2+x%
\stackrel{\rightharpoonup }{r}\cdot \stackrel{\rightharpoonup }{\rho }} 
\eqnum{30}
\end{equation}
In particular, this function is related to the spin-orbit coupling term in
the potential $V^t(\stackrel{\rightharpoonup }{\rho },\stackrel{%
\rightharpoonup }{\rho ^{\prime }}).$ The function $g(x,\stackrel{%
\rightharpoonup }{\rho })$ may appropriately be replaced by an interpolating
function such that 
\begin{equation}
g(x,\stackrel{\rightharpoonup }{\rho })\simeq \frac{e^{(1-\gamma )\frac{%
x\rho ^2}4}}{2\text{ }x}  \eqnum{31}
\end{equation}
To obtain the above expression, we have used the approximate expression of
the following integral 
\begin{equation}
f(x)=\int_0^xe^{-t^2}dt\approx xe^{-\gamma x^2}  \eqnum{32}
\end{equation}
As shown in Fig.5, when we take the parameter $\gamma =0.3$, the function on
the right hand side of Eq.(32) very approaches the real value of the
integral when $x$ is not too large. However, as shown in Fig.6, the above
value of $\gamma $ leads to worse P-wave phase shifts. In order to get
better P-wave phase shifts, we have to take some larger value of $\gamma $
which just plays the role of suppressing the effect of spin-orbital coupling.

Finally, let us discuss the inclusion of QCD renomalization effect. As
mentioned in Introduction, the OGEP is derived from the S-matrix or the B-S
irreducible interaction kernel in the tree diagram approximation. Obviously,
to improve our calculation, it is natural to consider the correction arising
from QCD renormalization. This can be done by replacing the QCD coupling
constant and quark masses in the OGEP with the effective ones which are
obtained by solving the renormalization group equations satisfied by the
renormalized coupling constant and quark masses. This procedure, as proved
in Ref.[27] and demonstrated in Appendix D, is equivalent to replacing the
free wave functions, the free propagators and the bare vertices in the tree
diagrams with the exact ones. In the calculation of this paper, we employ
the effective coupling constant and quark masses given in Ref.[23] which
were derived from QCD in the one-loop approximation and the mass-dependent
momentum space subtraction. These effective quantities are suitable to any
energy, particularly, to the low energy, unlike the results obtained in the
minimal subtraction [28] which actually are applicable only in the large
momentum limit. The effective fine structure constant used has the
expression like this [23]

\begin{equation}
\alpha _R(\lambda )=\frac{\alpha _R^0}{1+\frac{\alpha _R^0}{2\pi }G(\lambda )%
}  \eqnum{33}
\end{equation}
where $\alpha _R^o$ is a coupling constant and $G(\lambda )$ is a function
of variable $\lambda $ which has different expressions given by the
time-like momentum subtraction (the subtraction performed at time-like
renormalization point) and the space-like momentum subtraction (the
subtraction carried out at the space-like renormalization point). For the
time-like momentum subtraction, 
\begin{equation}
G(\lambda )=11\ln \lambda -\frac 23N_f[2+\sqrt{3}\pi -\frac 2{\lambda ^2}+(%
\frac 2{\lambda ^2}+1)\frac{\sqrt{\lambda ^2-4}}\lambda \ln \frac 12(\lambda
+\sqrt{\lambda ^2-4})]  \eqnum{34}
\end{equation}
where $N_f$ is the quark flavor number which will be taken to be three in
this paper. While, for the space-like momentum subtraction, 
\begin{equation}
\begin{tabular}{l}
$G(\lambda )=11\ln \lambda -\frac 23N_f[\frac 2{\lambda ^2}-2-(\frac 2{%
\lambda ^2}-1)\frac{\sqrt{\lambda ^2+4}}\lambda \ln \frac 12(\lambda +\sqrt{%
\lambda ^2+4})$ \\ 
$+\sqrt{5}\ln \frac 12(1+\sqrt{5})]$%
\end{tabular}
\eqnum{35}
\end{equation}
in which $\lambda $ is defined as $\lambda =\sqrt{q^2/\mu ^2}$ with $q$
being a momentum variable and $\mu $ the fixed scale parameter. The
expression in Eq.(33) with the function $G(\lambda )$ given either in
Eq.(34) or in Eq.(35) will immediately goes over to the result given in the
minimal subtraction in the large momentum limit. The latter subtraction was
performed at the space-like renormalization point. It would be noted that in
writing the above effective coupling constant, the mass difference between
different quarks is ignored for simplicity. The behaviors of the effective
coupling constants given in Eqs.(33)-(35) are described in Fig.7. From the
figures, we see that the effective coupling constants given by the time-like
and space-like momentum subtractions have different behaviors in the
low-energy regime. It is interesting to note that the effective coupling
constant given in the space-like momentum subtraction is almost the same as
given in the minimal subtraction in the regime. This situation happens only
in the case of ignoring the mass difference between different quarks and
taking $N_f=3$. In other cases, the difference between the results given by
the both subtractions will be manifest.

The effective quark mass is represented as 
\begin{equation}
m_R(\lambda )=m_Re^{-S(\lambda )}  \eqnum{36}
\end{equation}
where $m_R$ is the constant quark mass given at $\lambda =1$ which will
appropriately be chosen to be the constituent quark mass in the quark
potential model, $S(\lambda )$ is a function which also has different
expressions for the different subtractions. For the time-like momentum
subtraction, 
\begin{equation}
S(\lambda )=\frac{\alpha _R^0}\pi \frac{1-\lambda }\lambda \{2+(\frac 2{%
\lambda ^2}-\frac{1+\lambda }{\lambda ^2})\ln \mid 1-\lambda ^2\mid \} 
\eqnum{37}
\end{equation}
While, for the space-like momentum subtraction, 
\begin{equation}
S(\lambda )=S_1(\lambda )+iS_2(\lambda )  \eqnum{38}
\end{equation}
where 
\begin{equation}
S_1(\lambda )=\frac{\alpha _R^0}\pi [(\frac 3{\lambda ^2}+1)\ln (1+\lambda
^2)-4\sqrt{2}]  \eqnum{39}
\end{equation}
and 
\begin{equation}
S_2(\lambda )=\frac{2\alpha _R^0}\pi [\frac 1{\lambda ^3}\ln (1+\lambda ^2)-%
\frac 1\lambda +1-\ln \sqrt{2}]  \eqnum{40}
\end{equation}
The behaviors of the effective quark masse given by the time-like momentum
subtraction and the real part of the effective quark mass given in the
space-like momentum subtraction are depicted in Fig.8. The figures show that
at low energy the effective masses given in the both subtractions are not
different so much. For the interaction taking place in the t-channel, as
explained in Appendix D, the transfer momentum is space-like, while for the
interaction in the s-channel, the transfer momentum is time-like. Therefore,
for the t-channel OGEP, we will use the effective coupling constant and
quark masses given by the space-like momentum subtraction and in this case,
we only adopt the real part of the effective masses in our calculation;
while, for the s-channel OGEP, the effective coupling constant and the quark
masses given in the time-like momentum subtraction will be employed. The
variable $\lambda $ is usually defined as a ratio of the momentum related to
the process of quark-gluon interactions. In this paper, as an effective
treatment, we directly define it as $\lambda =k/\mu $ where $k$ is taken to
be the magnitude of the relative momentum of the two interacting particles $%
K $ and $N$ ( or $\overline{K}$ and $N$ ).

\section{Results and Discussions}

This section is used to present calculated results for the $KN$ ( $\overline{%
K}N$ ) elastic scattering phase shifts, discuss adjustments of the
theoretical parameters and analyze the effect of color octet and QCD
renormalization as well as the suppression of the spin-orbital coupling.
First, we focus our attention on the $KN$ scattering. The theoretical phase
shifts of the $KN$ scattering are depicted in Figs.1-3. In the figures, the
solid lines represent the final results obtained by considering the
contributions arising from the color octet, the QCD renormalization and the
spin-orbit suppression. To exhibit the effects of the color octet and the
QCD renormalization, in the figures, we also show the results without
considering these effects. Such results are calculated with the same
parameters as for the solid lines and represented by the dotted and dashed
lines respectively in Figs.1-3. The figures show us that the agreement
between the final calculated results and the experimental data is good for
the phase shifts of all S-waves, $P_{13}$ wave and $D_{13}$ wave in the
low-energy domain, particularly, in the region of the laboratory momentum
less than 600 MeV to which the nonrelativistic quark potential model is
considered to be applicable. For the other P-wave and D-wave phase shifts,
the agreement is qualitatively reasonable. In obtaining these results, we
used the parameters as follows: the QCD coupling constant $\alpha _s^0=$ $%
0.23$, the constituent quark masses $m_u=m_d=350$ MeV and $m_s=550$ MeV, the
size parameter of harmonic oscillator $b=0.255$ fm, the force strength of
confinement $\omega =0.2$ GeV , the color combination coefficient $\alpha
=0.915$, the scale parameter of QCD renormalization $\mu =0.195$ GeV and the
parameter of spin-orbital suppression $\gamma =0.45$. These parameters are
adjusted to give a better fit to the $KN$ elastic scattering experimental
data, mainly to the S-wave phase shifts because the $KN$ elastic scattering
data are available and rather sufficient [29,30]. In comparison with the
previous results given in Refs.[10, 11, 17, 18], our calculation achieves a
considerable improvement on the theoretical phase shifts for all the partial
waves not only in the magnitude, but also in the sign. Especially, for the $%
P_{13}$ wave phase shift, it now gets a right sign in our calculation,
opposite to the previous result which was given a wrong sign [17].

Now let us analyze the effects of the color octet, the QCD renormalization
and the suppression of the spin-orbital coupling. In adjusting the
theoretical parameters, we found that the calculated results are sensitive
to the parameters $\alpha $ and $b.$ Any small change would cause noticeable
influence on calculated values. For instance, when we let $\alpha $
increase, the absolute values of S-wave phase shifts decrease rather fast.
In particular, when the $\alpha $ tends to unity, i.e. the color octet is
absent, as denoted by the dotted lines in Fig.1, we obtain the S-wave phase
shifts similar to those given in Ref.[17]. In this case, certainly, we may
give a better fit of the calculated result to the experimental one by
adjusting the parameter $b$ and others, but, we failed to simultaneously get
a good result for another S-wave phase shift, as was demonstrated previously
in Ref.[18]. Only when the color octet is considered, it is possible to get
consistently good results for the both S-wave phase shifts as denoted by the
solid lines in Fig.1. This suggests that the introduction of the color octet
is necessary in our calculation. From Figs.2 and 3, we also see that the
inclusion of the color octet gives a appreciable effect on the P and D wave
phase shifts.

In this paper, the two kinds of spin-orbital terms in the t-channel OGEP:
the spin-symmetric term and the spin-antisymmetric one are all taken into
account. In this case, we still encountered the puzzling problem for the
P-wave phase shifts as was revealed originally in Ref.[11] and mentioned in
the Introduction. If the QCD renormalization effect is not considered,
except for the $P_{11}$ and $P_{13}$ wave phase shifts, it seems to be able
to reproduce the phase shifts for the other phase shifts by readjusting the
parameters involved. However, as exhibited in Fig.2, for the $P_{11}$ and $%
P_{13}$ waves, it is impossible to get a satisfactory result of their phase
shifts. Particularly, for the $P_{13}$ wave, its phase shift is always of a
wrong sign as the previous result given in Ref.(17). This problem can only
be resolved by taking the QCD renormalization effect into account in our
calculation. In this way, the $P_{11}$ and $P_{13}$ wave phase shifts are
accessible to the experimental values. As shown in Fig.3, the QCD
renormalization effect gives an essential improvement on the D-wave phase
shifts as well. Particularly, it renders the $D_{13}$ and $D_{15}$ wave
phase shifts to have the right signs. But, this effect is not noteworthy for
the S-wave phase shifts. This explains why the previous investigations could
give some rather reasonable results for the S-wave phase shift. However, as
shown in our calculation, in order to get the desirable phase shifts for all
partial waves, it is necessary to incorporate the QCD renormalization effect
into the model used. In addition, to achieve such results, as mentioned in
Sect.3, the spin-orbital coupling effect is necessary to be suppressed. The
necessity of the suppression is separately illustrated in Fig.6 for the
P-wave phase shifts only. This is because the spin-orbit term in the
effective potential gives no contribution to the S-wave scattering and it
mainly affect the P-wave scattering. Fig.6 indicates that when the parameter 
$\gamma $ is taken to be the value $\gamma =0.3$ which makes the function $%
g(x,\stackrel{\rightharpoonup }{\rho })$ reach its real values, the $P_{01}$
wave and the $P_{11}$ wave phase shifts are far from the experimental ones,
but, when the $\gamma $ is getting larger, the absolute values of the $%
P_{01} $ wave and the $P_{11}$ wave phase shifts become to be comparable
with the experimental results. For the other P-waves, the phase shifts
evaluated at $\gamma =0.45$ are also better than those given at $\gamma =0.3$%
.

Let us turn to the $\overline{K}N$ scattering. At present, the low energy
elastic and inelastic experimental data for the $\overline{K}N$ scattering
are insufficient [31-33]. Therefore, the detailed partial wave analysis for
the scattering is almost absent. But it is a common conclusion suggested in
the previous investigations that the $K^{-}p$ interaction is strongly
attractive [34]. For the $\overline{K}N$ scattering in the $I=1$channel,
i.e., for the $K^{-}-neutron$ scattering, there is almost no available data
and different theoretical models give different predictions. In view of this
situation, our calculated results for the $\overline{K}N$ phase shifts can
only be viewed as a theoretical prediction [35,36]. It would be noted that
unlike the $KN$ interaction for which the exchanged part of the effective
potential generated from the t-channel OGEP is dominant, for the $\overline{K%
}N$ interaction, there is no such an exchanged potential. Instead, the
direct part of the s-channel OGEP plays an essential role and leads to an
attractive interaction as seen from the positiveness of the phase shifts
plotted in Fig.4. In the figure, the prediction for the P and D wave phase
shifts are simultaneously given as well. all the phase shifts in the figure
are presented in the momentum region less than 200 MeV where any resonance
could not appear. Here we take the $K^{-}p$ S-wave phase shift given in this
paper as an example to estimate the reasonability of our calculation. From
the relation $\sqrt{\pi }\mid a_{eff}^l\mid \simeq \mid f_{+}^l\mid $ [37]
where $a_{eff}^l$ is the effective scattering length and $f_{+}^l$ is the $l$%
-th wave scattering amplitude and the formulas given in (C.2) and (C.15), we
have $\delta _l^{IJ}\simeq kRe(a_{eff}^l).$ By this relation and the
calculated S-wave phase shift, it is found that the real part of the
scattering length is above 0.58 which just lies in the range shown in
Refs.[33,35]. In addition, we note that our calculation predicts a weaker
attraction for the $K^{-}-neutron$ interaction which is different from
previous result (see nucl-th/0004021).

At last, it should be emphasized that the quark potential model used in this
paper was established in the nonrelativistic approximation of order $p^2/m^2$%
, therefore, the calculated results are only valid for the $KN$ and $%
\overline{K}N$ elastic scattering in the low energy domain. The model used
can not give a complete description for the inelastic scattering and the
production of resonances which would appear for some higher partial waves in
the higher energy regime. To explore the $KN$ and $\overline{K}N$ inelastic
scattering in the higher energy regime, it is necessary to apply a
relativistic approach or a nonperturbative theory. Anyway, the investigation
based on the constituent quark model is meaningful as it is not only helpful
to understand $KN$ and $\overline{K}N$ interactions from the underlying
dynamics, but also provides a firm basis of studying five-quark bound states.

\section{Acknowledgment}

The authors would like to express their thanks to professor Dick Arndt for
his kind help. He offers us the experimental data which are useful in our
calculation. This project was supported in part by the National Natural
Science Foundation of China.

\section{Appendix A: The color-flavor-spin wave functions}

In general, the color singlet color state of the five quark cluster $(q^4%
\overline{s})$ or $(q^3\overline{q}s)$ may be built up by the color singlets
of the $N$-cluster $(qqq)$ and $K$-cluster $(q\overline{s})$ (or the $%
\overline{K}$-cluster $(\overline{q}s)$) or the color octets of the two
subclusters. Correspondingly, for the five quark cluster, there are two
classes of color-flavor-spin wave functions denoted by $\Psi _{TM\frac 12%
m}^{(1)}(1,2,3,4,5)$ and $\Psi _{TM\frac 12m}^{(2)}(1,2,3,4,5)$ which are
color singlets as a whole, but associated respectively with the color
singlets and the color octets of the two subclusters. In the function $\Psi
_{TM\frac 12m}^{(1)}(1,2,3,4,5),$ the color-flavor-spin (CFS) wave function $%
\Psi _{\frac 12M_1\frac 12m_s}^{(1)}(1,2,3)_N$ for the $N$-cluster which is
totally antisymmetric ( of the symmetry denoted by the Young diagram $%
[1^3]_{cfs}$ ) is constructed from the C-G coupling of $[1^3]_C\times
[3]_{FS}$ where $[1^3]_C$ and $[3]_{FS}$ are the Young diagrams denoting the
antisymmetric color singlet and the symmetric flavor-spin states
respectively. In the function $\Psi _{TM\frac 12m}^{(2)}(1,2,3,4,5)$, the
antisymmetric CFS wave function $\Psi _{\frac 12M_1\frac 12%
m_s}^{(1)}(1,2,3)_N$ for the $N$-cluster is given by the C-G coupling of $%
[21]_C\times [21]_{FS}$ where $[21]_C$ and $[21]_{FS}$ represent the color
octet state and the flavor-spin state of mixed symmetry respectively. The
explicit expressions of the wave functions mentioned above can easily be
written out by the familiar method given in the group theory, as displayed
in the following.

The first class of the CFS wave function in Eq.(12) for the whole system is

\begin{equation}
\Psi _{TM\frac 12m}^{(1)}(1,2,3,4,5)=\sum\limits_{M_1M_2}C_{\frac 12M_1\frac 
12M_2}^{TM}\Psi _{\frac 12M_1\frac 12m}^{(1)}(1,2,3)_N\Psi _{\frac 12%
M_200}^{(1)}(4,5)_K  \eqnum{A.1}
\end{equation}
where $\Psi _{\frac 12M_1\frac 12m}^{(1)}(1,2,3)_N$ , as mentioned before,
is the CFS wave function for the $N$-cluster and $\Psi _{\frac 12%
M_200}^{(1)}(4,5)_K$ is the CFS wave function for the $K$-cluster. They are
represented separately as 
\begin{equation}
\Psi _{\frac 12M_1\frac 12m_s}^{(1)}(1,2,3)_N=\xi _c^0(1,2,3)\chi _{\frac 12%
M_1\frac 12m_s}^{(1)}(1,2,3)  \eqnum{A.2}
\end{equation}
where 
\begin{equation}
\xi _c^0(1,2,3)=\frac 1{\sqrt{6}}\epsilon _{abc}q^a(1)q^b(2)q^c(3) 
\eqnum{A.3}
\end{equation}
represents the color singlet wave function of the $N$-cluster and 
\begin{equation}
\chi _{\frac 12M_1\frac 12m_s}^{(1)}(1,2,3)=\frac 1{\sqrt{2}}[\chi _{\frac 12%
M_1}^a(1,2,3)\varphi _{\frac 12m_s}^a(1,2,3)+\chi _{\frac 12%
M_1}^b(1,2,3)\varphi _{\frac 12m_s}^b(1,2,3)]  \eqnum{A.4}
\end{equation}
is the isospin-spin wave function of the $N$-cluster in which the isospin
wave functions $\chi _{\frac 12M_1}^a(1,2,3)$ and $\chi _{\frac 12%
M_1}^b(1,2,3)$ and the spin wave functions $\varphi _{\frac 12m_s}^a(1,2,3)$
and $\varphi _{\frac 12m_s}^a(1,2,3)$ are expressed as follows 
\begin{equation}
\begin{tabular}{l}
$\chi _{\frac 12M_1}^a(1,2,3)=\sum\limits_{m,m_{3,}m_1,m_2}C_{1m\frac 12%
m_3}^{\frac 12M_1}C_{\frac 12m_1\frac 12m_2}^{1m}\chi _{\frac 12m_1}(1)\chi
_{\frac 12m_2}(2)\chi _{\frac 12m_3}(3)$ \\ 
$\chi _{\frac 12M_1}^b(1,2,3)=\sum\limits_{m_1,m_2,m_3}C_{00\frac 12M_1}^{%
\frac 12M_1}C_{\frac 12m_1\frac 12m_2}^{00}\chi _{\frac 12m_1}(1)\chi _{%
\frac 12m_2}(2)\chi _{\frac 12m_3}(3)$ \\ 
$\varphi _{\frac 12m_s}^a(1,2,3)=\sum\limits_{m,m_3,m_1,m_2}C_{1m\frac 12%
m_3}^{\frac 12m_s}C_{\frac 12m_1\frac 12m_2}^{1m}\varphi _{\frac 12%
m_2}(1)\varphi _{\frac 12m_2}(2)\varphi _{\frac 12m_3}(3)$ \\ 
$\varphi _{\frac 12m_s}^b(1,2,3)=\sum\limits_{m_1,m_2,m_3}C_{00\frac 12m_s}^{%
\frac 12m_s}C_{\frac 12m_1\frac 12m_2}^{00}\varphi _{\frac 12m_2}(1)\varphi
_{\frac 12m_2}(2)\varphi _{\frac 12m_3}(3)$%
\end{tabular}
\eqnum{A.5}
\end{equation}
The CFS wave function of the $K$- cluster is 
\begin{equation}
\Psi _{\frac 12M00}^{(1)}(4,5)_K=C_0(4,5)\chi _{\frac 12M}(4,5)\varphi
_{00}(4,5)  \eqnum{A.6}
\end{equation}
where $C_0(4,5)$, $\chi _{\frac 12M}(4,5)$ and $\varphi _{00}(4,5)$ are the
color, isospin and spin wave functions, respectively. Since there is no
identical particles in the cluster, these wave functions are of the forms 
\begin{equation}
C_0(4,5)=\frac 1{\sqrt{3}}q^a(4)\overline{q}^a(5)  \eqnum{A.7}
\end{equation}
and 
\begin{equation}
\begin{tabular}{l}
$\chi _{\frac 12M}(4,5)=\sum\limits_{m_1\text{ }m_2}C_{\frac 12m_100}^{\frac 
12M}\chi _{\frac 12m_1}(4)\chi _{00}(5)$ \\ 
$\varphi _{00}(4,5)=\sum\limits_{m_1\text{ }m_2}C_{\frac 12m_1\frac 12%
m_2}^{00}\varphi _{\frac 12m_1}(4)\varphi _{\frac 12m_2}(5)$%
\end{tabular}
\eqnum{A.8}
\end{equation}
For the second class of the CFS wave function in Eq.(12), it can be
represented as 
\begin{equation}
\Psi _{TM\frac 12m}^{(2)}(1,2,3,4,5)=\sum\limits_{M_1M_2}\sum\limits_cC_{%
\frac 12M_1\frac 12M_2}^{TM_T}\Psi _{\frac 12M_1\frac 12m}^{(2)c}(1,2,3)_N%
\Psi _{\frac 12M00}^{(2)c}(4,5)_K  \eqnum{A.9}
\end{equation}
where $\Psi _{\frac 12M_1\frac 12m}^{(2)c}(1,2,3)_N$ and $\Psi _{\frac 12%
M00}^{(2)c}(4,5)_K$ are the second class of CFS wave functions for the $N$%
-cluster and the $K$-cluster respectively. Their expressions are shown in
the following. 
\begin{equation}
\Psi _{\frac 12M_1\frac 12m_s}^{(2)C}(1,2,3)_N=\frac 1{\sqrt{2}}[\xi
_c^A(1,2,3)\chi _{\frac 12M_1\frac 12m_s}^{(2)B}(1,2,3)-\xi _c^B(1,2,3)\chi
_{\frac 12M_1\frac 12m_s}^{(2)A}(1,2,3)]  \eqnum{A.10}
\end{equation}
where $\xi _c^A(1,2,3)$ and $\xi _c^B(1,2,3)$ are the color octet wave
functions given respectively by the Young-Tableau [211] and the
Young-Tableau [121] and $\chi _{\frac 12M_1\frac 12m_s}^{(2)A}(1,2,3)$ and $%
\chi _{\frac 12M_1\frac 12m_s}^{(2)B}(1,2,3)$ are the corresponding
isospin-spin wave functions. Their expressions are 
\begin{equation}
\begin{tabular}{l}
$\xi _c^A(1,2,3)=\frac 12\epsilon
_{ijb}[q^a(1)q^i(2)q^j(3)+q^a(2)q^i(1)q^j(3)]$ \\ 
$\xi _c^B(1,2,3)=\frac 1{2\sqrt{3}}\epsilon
_{ijb}[q^a(1)q^i(2)q^j(3)-q^a(2)q^i(1)q^j(3)-2\text{ }q^a(3)q^i(1)q^j(2)]$
\\ 
$\chi _{\frac 12M_1\frac 12m_s}^{(2)A}(1,2,3)=\frac 1{\sqrt{2}}[\chi _{\frac 
12M_1}^a(1,2,3)\varphi _{\frac 12m_s}^a(1,2,3)-\chi _{\frac 12%
M_1}^b(1,2,3)\varphi _{\frac 12m_s}^b(1,2,3)]$ \\ 
$\chi _{\frac 12M_1\frac 12m_s}^{(2)B}(1,2,3)=-\frac 1{\sqrt{2}}[\chi _{%
\frac 12M_1}^a(1,2,3)\varphi _{\frac 12m_s}^b(1,2,3)+\chi _{\frac 12%
M_1}^b(1,2,3)\varphi _{\frac 12m_s}^a(1,2,3)]$%
\end{tabular}
\eqnum{A.11}
\end{equation}
The second class of the CFS wave function for the $K$(or $\overline{K}$%
)-cluster is as follows 
\begin{equation}
\Psi _{\frac 12M00}^{(2)c}(4,5)_\pi =C_a^b(4,5)\chi _{\frac 12M}(4,5)\varphi
_{00}(4,5)  \eqnum{A.12}
\end{equation}
where 
\begin{equation}
C_a^b(4,5)=q^b(4)q_a(5)-\frac 13\delta _a^bq^c(4)q_c(5)  \eqnum{A.13}
\end{equation}
is the color octet for the $K(\overline{K})$ cluster and the other two
functions $\chi _{\frac 12M}(4,5)$, $\varphi _{00}(4,5)$ are the same as in
(A.8).

\section{Appendix B: The effective $KN$ and $\overline{K}N$ interaction
potentials}

In this appendix, we show the nonlocal effective interaction potentials of
the $KN$ and $\overline{K}N$ systems which are derived from the interquark
potentials and the RGM.

The $KN$ nonlocal effective potential $V_t(\stackrel{\rightharpoonup }{\rho }%
,\stackrel{\rightharpoonup }{\rho ^{\prime }})$ which is derived from the
t-channel OGEP written in Eq.(4) is divided into two parts: the direct part $%
V_t^D(\stackrel{\rightharpoonup }{\rho },\stackrel{\rightharpoonup }{\rho
^{\prime }})$ and the exchanged part $V_t^{ex}(\stackrel{\rightharpoonup }{%
\rho },\stackrel{\rightharpoonup }{\rho ^{\prime }})$:

\begin{equation}
V_t(\stackrel{\rightharpoonup }{\rho },\stackrel{\rightharpoonup }{\rho
^{\prime }})=V_t^D(\stackrel{\rightharpoonup }{\rho },\stackrel{%
\rightharpoonup }{\rho ^{\prime }})-V_{ex}^t(\stackrel{\rightharpoonup }{%
\rho },\stackrel{\rightharpoonup }{\rho ^{\prime }})  \eqnum{B.1}
\end{equation}
where 
\begin{equation}
V_t^{ex}(\stackrel{\rightharpoonup }{\rho },\stackrel{\rightharpoonup }{\rho
^{\prime }})=V_t^{ex}(\stackrel{\rightharpoonup }{\rho },\stackrel{%
\rightharpoonup }{\rho ^{\prime }})^{14}+V_t^{ex}(\stackrel{\rightharpoonup 
}{\rho },\stackrel{\rightharpoonup }{\rho ^{\prime }})^{24}+V_t^{ex}(%
\stackrel{\rightharpoonup }{\rho },\stackrel{\rightharpoonup }{\rho ^{\prime
}})^{34}  \eqnum{B.2}
\end{equation}
here the superscript $ab=14,24$ or $34$ designate which pair of quarks
interchange. Each part of the potential contains several terms as shown in
the following 
\begin{equation}
\begin{tabular}{l}
$V_t^D(\stackrel{\rightharpoonup }{\rho },\stackrel{\rightharpoonup }{\rho
^{\prime }})=V_{15}^D(\stackrel{\rightharpoonup }{\rho },\stackrel{%
\rightharpoonup }{\rho ^{\prime }})+V_{25}^D(\stackrel{\rightharpoonup }{%
\rho },\stackrel{\rightharpoonup }{\rho ^{\prime }})+V_{35}^D(\stackrel{%
\rightharpoonup }{\rho },\stackrel{\rightharpoonup }{\rho ^{\prime }})$ \\ 
$+V_{14}^D(\stackrel{\rightharpoonup }{\rho },\stackrel{\rightharpoonup }{%
\rho ^{\prime }})+V_{24}^D(\stackrel{\rightharpoonup }{\rho },\stackrel{%
\rightharpoonup }{\rho ^{\prime }})+V_{34}^D(\stackrel{\rightharpoonup }{%
\rho },\stackrel{\rightharpoonup }{\rho ^{\prime }})$%
\end{tabular}
\eqnum{B.3}
\end{equation}
\begin{equation}
\begin{tabular}{l}
$V_t^{ex}(\stackrel{\rightharpoonup }{\rho },\stackrel{\rightharpoonup }{%
\rho ^{\prime }})^{ab}=V_{14}^{ex}(\stackrel{\rightharpoonup }{\rho },%
\stackrel{\rightharpoonup }{\rho ^{\prime }})^{ab}+V_{24}^{ex}(\stackrel{%
\rightharpoonup }{\rho },\stackrel{\rightharpoonup }{\rho ^{\prime }}%
)^{ab}+V_{34}^{ex}(\stackrel{\rightharpoonup }{\rho },\stackrel{%
\rightharpoonup }{\rho ^{\prime }})^{ab}+V_{12}^{ex}(\stackrel{%
\rightharpoonup }{\rho },\stackrel{\rightharpoonup }{\rho ^{\prime }}%
)^{ab}+V_{23}^{ex}(\stackrel{\rightharpoonup }{\rho },\stackrel{%
\rightharpoonup }{\rho ^{\prime }})^{ab}$ \\ 
$+V_{15}^{ex}(\stackrel{\rightharpoonup }{\rho },\stackrel{\rightharpoonup }{%
\rho ^{\prime }})^{ab}+V_{25}^{ex}(\stackrel{\rightharpoonup }{\rho },%
\stackrel{\rightharpoonup }{\rho ^{\prime }})^{ab}+V_{35}^{ex}(\stackrel{%
\rightharpoonup }{\rho },\stackrel{\rightharpoonup }{\rho ^{\prime }}%
)^{ab}+V_{45}^{ex}(\stackrel{\rightharpoonup }{\rho },\stackrel{%
\rightharpoonup }{\rho ^{\prime }})^{ab}+V_{13}^{ex}(\stackrel{%
\rightharpoonup }{\rho },\stackrel{\rightharpoonup }{\rho ^{\prime }})^{ab}$%
\end{tabular}
\eqnum{B.4}
\end{equation}
the subscript in each term on the right hand sides (RHS) of (B.3) and (B.4)
marks the two interacting quarks: one in the $N$-cluster and another in the $%
K$-cluster. The terms $V_{ij}^D(\stackrel{\rightharpoonup }{\rho },\stackrel{%
\rightharpoonup }{\rho ^{\prime }})$ and $V_{ij}^{ex}(\stackrel{%
\rightharpoonup }{\rho },\stackrel{\rightharpoonup }{\rho ^{\prime }})^{ab}$
are derived by the RGM in such a way 
\begin{equation}
V_{ij}^D(\stackrel{\rightharpoonup }{\rho },\stackrel{\rightharpoonup }{\rho
^{\prime }})=\int \prod_{i=1}^5\frac{d\stackrel{\rightharpoonup }{p_k}}{%
(2\pi )^3}\frac{d\stackrel{\rightharpoonup }{p_k^{\prime }}}{(2\pi )^3}%
\langle R(\stackrel{\rightharpoonup }{p_1},\stackrel{\rightharpoonup }{p_2},%
\stackrel{\rightharpoonup }{p_3},\stackrel{\rightharpoonup }{p_4},\stackrel{%
\rightharpoonup }{p_5};\stackrel{\rightharpoonup }{\rho })\mid V_{ij}^t\mid
R(\stackrel{\rightharpoonup }{p_1^{\prime }},\stackrel{\rightharpoonup }{%
p_2^{\prime }},\stackrel{\rightharpoonup }{p_3^{\prime }},\stackrel{%
\rightharpoonup }{p_4^{\prime }},\stackrel{\rightharpoonup }{p_5^{\prime }};%
\stackrel{\rightharpoonup }{\rho ^{\prime }})\rangle  \eqnum{B.5}
\end{equation}
and 
\begin{equation}
V_{ij}^{ex}(\stackrel{\rightharpoonup }{\rho },\stackrel{\rightharpoonup }{%
\rho ^{\prime }})^{ab}=\int \prod_{i=1}^5\frac{d\stackrel{\rightharpoonup }{%
p_k}}{(2\pi )^3}\frac{d\stackrel{\rightharpoonup }{p_k^{\prime }}}{(2\pi )^3}%
\langle R(\stackrel{\rightharpoonup }{p_1},\stackrel{\rightharpoonup }{p_2},%
\stackrel{\rightharpoonup }{p_3},\stackrel{\rightharpoonup }{p_4},\stackrel{%
\rightharpoonup }{p_5};\stackrel{\rightharpoonup }{\rho })\mid
V_{ij}^tP_{ab}\mid R(\stackrel{\rightharpoonup }{p_1^{\prime }},\stackrel{%
\rightharpoonup }{p_2^{\prime }},\stackrel{\rightharpoonup }{p_3^{\prime }},%
\stackrel{\rightharpoonup }{p_4^{\prime }},\stackrel{\rightharpoonup }{%
p_5^{\prime }};\stackrel{\rightharpoonup }{\rho ^{\prime }})\rangle \text{ ,}
\eqnum{B.6}
\end{equation}
where the quark potential $V_{ij}^t$was denoted in Eq.(4) and the position
space wave function was given in Eq.(14).

First we describe the ten terms on the RHS of (B.4). By introducing the
following functions: 
\begin{equation}
\begin{tabular}{l}
$g(x,\stackrel{\rightharpoonup }{\rho })=\int \frac{d^3r}{4\pi r}e^{-xr^2+x%
\stackrel{\rightharpoonup }{r}\cdot \stackrel{\rightharpoonup }{\rho }}$ \\ 
$f_{24}^t(\stackrel{\rightharpoonup }{\rho },\stackrel{\rightharpoonup }{%
\rho ^{\prime }})_{ex}=e^{-\frac{\stackrel{\rightharpoonup }{\rho }\cdot 
\stackrel{\rightharpoonup }{\rho ^{\prime }}}{2b^2}-\frac{\stackrel{%
\rightharpoonup }{\rho }^2}{8b^2}-\frac{3\text{ }\beta _2}{4b^2}(\stackrel{%
\rightharpoonup }{\rho }-\stackrel{\rightharpoonup }{\rho ^{\prime }})^2}$
\\ 
$f_{25}^t(\stackrel{\rightharpoonup }{\rho },\stackrel{\rightharpoonup }{%
\rho ^{\prime }})_{ex}=e^{-\frac{\stackrel{\rightharpoonup }{\rho }\cdot 
\stackrel{\rightharpoonup }{\rho ^{\prime }}}{2b^2}-\frac{\alpha _2}{4b^2}(%
\stackrel{\rightharpoonup }{\rho }+\stackrel{\rightharpoonup }{\rho ^{\prime
}})^2-\frac{3\beta _2}{4\text{ }b^2}(\stackrel{\rightharpoonup }{\rho }-%
\stackrel{\rightharpoonup }{\rho ^{\prime }})^2}$ \\ 
$f_{15}^t(\stackrel{\rightharpoonup }{\rho },\stackrel{\rightharpoonup }{%
\rho ^{\prime }})_{ex}=e^{-\frac{\stackrel{\rightharpoonup }{\rho }\cdot 
\stackrel{\rightharpoonup }{\rho ^{\prime }}}{2b^2}-\frac{\alpha _2}{4b^2}%
\stackrel{\rightharpoonup }{\rho }^2-\frac{3\beta _2}{4b^2}(\stackrel{%
\rightharpoonup }{\rho }-\stackrel{\rightharpoonup }{\rho ^{\prime }})^2}$
\\ 
$f_{14}^t(\stackrel{\rightharpoonup }{\rho },\stackrel{\rightharpoonup }{%
\rho ^{\prime }})_{ex}=e^{-\frac{\stackrel{\rightharpoonup }{\rho }\cdot 
\stackrel{\rightharpoonup }{\rho ^{\prime }}}{2b^2}-\frac{\text{ }1}{8b^2}(%
\stackrel{\rightharpoonup }{\rho }-\stackrel{\rightharpoonup }{\rho ^{\prime
}})^2-\frac{3\beta _2}{4b^2}(\stackrel{\rightharpoonup }{\rho }-\stackrel{%
\rightharpoonup }{\rho ^{\prime }})^2}$ \\ 
$f_{23}^t(\stackrel{\rightharpoonup }{\rho },\stackrel{\rightharpoonup }{%
\rho ^{\prime }})_{ex}=e^{-\frac{\stackrel{\rightharpoonup }{\rho }\cdot 
\stackrel{\rightharpoonup }{\rho ^{\prime }}}{2b^2}-\frac{3\beta _2}{4\text{ 
}b^2}(\stackrel{\rightharpoonup }{\rho }-\stackrel{\rightharpoonup }{\rho
^{\prime }})^2}$%
\end{tabular}
\eqnum{B.7}
\end{equation}
the exchanged terms of the potential $V_t^{ex}(\stackrel{\rightharpoonup }{%
\rho },\stackrel{\rightharpoonup }{\rho ^{\prime }})^{14}$ can be written
as: 
\begin{equation}
\begin{tabular}{l}
$V_{24}^{ex}(\stackrel{\rightharpoonup }{\rho },\stackrel{\rightharpoonup }{%
\rho ^{\prime }})^{14}=\frac{4\pi \alpha _sC_{24}^t}{(2\pi b^2)^{3/2}}%
f_{24}^t(\stackrel{\rightharpoonup }{\rho },\stackrel{\rightharpoonup }{\rho
^{\prime }})_{ex}\{g(\frac 1{2b^2},\stackrel{\rightharpoonup }{\rho })-\frac 
1{4m_1^2}(1+\stackrel{\rightharpoonup }{\sigma }_2\cdot \stackrel{%
\rightharpoonup }{\sigma }_4)-\frac 1{4m_1^2b^2}[\frac{\stackrel{%
\rightharpoonup }{\rho }^2}{4b^2}$ \\ 
$-\frac 1{b^2}(\frac{2\beta _2-1}2\stackrel{\rightharpoonup }{\rho }-\beta _2%
\stackrel{\rightharpoonup }{\rho ^{\prime }})^2-\frac 12\stackrel{%
\rightharpoonup }{\sigma }_2\cdot \stackrel{\rightharpoonup }{\sigma }_4+i%
\frac{\beta _2}{8\text{ }b^2}(1+\gamma )(\stackrel{\rightharpoonup }{\rho }%
\times \stackrel{\rightharpoonup }{\rho ^{\prime }})\cdot (\stackrel{%
\rightharpoonup }{\sigma }_2-\stackrel{\rightharpoonup }{\sigma }_4)$ \\ 
$+\frac 1{16b^2}(1+\gamma ^2)\stackrel{\rightharpoonup }{\rho }\cdot 
\stackrel{\rightharpoonup }{\sigma }_2\stackrel{\rightharpoonup }{\rho }%
\cdot \stackrel{\rightharpoonup }{\sigma }_4]g(\frac 1{2b^2},\stackrel{%
\rightharpoonup }{\rho })\}$%
\end{tabular}
\eqnum{B.8}
\end{equation}
\begin{equation}
\begin{tabular}{l}
$V_{34}^{ex}(\stackrel{\rightharpoonup }{R},\stackrel{\rightharpoonup }{%
R^{\prime }})^{14}=\frac{4\pi \alpha _sC_{34}^t}{(2\pi b^2)^{3/2}}f_{24}^t(%
\stackrel{\rightharpoonup }{\rho },\stackrel{\rightharpoonup }{\rho ^{\prime
}})_{ex}\{g(\frac 1{2\text{ }b^2},\stackrel{\rightharpoonup }{\rho })-\frac 1%
{4m_1^2}(1+\stackrel{\rightharpoonup }{\sigma }_3\cdot \stackrel{%
\rightharpoonup }{\sigma }_4)-\frac 1{4m_1^2b^2}[\frac{\stackrel{%
\rightharpoonup }{\rho }^2}{4b^2}$ \\ 
$-\frac 1{b^2}(\frac{2\beta _2-1}2\stackrel{\rightharpoonup }{\rho }-\beta _2%
\stackrel{\rightharpoonup }{\rho ^{\prime }})^2-\frac 12\stackrel{%
\rightharpoonup }{\sigma }_3\cdot \stackrel{\rightharpoonup }{\sigma }_4+i%
\frac{\beta _2}{8b^2}(1+\gamma )(\stackrel{\rightharpoonup }{\rho }\times 
\stackrel{\rightharpoonup }{\rho ^{\prime }})\cdot (\stackrel{%
\rightharpoonup }{\sigma }_3-\stackrel{\rightharpoonup }{\sigma }_4)$ \\ 
$+\frac 1{16b^2}(1+\gamma ^2)\stackrel{\rightharpoonup }{\rho }\cdot 
\stackrel{\rightharpoonup }{\sigma }_3\stackrel{\rightharpoonup }{\rho }%
\cdot \stackrel{\rightharpoonup }{\sigma }_4]g(\frac 1{2b^2},\stackrel{%
\rightharpoonup }{\rho })\}$%
\end{tabular}
\eqnum{B.9}
\end{equation}
\begin{equation}
\begin{tabular}{l}
$V_{14}^{ex}(\stackrel{\rightharpoonup }{\rho },\stackrel{\rightharpoonup }{%
\rho ^{\prime }})^{14}=\frac{4\pi \alpha _sC_{14}^t}{(2\pi b^2)^{3/2}}%
f_{14}^t(\stackrel{\rightharpoonup }{\rho },\stackrel{\rightharpoonup }{\rho
^{\prime }})_{ex}\{g(\frac 1{2b^2},\stackrel{\rightharpoonup }{\rho }-%
\stackrel{\rightharpoonup }{\rho ^{\prime }})-\frac 1{4m_1^2}(1+\stackrel{%
\rightharpoonup }{\sigma }_1\cdot \stackrel{\rightharpoonup }{\sigma }_4)$
\\ 
$-\frac 1{4m_1^2b^2}[-\frac{(\beta _2-\beta _1)^2}{4b^2}(\stackrel{%
\rightharpoonup }{\rho }-\stackrel{\rightharpoonup }{\rho ^{\prime }})^2+%
\frac{(\stackrel{\rightharpoonup }{\rho }+\stackrel{\rightharpoonup }{\rho
^{\prime }})^2}{b^2}-9-\frac 12\stackrel{\rightharpoonup }{\sigma }_1\cdot 
\stackrel{\rightharpoonup }{\sigma }_4+i\frac 1{b^2}(1+\gamma )(\stackrel{%
\rightharpoonup }{\rho }\times \stackrel{\rightharpoonup }{\rho ^{\prime }}%
)\cdot (\stackrel{\rightharpoonup }{\sigma }_1+\stackrel{\rightharpoonup }{%
\sigma }_4)$ \\ 
$+\frac 1{16b^2}(1+\gamma ^2)(\stackrel{\rightharpoonup }{\rho }-\stackrel{%
\rightharpoonup }{\rho ^{\prime }})\cdot \stackrel{\rightharpoonup }{\sigma }%
_1(\stackrel{\rightharpoonup }{\rho }-\stackrel{\rightharpoonup }{\rho
^{\prime }})\cdot \stackrel{\rightharpoonup }{\sigma }_4]g(\frac 1{2b^2},%
\stackrel{\rightharpoonup }{\rho }-\stackrel{\rightharpoonup }{\rho ^{\prime
}})\}$%
\end{tabular}
\eqnum{B.10}
\end{equation}
\begin{equation}
\begin{tabular}{l}
$V_{12}^{ex}(\stackrel{\rightharpoonup }{\rho },\stackrel{\rightharpoonup }{%
\rho ^{\prime }})^{14}=\frac{4\pi \alpha _sC_{12}^t}{(2\pi b^2)^{3/2}}%
f_{24}^t(\stackrel{\rightharpoonup }{\rho ^{\prime }},\stackrel{%
\rightharpoonup }{\rho })_{ex}\{g(\frac 1{2b^2},\stackrel{\rightharpoonup }{%
\rho ^{\prime }})-\frac 1{4m_1^2}(1+\stackrel{\rightharpoonup }{\sigma }%
_1\cdot \stackrel{\rightharpoonup }{\sigma }_2)-\frac 1{4m_1^2b^2}[\frac{%
\stackrel{\rightharpoonup }{\rho ^{\prime }}^2}{4b^2}$ \\ 
$-\frac 1{b^2}(\frac{2\beta _2-1}2\stackrel{\rightharpoonup }{\rho ^{\prime }%
}-\beta _2\stackrel{\rightharpoonup }{\rho })^2-\frac 12\stackrel{%
\rightharpoonup }{\sigma }_1\cdot \stackrel{\rightharpoonup }{\sigma }_2+i%
\frac{\beta _2}{8b^2}(1+\gamma )(\stackrel{\rightharpoonup }{\rho ^{\prime }}%
\times \stackrel{\rightharpoonup }{\rho })\cdot (\stackrel{\rightharpoonup }{%
\sigma }_1-\stackrel{\rightharpoonup }{\sigma }_2)$ \\ 
$+\frac 1{16b^2}(1+\gamma ^2)\stackrel{\rightharpoonup }{\rho ^{\prime }}%
\cdot \stackrel{\rightharpoonup }{\sigma }_1\stackrel{\rightharpoonup }{\rho
^{\prime }}\cdot \stackrel{\rightharpoonup }{\sigma }_2]g(\frac 1{2b^2},%
\stackrel{\rightharpoonup }{\rho ^{\prime }})\}$%
\end{tabular}
\eqnum{B.11}
\end{equation}
\begin{equation}
\begin{tabular}{l}
$V_{13}^{ex}(\stackrel{\rightharpoonup }{\rho },\stackrel{\rightharpoonup }{%
\rho ^{\prime }})^{14}=\frac{4\pi \alpha _sC_{13}^t}{(2\pi b^2)^{3/2}}%
f_{24}^t(\stackrel{\rightharpoonup }{\rho ^{\prime }},\stackrel{%
\rightharpoonup }{\rho })_{ex}\{g(\frac 1{2\text{ }b^2},\stackrel{%
\rightharpoonup }{\rho ^{\prime }})-\frac 1{4m_1^2}(1+\stackrel{%
\rightharpoonup }{\sigma }_1\cdot \stackrel{\rightharpoonup }{\sigma }_3)$
\\ 
$-\frac 1{b^2}(\frac{2\beta _2-1}2\stackrel{\rightharpoonup }{\rho ^{\prime }%
}-\beta _2\stackrel{\rightharpoonup }{\rho })^2-\frac 12\stackrel{%
\rightharpoonup }{\sigma }_1\cdot \stackrel{\rightharpoonup }{\sigma }_3+i%
\frac{\beta _2}{8b^2}(1+\gamma )(\stackrel{\rightharpoonup }{\rho ^{\prime }}%
\times \stackrel{\rightharpoonup }{\rho })\cdot (\stackrel{\rightharpoonup }{%
\sigma }_1-\stackrel{\rightharpoonup }{\sigma }_3)$ \\ 
$+\frac 1{16b^2}(1+\gamma ^2)\stackrel{\rightharpoonup }{\rho ^{\prime }}%
\cdot \stackrel{\rightharpoonup }{\sigma }_1\stackrel{\rightharpoonup }{\rho
^{\prime }}\cdot \stackrel{\rightharpoonup }{\sigma }_3]g(\frac 1{2b^2},%
\stackrel{\rightharpoonup }{\rho ^{\prime }})\}$%
\end{tabular}
\eqnum{B.12}
\end{equation}

\begin{equation}
V_{23}^{ex}(\stackrel{\rightharpoonup }{\rho },\stackrel{\rightharpoonup }{%
\rho ^{\prime }})^{14}=\frac{4\pi \alpha _sC_{23}^t}{(2\pi b^2)^{3/2}}%
f_{23}^t(\stackrel{\rightharpoonup }{\rho },\stackrel{\rightharpoonup }{\rho
^{\prime }})_{ex}[b^2-\frac 1{4m_1^2}-\frac{\stackrel{\rightharpoonup }{%
\sigma }_2\cdot \stackrel{\rightharpoonup }{\sigma }_3}{6m_1^2}+\frac{\beta
_2^2}{4m_1^2b^2}(1+\gamma ^2)(\stackrel{\rightharpoonup }{\rho }-\stackrel{%
\rightharpoonup }{\rho ^{\prime }})^2]  \eqnum{B.13}
\end{equation}
\begin{equation}
\begin{tabular}{l}
$V_{15}^{ex}(\stackrel{\rightharpoonup }{\rho },\stackrel{\rightharpoonup }{%
\rho ^{\prime }})^{14}=\frac{4\pi \alpha _sC_{15}^t}{(\pi b^2/\alpha
_1)^{3/2}}f_{15}^t(\stackrel{\rightharpoonup }{\rho },\stackrel{%
\rightharpoonup }{\rho ^{\prime }})_{ex}\{g(\frac{\alpha _2}{b^2},\stackrel{%
\rightharpoonup }{\rho })-(\frac{m_1^2+m_2^2}{8m_1^2m_2^2}+\frac{\stackrel{%
\rightharpoonup }{\sigma }_1\cdot \stackrel{\rightharpoonup }{\sigma }_5}{%
4m_1m_2})-\frac 1{4(m_1+m_2)^2b^2}[\frac 6{\alpha _1}$ \\ 
$-\frac{(\alpha _1\stackrel{\rightharpoonup }{\rho }-\beta _1\stackrel{%
\rightharpoonup }{\rho }+\beta _1\stackrel{\rightharpoonup }{\rho ^{\prime }}%
)^2}{\alpha _1^2b^2}-\frac{\alpha _2\beta _1}{4\alpha _1b^2}i(1-\gamma )(%
\stackrel{\rightharpoonup }{\rho }\times \stackrel{\rightharpoonup }{\rho
^{\prime }})\cdot [(1+\frac{m_2}{m_1})\stackrel{\rightharpoonup }{\sigma }%
_1-(1+\frac{m_1}{m_2})\stackrel{\rightharpoonup }{\sigma }_5]]g(\frac{\alpha
_2}{b^2},\stackrel{\rightharpoonup }{\rho })$ \\ 
$-\frac 1{4m_1m_2b^2}[6\alpha _2-\frac{\alpha _2(\alpha _1-\alpha _2)\beta _1%
}{\alpha _1b^2}\stackrel{\rightharpoonup }{\rho }\cdot \stackrel{%
\rightharpoonup }{\rho ^{\prime }}-\frac{\alpha _2(\alpha _1-\beta
_1)^2-\alpha _2\beta _1\beta _2}{\alpha _1b^2}\stackrel{\rightharpoonup }{%
\rho }^2-\alpha _2\stackrel{\rightharpoonup }{\sigma }_1\cdot \stackrel{%
\rightharpoonup }{\sigma }_5$ \\ 
$-\frac{\alpha _2^2}{4\text{ }b^2}(1-\gamma ^2)\stackrel{\rightharpoonup }{%
\rho }\cdot \stackrel{\rightharpoonup }{\sigma }_1\stackrel{\rightharpoonup 
}{\rho }\cdot \stackrel{\rightharpoonup }{\sigma }_5]g(\frac{\alpha _2}{b^2},%
\stackrel{\rightharpoonup }{\rho })\}$%
\end{tabular}
\eqnum{B.14}
\end{equation}
\begin{equation}
\begin{tabular}{l}
$V_{25}^{ex}(\stackrel{\rightharpoonup }{\rho },\stackrel{\rightharpoonup }{%
\rho ^{\prime }})^{14}=\frac{4\pi \alpha _sC_{25}^t}{(\pi b^2/\alpha
_1)^{3/2}}f_{25}^t(\stackrel{\rightharpoonup }{\rho },\stackrel{%
\rightharpoonup }{\rho ^{\prime }})_{ex}\{g(\frac{\alpha _2}{b^2},\stackrel{%
\rightharpoonup }{\rho }+\stackrel{\rightharpoonup }{\rho ^{\prime }})-(%
\frac{m_1^2+m_2^2}{8m_1^2m_2^2}+\frac{\stackrel{\rightharpoonup }{\sigma }%
_2\cdot \stackrel{\rightharpoonup }{\sigma }_5}{4m_1m_2})$ \\ 
$-\frac 1{4(m_1+m_2)^2b^2}[\frac 6{\alpha _1}-\frac{(\beta _2-\alpha _2)^2}{%
\alpha _1^2b^2}(\stackrel{\rightharpoonup }{\rho }-\stackrel{\rightharpoonup 
}{\rho ^{\prime }})^2+\frac{\alpha _2(\alpha _1-\beta _1)}{2\alpha _1b^2}%
i(1-\gamma )(\stackrel{\rightharpoonup }{\rho }\times \stackrel{%
\rightharpoonup }{\rho ^{\prime }})\cdot [(1+\frac{m_2}{m_1})\stackrel{%
\rightharpoonup }{\sigma }_2$ \\ 
$-(1+\frac{m_1}{m_2})\stackrel{\rightharpoonup }{\sigma }_5]]g(\frac{\alpha
_2}{b^2},\stackrel{\rightharpoonup }{\rho }+\stackrel{\rightharpoonup }{\rho
^{\prime }})]]g(\frac{\alpha _2}{b^2},\stackrel{\rightharpoonup }{\rho }+%
\stackrel{\rightharpoonup }{\rho ^{\prime }})-\frac 1{4m_1m_2b^2}[6\alpha _2+%
\frac{\alpha _2}{b^2}(\alpha _1-\frac{\beta _1}{\alpha _1}$ \\ 
$-2\alpha _2\beta _2)(\stackrel{\rightharpoonup }{\rho }-\stackrel{%
\rightharpoonup }{\rho ^{\prime }})^2-\alpha _2\stackrel{\rightharpoonup }{%
\sigma }_2\cdot \stackrel{\rightharpoonup }{\sigma }_5+\frac{\alpha _2^2}{%
2b^2}i(1+\gamma )(\stackrel{\rightharpoonup }{\rho }\times \stackrel{%
\rightharpoonup }{\rho ^{\prime }})\cdot [(2+\frac{m_2}{m_1})\stackrel{%
\rightharpoonup }{\sigma }_2$ \\ 
$+(2+\frac{m_1}{m_2})\stackrel{\rightharpoonup }{\sigma }_5]-\frac{\alpha
_2^2}{4b^2}(1-\gamma ^2)(\stackrel{\rightharpoonup }{\rho }+\stackrel{%
\rightharpoonup }{\rho ^{\prime }})\cdot \stackrel{\rightharpoonup }{\sigma }%
_2(\stackrel{\rightharpoonup }{\rho }+\stackrel{\rightharpoonup }{\rho
^{\prime }})\cdot \stackrel{\rightharpoonup }{\sigma }_5]g(\frac{\alpha _2}{%
b^2},\stackrel{\rightharpoonup }{\rho }+\stackrel{\rightharpoonup }{\rho
^{\prime }})\}$%
\end{tabular}
\eqnum{B.15}
\end{equation}
\begin{equation}
\begin{tabular}{l}
$V_{35}^{ex}(\stackrel{\rightharpoonup }{\rho },\stackrel{\rightharpoonup }{%
\rho ^{\prime }})^{14}=\frac{4\pi \alpha _sC_{35}^t}{(\pi b^2/\alpha
_1)^{3/2}}f_{25}^t(\stackrel{\rightharpoonup }{\rho },\stackrel{%
\rightharpoonup }{\rho ^{\prime }})_{ex}\{g(\frac{\alpha _2}{b^2},\stackrel{%
\rightharpoonup }{\rho }+\stackrel{\rightharpoonup }{\rho ^{\prime }})-(%
\frac{m_1^2+m_2^2}{8m_1^2m_2^2}+\frac{\stackrel{\rightharpoonup }{\sigma }%
_3\cdot \stackrel{\rightharpoonup }{\sigma }_5}{4m_1m_2})$ \\ 
$-\frac 1{4(m_1+m_2)^2b^2}[\frac 6{\alpha _1}-\frac{(\beta _2-\alpha _2)^2}{%
\alpha _1^2b^2}(\stackrel{\rightharpoonup }{\rho }-\stackrel{\rightharpoonup 
}{\rho ^{\prime }})^2+\frac{\alpha _2(\alpha _1-\beta _1)}{2\alpha _1b^2}%
i(1-\gamma )(\stackrel{\rightharpoonup }{\rho }\times \stackrel{%
\rightharpoonup }{\rho ^{\prime }})\cdot [(1+\frac{m_2}{m_1})\stackrel{%
\rightharpoonup }{\sigma }_3$ \\ 
$-(2+\frac{m_1}{m_2})\stackrel{\rightharpoonup }{\sigma }_5]]g(\frac{\alpha
_2}{b^2},\stackrel{\rightharpoonup }{\rho }+\stackrel{\rightharpoonup }{\rho
^{\prime }})-\frac 1{4m_1m_2b^2}[6\alpha _2+\frac{\alpha _2}{b^2}(\alpha _1-%
\frac{\beta _1}{\alpha _1}-2\alpha _2\beta _2)(\stackrel{\rightharpoonup }{%
\rho }-\stackrel{\rightharpoonup }{\rho ^{\prime }})^2$ \\ 
$-\alpha _2\stackrel{\rightharpoonup }{\sigma }_3\cdot \stackrel{%
\rightharpoonup }{\sigma }_5+\frac{\alpha _2^2}{2\text{ }b^2}i(1+\gamma )(%
\stackrel{\rightharpoonup }{\rho }\times \stackrel{\rightharpoonup }{\rho
^{\prime }})\cdot [(2+\frac{m_2}{m_1})\stackrel{\rightharpoonup }{\sigma }%
_3+(2+\frac{m_1}{m_2})\stackrel{\rightharpoonup }{\sigma }_5]$ \\ 
$-\frac{\alpha _2^2}{4b^2}(1-\gamma ^2)(\stackrel{\rightharpoonup }{\rho }+%
\stackrel{\rightharpoonup }{\rho ^{\prime }})\cdot \stackrel{\rightharpoonup 
}{\sigma }_3(\stackrel{\rightharpoonup }{\rho }+\stackrel{\rightharpoonup }{%
\rho ^{\prime }})\cdot \stackrel{\rightharpoonup }{\sigma }_5]g(\frac{\alpha
_2}{b^2},\stackrel{\rightharpoonup }{\rho }+\stackrel{\rightharpoonup }{\rho
^{\prime }})\}$%
\end{tabular}
\eqnum{B.16}
\end{equation}
\begin{equation}
\begin{tabular}{l}
$V_{45}^{ex}(\stackrel{\rightharpoonup }{\rho },\stackrel{\rightharpoonup }{%
\rho ^{\prime }})^{14}=\frac{4\pi \alpha _sC_{45}^t}{(\pi b^2/\alpha
_1)^{3/2}}f_{15}^t(\stackrel{\rightharpoonup }{\rho ^{\prime }},\stackrel{%
\rightharpoonup }{\rho })_{ex}\{g(\frac{\alpha _2}{b^2},\stackrel{%
\rightharpoonup }{\rho ^{\prime }})-(\frac{m_1^2+m_2^2}{8m_1^2m_2^2}+\frac{%
\stackrel{\rightharpoonup }{\sigma }_4\cdot \stackrel{\rightharpoonup }{%
\sigma }_5}{4m_1m_2})-\frac 1{4(m_1+m_2)^2b^2}[\frac 6{\alpha _1}$ \\ 
$-\frac{(\alpha _1\stackrel{\rightharpoonup }{\rho ^{\prime }}-\beta _1%
\stackrel{\rightharpoonup }{\rho ^{\prime }}+\beta _1\stackrel{%
\rightharpoonup }{\rho })^2}{\alpha _1^2b^2}-\frac{\alpha _2\beta _1}{%
4\alpha _1b^2}i(1-\gamma )(\stackrel{\rightharpoonup }{\rho ^{\prime }}%
\times \stackrel{\rightharpoonup }{\rho })\cdot [(1+\frac{m_2}{m_1})%
\stackrel{\rightharpoonup }{\sigma }_4-(1+\frac{m_1}{m_2})\stackrel{%
\rightharpoonup }{\sigma }_5]]g(\frac{\alpha _2}{b^2},\stackrel{%
\rightharpoonup }{\rho ^{\prime }})$ \\ 
$-\frac 1{4m_1m_2b^2}[6\alpha _2-\frac{\alpha _2(\alpha _1-\alpha _2)\beta _1%
}{\alpha _1b^2}\stackrel{\rightharpoonup }{\rho }\cdot \stackrel{%
\rightharpoonup }{\rho ^{\prime }}-\frac{\alpha _2(\alpha _1-\beta
_1)^2-\alpha _2\beta _1\beta _2}{\alpha _1b^2}\stackrel{\rightharpoonup }{%
\rho ^{\prime }}^2-\alpha _2\stackrel{\rightharpoonup }{\sigma }_4\cdot 
\stackrel{\rightharpoonup }{\sigma }_5$ \\ 
$-\frac{\alpha _2^2}{4b^2}(1-\gamma ^2)\stackrel{\rightharpoonup }{\rho
^{\prime }}\cdot \stackrel{\rightharpoonup }{\sigma }_4\stackrel{%
\rightharpoonup }{\rho ^{\prime }}\cdot \stackrel{\rightharpoonup }{\sigma }%
_5]g(\frac{\alpha _2}{b^2},\stackrel{\rightharpoonup }{\rho ^{\prime }})\}$%
\end{tabular}
\eqnum{B.17}
\end{equation}
in which $\alpha _1=\frac{m_1}{m_1+m_2},\alpha _2=\frac{m_2}{m_1+m_2}$. The
other exchanged potential $V_t^{ex}(\stackrel{\rightharpoonup }{\rho },%
\stackrel{\rightharpoonup }{\rho ^{\prime }})^{24}$ and $V_t^{ex}(\stackrel{%
\rightharpoonup }{\rho },\stackrel{\rightharpoonup }{\rho ^{\prime }})^{34}$
can directly be written out by changing the superscripts 14 to 24 and 34.

The terms of the direct part of the potential in (B.3) are shown below 
\begin{equation}
\begin{tabular}{l}
$V_{24}^D(\stackrel{\rightharpoonup }{\rho },\stackrel{\rightharpoonup }{%
\rho ^{\prime }})=\frac{4\pi \alpha _sC_{24}^t}{(2\pi b^2)^{3/2}}f_{24}^t(%
\stackrel{\rightharpoonup }{\rho },\stackrel{\rightharpoonup }{\rho ^{\prime
}})_D\{g(\frac 1{2b^2},\stackrel{\rightharpoonup }{\rho }+\stackrel{%
\rightharpoonup }{\rho ^{\prime }})+\frac 1{4m_1^2}(1-\stackrel{%
\rightharpoonup }{\sigma }_2\cdot \stackrel{\rightharpoonup }{\sigma }_4)-%
\frac 1{4m_1^2b^2}[6-\frac 12\stackrel{\rightharpoonup }{\sigma }_2\cdot 
\stackrel{\rightharpoonup }{\sigma }_4$ \\ 
$-\frac{(\beta _2-\beta _1)^2}{4b^2}\stackrel{\rightharpoonup }{\rho }^2+i%
\frac 3{8b^2}(1+\gamma )(\stackrel{\rightharpoonup }{\rho }\times \stackrel{%
\rightharpoonup }{\rho ^{\prime }})\cdot (\stackrel{\rightharpoonup }{\sigma 
}_2+\stackrel{\rightharpoonup }{\sigma }_4)-\frac 1{4b^2}(\stackrel{%
\rightharpoonup }{\rho }-\stackrel{\rightharpoonup }{\rho ^{\prime }})^2+i%
\frac{\beta _2-\beta _1}{8b^2}(1-\gamma )(\stackrel{\rightharpoonup }{\rho }%
\times \stackrel{\rightharpoonup }{\rho ^{\prime }})$ \\ 
$\cdot (\stackrel{\rightharpoonup }{\sigma }_2-\stackrel{\rightharpoonup }{%
\sigma }_4)+\frac 1{16b^2}(1+\gamma ^2)(\stackrel{\rightharpoonup }{\rho }+%
\stackrel{\rightharpoonup }{\rho ^{\prime }})\cdot \stackrel{\rightharpoonup 
}{\sigma }_2(\stackrel{\rightharpoonup }{\rho }+\stackrel{\rightharpoonup }{%
\rho ^{\prime }})\cdot \stackrel{\rightharpoonup }{\sigma }_4]g(\frac 1{2b^2}%
,\stackrel{\rightharpoonup }{\rho }+\stackrel{\rightharpoonup }{\rho
^{\prime }})\}$%
\end{tabular}
\eqnum{B.18}
\end{equation}
where 
\begin{equation}
f_{24}^t(\stackrel{\rightharpoonup }{\rho },\stackrel{\rightharpoonup }{\rho
^{\prime }})_D=e^{-\frac{3\beta _2+1}{4b^2}(\stackrel{\rightharpoonup }{\rho 
}-\stackrel{\rightharpoonup }{\rho ^{\prime }})^2-\frac{\text{ }3}{16b^2}(%
\stackrel{\rightharpoonup }{\rho }+\stackrel{\rightharpoonup }{\rho ^{\prime
}})^2}  \eqnum{B.19}
\end{equation}
The terms $V_{14}^D(\stackrel{\rightharpoonup }{\rho },\stackrel{%
\rightharpoonup }{\rho ^{\prime }})$ and $V_{34}^D(\stackrel{\rightharpoonup 
}{\rho },\stackrel{\rightharpoonup }{\rho ^{\prime }})$ in (B.3) have the
same form as shown above except for the subscripts 24 being changed to 14
and 34.

The term $V_{25}^D(\stackrel{\rightharpoonup }{\rho },\stackrel{%
\rightharpoonup }{\rho ^{\prime }})$ in (B.3) is of the form 
\begin{equation}
\begin{tabular}{l}
$V_{25}^D(\stackrel{\rightharpoonup }{\rho },\stackrel{\rightharpoonup }{%
\rho ^{\prime }})=\frac{4\pi \alpha _sC_{25}^t}{(\pi b^2/\alpha _1)^{3/2}}%
f_{25}^t(\stackrel{\rightharpoonup }{\rho },\stackrel{\rightharpoonup }{\rho
^{\prime }})_D\{g(\frac{\alpha _2}{b^2},\stackrel{\rightharpoonup }{\rho }+%
\stackrel{\rightharpoonup }{\rho ^{\prime }})-(\frac{m_1^2+m_2^2}{8m_1^2m_2^2%
}+\frac{\stackrel{\rightharpoonup }{\sigma }_2\cdot \stackrel{%
\rightharpoonup }{\sigma }_5}{4m_1m_2})$ \\ 
$-\frac 1{4(m_1+m_2)^2b^2}[\frac 6{\alpha _1}-\frac{(\beta _2-\alpha _2)^2}{%
\alpha _1^2b^2}(\stackrel{\rightharpoonup }{\rho }-\stackrel{\rightharpoonup 
}{\rho ^{\prime }})^2+\frac{\alpha _1-\beta _1}{2\alpha b^2}i(1+\gamma )(%
\stackrel{\rightharpoonup }{\rho }\times \stackrel{\rightharpoonup }{\rho
^{\prime }})\cdot [(1+\frac{m_2}{m_1})\stackrel{\rightharpoonup }{\sigma }_2$
\\ 
$-(1+\frac{m_1}{m_2})\stackrel{\rightharpoonup }{\sigma }_5]]g(\frac{\alpha
_2}{b^2},\stackrel{\rightharpoonup }{\rho }+\stackrel{\rightharpoonup }{\rho
^{\prime }})-\frac 1{4m_1m_2b^2}[6\alpha _2+\frac{\alpha _2}{b^2}(\alpha _1-%
\frac{\beta _1}{\alpha _1}-2\alpha _2\beta _2)(\stackrel{\rightharpoonup }{%
\rho }-\stackrel{\rightharpoonup }{\rho ^{\prime }})^2$ \\ 
$-\alpha _2\stackrel{\rightharpoonup }{\sigma }_2\cdot \stackrel{%
\rightharpoonup }{\sigma }_5+\frac{\alpha _2^2}{2b^2}i(1+\gamma )(\stackrel{%
\rightharpoonup }{\rho }\times \stackrel{\rightharpoonup }{\rho ^{\prime }}%
)\cdot [(2+\frac{m_2}{m_1})\stackrel{\rightharpoonup }{\sigma }_2+(2+\frac{%
m_1}{m_2})\stackrel{\rightharpoonup }{\sigma }_5]$ \\ 
$-\frac{\alpha _2^2}{4b^2}(1-\gamma ^2)(\stackrel{\rightharpoonup }{\rho }+%
\stackrel{\rightharpoonup }{\rho ^{\prime }})\cdot \stackrel{\rightharpoonup 
}{\sigma }_2(\stackrel{\rightharpoonup }{\rho }+\stackrel{\rightharpoonup }{%
\rho ^{\prime }})\cdot \stackrel{\rightharpoonup }{\sigma }_5]g(\frac{\alpha
_2}{b^2},\stackrel{\rightharpoonup }{\rho }+\stackrel{\rightharpoonup }{\rho
^{\prime }})\}$%
\end{tabular}
\eqnum{B.20}
\end{equation}
where 
\begin{equation}
f_{25}^t(\stackrel{\rightharpoonup }{\rho },\stackrel{\rightharpoonup }{\rho
^{\prime }})_D=e^{-\frac{3\beta _2-\beta _1^2}{4b^2}(\stackrel{%
\rightharpoonup }{\rho }-\stackrel{\rightharpoonup }{\rho ^{\prime }})^2-%
\frac{3\alpha _2}{8b^2}(\stackrel{\rightharpoonup }{\rho }+\stackrel{%
\rightharpoonup }{\rho ^{\prime }})^2}  \eqnum{B.21}
\end{equation}
The remaining two terms in (B.3) are of the same form as given above except
for the subscripts 25 being replaced by 15 and 35.

Let us turn to the $\overline{K}N$ interaction potential. For the $\overline{%
K}N$ interaction, there is only a direct part of the potential coming from
the t-channel OGEP as represented in (B.3), (B.18) and (B.20) because there
are no identical particles between the $N$-cluster $(qqq)$ and the $%
\overline{K}$-cluster $(\overline{q}s)$. In addition, the nonlocal effective
potential derived from the s-channel OGEP plays an essential role in the $%
\overline{K}N$ interaction. This potential can be written as

\begin{equation}
V^s(\stackrel{\rightharpoonup }{\rho },\stackrel{\rightharpoonup }{\rho
^{\prime }})=V_{14}^{s\_d}(\stackrel{\rightharpoonup }{\rho },\stackrel{%
\rightharpoonup }{\rho ^{\prime }})+V_{24}^{s\_d}(\stackrel{\rightharpoonup 
}{\rho },\stackrel{\rightharpoonup }{\rho ^{\prime }})+V_{34}^{s\_d}(%
\stackrel{\rightharpoonup }{\rho },\stackrel{\rightharpoonup }{\rho ^{\prime
}})  \eqnum{B.22}
\end{equation}
here $V_{14}^{s\_d}(\stackrel{\rightharpoonup }{\rho },\stackrel{%
\rightharpoonup }{\rho ^{\prime }})$ denotes the direct term of the
potential generated from the interaction of the quark 1 and the antiquark 4 
\begin{equation}
\begin{tabular}{l}
$V_{14}^{s\_d}(\stackrel{\rightharpoonup }{\rho },\stackrel{\rightharpoonup 
}{\rho ^{\prime }})=\frac{4\pi \alpha _sF_{14}^aC_{14}^a}{(2\pi b^2)^{3/2}}%
f_{14}^s(\stackrel{\rightharpoonup }{\rho },\stackrel{\rightharpoonup }{\rho
^{\prime }})\{(3+\stackrel{\rightharpoonup }{\sigma }_1\cdot \stackrel{%
\rightharpoonup }{\sigma }_4)-\frac 1{4m_1^2b^2}[3-\frac{(\beta _2-\beta
_1)^2}{4b^2}(\stackrel{\rightharpoonup }{\rho }-\stackrel{\rightharpoonup }{%
\rho ^{\prime }})^2]$ \\ 
$-\frac 1{m_1^2b^2}(2+\stackrel{\rightharpoonup }{\sigma }_1\cdot \stackrel{%
\rightharpoonup }{\sigma }_4)(3-\frac{\stackrel{\rightharpoonup }{\rho }^2+%
\stackrel{\rightharpoonup }{\rho ^{\prime }}^2}{4b^2})-i\frac{\beta _2-\beta
_1}{8m_1^2b^2}(\stackrel{\rightharpoonup }{\rho }\times \stackrel{%
\rightharpoonup }{\rho ^{\prime }})\cdot (\stackrel{\rightharpoonup }{\sigma 
}_1-\stackrel{\rightharpoonup }{\sigma }_4)$ \\ 
$+\frac{\stackrel{\rightharpoonup }{\sigma }_1\cdot \stackrel{%
\rightharpoonup }{\sigma }_4}{m_1^2b^2}-\frac 1{4m_1^2b^2}[\stackrel{%
\rightharpoonup }{\rho }\cdot \stackrel{\rightharpoonup }{\sigma }_1%
\stackrel{\rightharpoonup }{\rho }\cdot \stackrel{\rightharpoonup }{\sigma }%
_4+\stackrel{\rightharpoonup }{\rho ^{\prime }}\cdot \stackrel{%
\rightharpoonup }{\sigma }_1\stackrel{\rightharpoonup }{\rho ^{\prime }}%
\cdot \stackrel{\rightharpoonup }{\sigma }_4]\}$%
\end{tabular}
\eqnum{B.23}
\end{equation}
where

\begin{equation}
f_{14}^s(\stackrel{\rightharpoonup }{\rho },\stackrel{\rightharpoonup }{\rho
^{\prime }})_D=e^{\frac{3\beta _2-1/2}{2b_1^2}\stackrel{\rightharpoonup }{%
\rho }\cdot \stackrel{\rightharpoonup }{\rho ^{\prime }}-\frac{3\beta _2+1/2%
}{4b_1^2}(\stackrel{\rightharpoonup }{\rho }^2+\stackrel{\rightharpoonup }{%
\rho ^{\prime }}^2)}  \eqnum{B.24}
\end{equation}
The other two terms $V_{24}^{s\_d}(\stackrel{\rightharpoonup }{\rho },%
\stackrel{\rightharpoonup }{\rho ^{\prime }})$ and $V_{34}^{s\_d}(\stackrel{%
\rightharpoonup }{\rho },\stackrel{\rightharpoonup }{\rho ^{\prime }})$ can
be written out from the above expression by the substitution of the
subscripts 24 and 34 for 14.

The effective $KN$ ($\overline{K}N$) potential derived from the interquark
harmonic oscillator confining potential and quark interchanges are of simple
expressions. They are represented as 
\begin{equation}
\begin{tabular}{l}
$V_c^{ex}(\stackrel{\rightharpoonup }{\rho },\stackrel{\rightharpoonup }{%
\rho ^{\prime }})=-12b_1^2\omega ^2\{C_s^{24}\mu _{24}+C_s^{34}\mu
_{34}+C_s^{12}\mu _{12}+C_s^{13}\mu _{13}+C_s^{14}\mu _{14}$ \\ 
$+\frac 1{2\alpha _2}[C_s^{25}\mu _{25}+C_s^{35}\mu _{35}+C_s^{15}\mu
_{15}+C_s^{45}\mu _{45}]\}e^{^{-\frac{\stackrel{\rightharpoonup }{\rho }%
\cdot \stackrel{\rightharpoonup }{\rho ^{\prime }}}{2b^2}-\frac{3\beta _2}{%
4b^2}(\stackrel{\rightharpoonup }{\rho }-\stackrel{\rightharpoonup }{\rho
^{\prime }})^2}}$%
\end{tabular}
\eqnum{B.25}
\end{equation}
where the $\mu _{ij}$ denotes the reduced mass of interacting quarks $i$ and 
$j$ and the color factors are the same as the ones appearing in the
t-channel potentials.

Apart from the potentials listed above, there are additional terms in the $%
KN $ and $\overline{K}N$ potentials occurring in the resonating group
equation which arise from the kinetic term and the normalization term in the
equation due to the effect of quark rearrangement. They are written as
follows 
\begin{equation}
\begin{tabular}{l}
$T^{ex}(\stackrel{\rightharpoonup }{\rho },\stackrel{\rightharpoonup }{\rho
^{\prime }})=(-3)\alpha ^{3/2}\{\frac 1{2\mu }[\frac{\gamma _1}{2b^2}-\frac{%
(\gamma _1\stackrel{\rightharpoonup }{\rho }-\gamma _1\stackrel{%
\rightharpoonup }{\rho ^{\prime }}+\stackrel{\rightharpoonup }{\rho ^{\prime
}})^2}{4b^4}]+\frac 1{2\mu _1}[\frac 3{4b^2}-\frac{\stackrel{\rightharpoonup 
}{\rho ^{\prime }}^2}{16b^4}]$ \\ 
$+\frac 1{2\mu _2}[\frac 1{4b^2}-\frac{\stackrel{\rightharpoonup }{\rho
^{\prime }}^2}{144b^4}]+\frac 1{2\mu _3}[\frac{3\alpha _2}{2\text{ }b^2}-%
\frac{\alpha _2^2\stackrel{\rightharpoonup }{\rho ^{\prime }}^2}{4b^4}%
]\}f_T^{ex}(\stackrel{\rightharpoonup }{\rho },\stackrel{\rightharpoonup }{%
\rho ^{\prime }})$%
\end{tabular}
\eqnum{B.26}
\end{equation}
\qquad where 
\begin{equation}
f_T^{ex}(\stackrel{\rightharpoonup }{\rho },\stackrel{\rightharpoonup }{\rho
^{\prime }})=e^{-\frac{\stackrel{\rightharpoonup }{\rho }\cdot \stackrel{%
\rightharpoonup }{\rho ^{\prime }}}{2b^2}-\frac{\gamma _1}{4b^2}(\stackrel{%
\rightharpoonup }{\rho }-\stackrel{\rightharpoonup }{\rho ^{\prime }})^2} 
\eqnum{B.27}
\end{equation}
\begin{equation}
\mu =\frac{3m_1(m_1+m_2)}{4m_1+m_2},\mu _1=\frac{m_1}2,\mu _2=\frac{2m_1}3%
,\mu _3=\frac{m_1m_2}{m_1+m_2},\gamma _1=3/(1/3+\alpha _1).  \eqnum{B.28}
\end{equation}
and

\begin{equation}
N^{ex}(\stackrel{\rightharpoonup }{\rho },\stackrel{\rightharpoonup }{\rho
^{\prime }})=-3E_re^{-\frac{\stackrel{\rightharpoonup }{\rho }\cdot 
\stackrel{\rightharpoonup }{\rho ^{\prime }}}{2b^2}-\frac{(\beta _1^2+\beta
_2^2)}{2b^2}(\stackrel{\rightharpoonup }{\rho }-\stackrel{\rightharpoonup }{%
\rho ^{\prime }})^2}  \eqnum{B.29}
\end{equation}
here $E_r$ is the relative energy of two clusters.

The color-spin-isospin matrix elements of the above potentials are easily
evaluated by using the CFS wave functions given in Appendix A.

\section{Appendix C: Derivation of phase shift formula}

In this appendix, we briefly describe the derivation of the formula used to
compute the phase shifts with a comment on the Born approximation. One of
the authors of this paper, in his previous work on the relativistic
Pauli-Schr\"odinger equation for two-body scattering states, which was
proved to be equivalent to the corresponding Bethe Salpeter equation [38],
proved that in the relativistic case, the outgoing state wave function of
the system under consideration may be written, in the position space, as 
\begin{equation}
\psi (\stackrel{\rightharpoonup }{r})=\varphi ^0(\stackrel{\rightharpoonup }{%
r})+f(\Omega _k)\frac{e^{ikr}}r  \eqnum{C.1}
\end{equation}
where $\varphi ^0(\stackrel{\rightharpoonup }{r})$ is the wave function for
free particles and $f(\Omega _k)$ the probability amplitude of the outgoing
spherical wave . For a boson system, the wave function in (C.1) is scalar,
while, for a fermion system, the wave function is represented in the Pauli
spinor space. In the unequal mass case, the amplitude $f(\Omega _k)$ is
related to the transition amplitude $T_{fi}$ in such a fashion 
\begin{equation}
f(\Omega _k)=-\frac{M(E)}{2\pi }T_{fi}  \eqnum{C.2}
\end{equation}
where 
\begin{equation}
M(E)=\frac{E^4-(m_K^2-m_N^2)^2}{4E^3}  \eqnum{C.3}
\end{equation}
with $E,$ $m_K$ and $m_N$ being the total energy of the system, the kaon
mass and nucleon mass and $T_{fi}$ is the exact transition amplitude. The
amplitude is defined by 
\begin{equation}
T_{fi}=\langle \varphi _f^0\mid V\mid \psi _i\rangle  \eqnum{C.4}
\end{equation}
where $\varphi _f^0$ is the plane wave function of final state, $V$ stands
for the interaction Hamiltonian operator and $\psi _i$ is the exact initial
wave function which is determined by the following equation 
\begin{equation}
\psi _i=\varphi _i^0+G^0V\psi _i  \eqnum{C.5}
\end{equation}
in which $\varphi _i^0$ is the plane wave function of initial state and 
\begin{equation}
G^0=\frac 1{E-H_0+i\varepsilon },\varepsilon \rightarrow 0^{+}  \eqnum{C.6}
\end{equation}
is the Green's function with $H_0$ being the free Hamiltonian. The solution
of equation (C.5) can formally be represented as 
\begin{equation}
\psi _i=\frac 1{1-G^0V}\varphi _i^0=\sum_{n=0}^\infty [G^0V]^n\varphi _i^0 
\eqnum{C.7}
\end{equation}
In the lowest order Born approximation, $\psi _i=\varphi _i^0$ and,
correspondingly, the transition amplitude in (C.4) takes the form as given
in Eq.(24).

Now we proceed to derive the formula written in Eq.(26). Let us expand the
wave functions in (C.1) in partial waves 
\begin{equation}
\psi _i=\sum\limits_l\psi _lP_l(\cos \theta )  \eqnum{C.8}
\end{equation}
\begin{equation}
f(\theta )=\sum\limits_lf_lP_l(\cos \theta )  \eqnum{C.9}
\end{equation}
and 
\begin{equation}
\varphi _i^0=e^{i\stackrel{\rightharpoonup }{k}\cdot \stackrel{%
\rightharpoonup }{r}}=\sum\limits_l(2l+1)i^lj_l(kr)P_l(\cos \theta ) 
\eqnum{C.10}
\end{equation}
where $P_l(\cos \theta )$ is the Legendre function of rank $l$. With these
expansions, noticing the asymptotic behaviors of the spherical Bessel
function $j_l(kr)$ and the function $\psi _l(r)$%
\begin{equation}
j_l(k)_{r\rightarrow \infty }^{\longrightarrow }\frac 1{kr}\sin (kr-\frac{%
l\pi }2)  \eqnum{C.11}
\end{equation}
\begin{equation}
\psi _l(r)_{r\rightarrow \infty }^{\longrightarrow }\frac{C_l}r\sin (kr-%
\frac{l\pi }2+\delta _l)  \eqnum{C.12}
\end{equation}
One may find a result from (C.1) such that 
\begin{equation}
f_l=\frac{(2l+1)}{2ik}(e^{2i\delta _l}-1)  \eqnum{C.13}
\end{equation}
On the other hand, the transition amplitude may also be expanded in partial
waves 
\begin{equation}
T_{fi}=4\pi \sum\limits_l(2l+1)T_lP_l(\cos \theta )  \eqnum{C.14}
\end{equation}
where 
\begin{equation}
T_l=\frac 1{8\pi }\int_{-1}^1dxP_l(x)T_{fi}  \eqnum{C.15}
\end{equation}
Combining (C.2), (C.9), (C.13) and (C.14), it is easy to derive the
following relation 
\begin{equation}
e^{2i\delta _l}=1-4iM(E)kT_l  \eqnum{C.16}
\end{equation}

When (C.7) is substituted into (C.4), we can write 
\begin{equation}
T_{fi}=\sum_{n=0}^\infty T_{fi}^n  \eqnum{C.17}
\end{equation}
where 
\begin{equation}
T_{fi}^n=\langle \varphi _f^0\mid (VG^0)^nV\mid \phi _i^0\rangle 
\eqnum{C.18}
\end{equation}
On inserting (C.17) into (C.15), we have 
\begin{equation}
T_l=\sum_{n=0}^\infty T_l^n  \eqnum{C.19}
\end{equation}
where 
\begin{equation}
T_l^n=\frac 1{8\pi }\int_{-1}^1dxP_l(x)T_{fi}^n  \eqnum{C.20}
\end{equation}
Upon substituting (C.19) and the Taylor expansion of $e^{2i\delta _l}$ into
(C.16), one may find 
\begin{equation}
\sum_{n=1}^\infty \frac 1{n!}(2i\delta )^n=-i4M(E)k\sum_{n=0}^\infty T_l^n 
\eqnum{C.21}
\end{equation}
From the above equality, it is clearly seen that there exists an one-to-one
correspondence between the terms of the same order in the both series. When
only the first term in each series is considered, we obtain 
\begin{equation}
\delta _l=-2MkT_l(k)  \eqnum{C.22}
\end{equation}
which is proportional to the interaction Hamiltonian $V$. This just is the
formula for the phase shift which is given in the so-called Born
approximation. If the phase shift is really proportional to the interaction
Hamiltonian $V$ as in the Born approximation, from the corresponding higher
order terms of the both series in (C.21), we can write 
\begin{equation}
(\delta _l)^{n+1}\approx -\frac{(n+1)!Mk}{2^{2-1}i^n}T_l^n  \eqnum{C.23}
\end{equation}
which is proportional to $V^{n+1}$ and means that the $n$-th terms in the
both series in (C.21) are approximately equal to each other as if (C.23) is
approximately given by taking the equality in (C.22) to the $(n+1)$-th
power. If the relation in (C.23) holds, we see, the formula in (C.22)
appears to be a good approximation.

\section{Appendix D: Notes on QCD renormalization}

To help understanding of the renormalization formulas written in Sect.3, in
this appendix, we give some explanations of the QCD renormalization. It is
well-known that the t-channel OGEP in Eq.(4) and the s-channel OGEP in
Eq.(6) are usually derived from the lowest-order S-matrix elements given by
the tree Feynman diagrams representing respectively the quark-quark(or
quark-antiquark) scattering and quark-antiquark annihilation processes in
the nonrelativistic approximation of the order $p^2/m^2$. As an example, we
write the lowest order S-matrix element for two-quark scattering as follows

\begin{equation}
S_{fi}^{(0)}=\overline{u}^{(0)}(p_1)ig\gamma ^\mu T^au^{(0)}(q_1)iD_{\mu \nu
}^{(0)ab}(k)\overline{u}^{(0)}(p_2)ig\gamma ^\nu T^bu^{(0)}(q_2)  \eqnum{D.1}
\end{equation}
where $u^{(0)}(p)$ is the free quark wave function, $\overline{u}^{(0)}(p_1)$
is its Dirac conjugate (here the spin indices of the spinor functions are
suppressed for simplicity), $iD_{\mu \nu }^{(0)ab}(k)$ is the gluon free
propagator with $k=p_1-q_1=q_2-p_{2\text{ }}$and $ig\gamma ^\mu T^a$ is the
bare vertex with $g$ being the coupling constant and $T^a$ the color matrix.
Correspondingly, the exact S-matrix element given by the one-gluon exchange
interaction is represented as 
\begin{equation}
S_{fi}=\overline{u}(p_1)\Gamma ^{a\mu }(p_1,q_1)u(q_1)iD_{\mu \nu }^{ab}(k)%
\overline{u}(p_2)\Gamma ^{a\nu }(p_2,q_2)u(q_2)  \eqnum{D.2}
\end{equation}
where $u(p)$, $iD_{\mu \nu }^{ab}(k)$ and $\Gamma ^{a\mu }(p,q)$ denote the
full (or say, dressed) quark wave function, gluon propagator and quark-gluon
vertex respectively in which all higher order perturbative corrections are
included. According to the well-known renormalization relations 
\begin{equation}
\begin{tabular}{l}
$u(p)=\sqrt{Z_2}u_R(p),\overline{u}(p)=\sqrt{Z_2}\overline{u}_R(p),$ \\ 
$D_{\mu \nu }^{ab}(k)=Z_3D_{R\mu \nu }^{ab}(k),\Gamma ^{a\mu }(p,q)=Z_\Gamma
\Gamma _R^{a\mu }(p,q)$%
\end{tabular}
\eqnum{D.3}
\end{equation}
where the subscript R marks the renormalized quantities, $\sqrt{Z_2}$, $Z_3$
and $Z_\Gamma =Z_2^{-1}Z_3^{-\frac 12}$ are the renormalization constants
for the quark wave function, the gluon propagator and the quark-gluon vertex
respectively, one can get from (D.2) that 
\begin{equation}
S_{fi}=\overline{u}_R(p_1)\Gamma _R^{a\mu }(p_1,q_1)u_R(q_1)iD_{R\mu \nu
}^{ab}(k)\overline{u}_R(p_2)\Gamma _R^{a\nu }(p_2,q_2)u_R(q_2)  \eqnum{D.4}
\end{equation}
As shown in Ref (27), the renormalized quantities in the above can be
determined by their renormalization group equations and their
renormalization boundary conditions. The results given by solving the
renormalization group equations are 
\begin{equation}
\begin{tabular}{l}
$u_R(p)=e^{\frac 12\int_1^\lambda \frac{d\lambda }\lambda \gamma _2(\lambda
)}u_R^{(0)}(p)$ \\ 
$\overline{u}_R(p)=e^{\frac 12\int_1^\lambda \frac{d\lambda }\lambda \gamma
_2(\lambda )}\overline{u}_R^{(0)}(p)$ \\ 
$D_{R\mu \nu }^{ab}(k)=e^{\int \frac{d\lambda }\lambda \gamma _3(\lambda
)}iD_{R\mu \nu }^{(0)ab}(k)$ \\ 
$\Gamma _R^{a\mu }(p,q)=e^{\int \frac{d\lambda }\lambda \gamma _\Gamma
(\lambda )}ig_R(\lambda )\gamma ^\mu T^a$%
\end{tabular}
\eqnum{D.5}
\end{equation}
where $u_R^{(0)}(p)$ and $\overline{u}_R^{(0)}(p)$ are of the same forms as
the free wave functions except that the quark mass in them becomes the
effective (running) one, $iD_{R\mu \nu }^{(0)ab}(k)$ formally is the same as
the free propagator but the gauge parameter in it is replaced by the
effective one, $g_R(\lambda )$ in the bare vertex $ig_R(\lambda )\gamma ^\mu
T^a$ is the effective coupling constant, $\frac 12\gamma _2(\lambda )$, $%
\gamma _3(\lambda )$ and $\gamma _\Gamma (\lambda )$ are the anomalous
dimensions defined by 
\begin{equation}
\begin{tabular}{l}
$\frac 12\gamma _2(\lambda )=\lambda \frac d{d\lambda }\ln \sqrt{Z_2(\lambda
)},\gamma _3(\lambda )=\lambda \frac d{d\lambda }\ln Z_3(\lambda ),$ \\ 
$\gamma _\Gamma (\lambda )=\lambda \frac d{d\lambda }\ln Z_\Gamma (\lambda
)=-\gamma _2(\lambda )-\frac 12\gamma _3(\lambda )$%
\end{tabular}
\eqnum{D.6}
\end{equation}
On inserting (D.5) into (D.4), we see that the anomalous dimensions are all
cancelled out with each other. As a result, we have

\begin{equation}
S_{fi}=\overline{u}_R^{(0)}(p_1)ig_R(\lambda )\gamma ^\mu
T^au_R^{(0)}(q_1)iD_{R\mu \nu }^{(0)ab}(k)\overline{u}_R^{(0)}(p_2)ig_R(%
\lambda )\gamma ^\nu T^bu_R^{(0)}(q_2)  \eqnum{D.7}
\end{equation}
This S-matrix element is completely the same as the one shown in (D.1)
except that the quark mass, the gauge parameter and the coupling constant
are replaced by the running ones. In the nonrelativistic approximation of
order $p^2/m^2$, one may derive a OGEP from (D.7) which in the Feynman gauge
is just as that written in Eq.(4) with the coupling constant and quark mass
being the effective ones. These effective quantities precisely represent the
renornalization effect and at one-loop level, they are of the forms as given
in Sect.3. For the s-channel OGEP, the discussion is completely the same.

Next, we would like to address the renormalization point. In the ordinary
QCD renormalization performed in the minimal subtraction scheme, which is
suitable in the large momentum limit because only in this limit the quark
mass can be set to be zero, the renormalization point is chosen to be
space-like. This choice is suitable to the scattering process because in
this case the transfer momentum, i.e. the variable of the gluon propagator
in (D.2) is space-like. This can be seen from the following derivation, 
\begin{equation}
k^2=(p_1-q_1)^2=2m^2-2\sqrt{\overrightarrow{p_1}^2+m^2}\sqrt{\overrightarrow{%
q_1}^2+m^2}+2\left| \overrightarrow{p_1}\right| \left| \overrightarrow{q_1}%
\right| \cos \theta  \eqnum{D.8}
\end{equation}
in the high energy limit, we can set $m\approx 0$, therefore 
\begin{equation}
k^2\approx -2\left| \overrightarrow{p_1}\right| \left| \overrightarrow{q_1}%
\right| (1-\cos \theta )\leq 0  \eqnum{D.9}
\end{equation}
In the low energy domain, since 
\begin{equation}
\sqrt{\overrightarrow{p}^2+m^2}\approx m+\frac{\overrightarrow{p}^2}{2m^2} 
\eqnum{D.10}
\end{equation}
(D.8) can be approximated as 
\begin{equation}
k^2\approx -(\overrightarrow{p_1}-\overrightarrow{q_1})^2\leq 0  \eqnum{D.11}
\end{equation}
So, for the t-channel OGEP, it is suitable to use the effective coupling
constant and quark mass given by the subtraction performed at space-like
renormalization point . While, for the annihilation process, the momentum in
the gluon propagator is time-like because in this case $k=p_1+p_2=q_1+q_2$, 
\begin{equation}
k^2=(p_1+p_2)^2=2m^2++2\sqrt{\overrightarrow{p_1}^2+m^2}\sqrt{%
\overrightarrow{p_2}^2+m^2}-2\left| \overrightarrow{p_1}\right| \left| 
\overrightarrow{p_2}\right| \cos \theta  \eqnum{D.12}
\end{equation}
In the large momentum limit, we can write 
\begin{equation}
k^2\approx 2\left| \overrightarrow{p_1}\right| \left| \overrightarrow{p_2}%
\right| (1-\cos \theta )\geq 0  \eqnum{D.13}
\end{equation}
in the low energy regime, we have 
\begin{equation}
k^2\approx (\overrightarrow{p_1}-\overrightarrow{p_2})^2\geq 0  \eqnum{D.14}
\end{equation}
Therefore, for the s-channel OGEP, it is appropriate to use the effective
quantities given by the subtraction carried out at time-like renormalization
point.

\section{Figure Captions}

Fig.1: The theoretical $KN$ S-wave phase shifts in the $I=0$ and $1$
channels. The solid lines represent the phase shifts with considering the
effects of the color octet, the QCD renormalization and the spin-orbit
suppression. The dotted lines denotes the result without considering the
color octet and the dashed line shows the result without considering the QCD
renormalization. The experimental phase shifts [29, 30] are shown by black
squares with error bars.

Fig.2: The theoretical $KN$ P-wave phase shifts in the $I=0$ and $1$
channels. The solid lines represent the phase shifts with considering the
effects of the color octet, the QCD renormalization and the spin-orbit
suppression. The dotted and dashed lines denote the results without
considering the color octet and the QCD renormalization respectively. The
experimental phase shifts [29, 30] are shown by black squares with error
bars.

Fig.3: The theoretical $KN$ D-wave phase shifts in the $I=0$ and $1$
channels. The solid lines represent the phase shifts with considering the
effects of the color octet, the QCD renormalization and the spin-orbit
suppression. The dotted and dashed lines denote the results without
considering the color octet and the QCD renormalization respectively. The
experimental phase shifts [29, 30] are shown by black squares with error
bars.

Fig.4: The theoretical predictions for the $\overline{K}N$ S, P and D-wave
phase shifts.

Fig.5: The approximate expressions of the integral $f(x)$ in Eq.(32) given
by different values of the parameter $\gamma $ which are plotted with the
dotted lines. The real values of the integral is represented by the solid
line.

Fig.6: Illustration of the effect of the spin-orbit suppression on the
P-wave phase shifts. The solid lines represent the final results given by
taking the parameter $\gamma =0.45.$ The dotted lines denote the results
given by $\gamma =0.30$.

Fig.7: The QCD effective coupling constants obtained from the one-loop
renormalization. The solid, dashed and dotted lines represent the results
given by the time-like momentum subtraction, the space-like momentum
subtraction and the minimal subtraction respectively.

Fig.8: The QCD effective quark masses obtained from the one-loop
renormalization. The solid line represents the result given by the time-like
momentum subtraction. The dashed line shows the real part of the effective
quark mass given in the space-like momentum subtraction.


\section{References}

\begin{references}
\bibitem{}  A. De R\'ujula, H. Georgi and S. L. Glashow, Phys. Rev.{\bf \ }%
D12{\bf , }147 (1975).

\bibitem{}  N. Isgur and G. Karl, Phys. Rev.{\bf \ }D18, 4187 (1978);{\bf \ }%
D19, 2653 (1979).

\bibitem{}  W. Lucha, F. F. Schoberl and D. Gromes, Phys. Rep. 200{\bf , }%
127 (1991), many references therein

\bibitem{}  J. J. Griffin and J. A. Wheeler, Nucl. Phys{\it . }108, 331
(1957).

\bibitem{}  M. Harvey, Nucl. Phys. A352, 326 (1981).

\bibitem{}  M. Oka and K. Yazaki, Prog. Theor. Phys. 66, 556 (1981).

\bibitem{}  A. Faessler, E. Fernandez, G. Lubeck and K. Shimizu, Nucl. Phys.
A402, 555 (1983).

\bibitem{}  I. Bender and H. G. Dosch, Z. Phys. C13, 69 (1982).

\bibitem{}  H. J. Pirner and B. Povh, Phys. Lett.\ B114{\bf , }308 (1982).

\bibitem{}  I. Bender, H. G. Dosch, H. J. Priner and H. G. Kruse, Nucl. Phys.%
{\bf \ }A414, 359 (1984).

\bibitem{}  D. Mukhopadhyay and H. J. Pirner,{\it \ }Nucl. Phys. A442{\bf , }%
605 (1985).

\bibitem{}  J. Weinstein and N. Isgur, Phys. Rev. D27{\bf , }588 (1983); D41%
{\bf , }2236 (1990); D43, 95 (1991).

\bibitem{}  R. M\"uller, T. Schmeidl and H. M. Hofmann, Z. Phys. A334, 451
(1989).

\bibitem{}  T. Barnes, E. S. Swanson and J. Weinstein, Phys. Rev. D46{\bf ,}
4868 (1992).

\bibitem{}  T. Barnes and E. S. Swanson, Phys. Rev. D46{\bf ,} 131 (1992);
Phys. Rev. C49, 1166 (1994).

\bibitem{}  T. Barnes, N. Black and E. S. Swanson, Phys.Rev. C63, 025204
(2001).

\bibitem{}  N. Black, J. Phys. G: Nucl. Part. Phys. 28{\bf , }1953 (2002).

\bibitem{}  S. Lemaire, J. Labarsouque and B. Slivestre-Brac, Nucl. Phys.
A696{\bf ,} 497 (2001).

\bibitem{}  G. Q. Zhao, X. G. Jing, and J. C. Su, Phys. Rev.{\it \ }D58{\bf %
, }117503 (1998).

\bibitem{}  J. X. Chen, Y. H. Cao and J. C. Su, Phys. Rev. C64{\bf , }065201
(2001).

\bibitem{}  J .C. Su, Y. B .Dong and S. S. Wu, J. Phys. G: Nucl. Part. Phys.
18{\bf , }1347 (1992).

\bibitem{}  J. C. Su and S. S. Wu, Chinese Phys. 8, 978 (1989); J. C. Su, Z.
Q. Chen and S. S. Wu, Nucl. Phys. A254, 615{\bf \ }(1991).

\bibitem{}  J. C. Su, L. Y. Shan, Y. H. Cao, Commun. Theor. Phys. 36, 665
(2001).

\bibitem{}  E. Braaten and Y. Q. Chen, Phys. Rev. Lett. 76, 730 (1996).

\bibitem{}  Feng Yuan, Cong-Feng Qiao and Kuang-Ta Chao, Phys.Rev. D56, 321
(1997).

\bibitem{}  Y. B. Dong, J. C. Su and S. S. Wu, J. Phys. G: Nucl. Part. Phys.
18, 75 (1992).

\bibitem{}  J. C. Su, X. X. Yi and Y. H. Cao, J. Phys. G: Nucl. Part. Phys.
25{\bf , }2325 (1999).

\bibitem{}  G. t'Hooft, Nucl. Phys. B61, 455 (1973); W. A. Bardeen, A. J.
Buras, D. W. Duke and T. Muta, Phys. Rev. D18, 3998 (1978).

\bibitem{}  J. S. Hyslop, R. A. Arndt, L. D. Roper and R. L. Workman, D46%
{\bf , }961 (1992).

\bibitem{}  K. Hashimoto, Phys. Rev. C29{\bf ,} 1377 (1984).

\bibitem{}  J. D. Davies, G. J. Pyle, G. T. A. Squier, C. J. Batty, S. F.
Biagi, S. D. Hoath, P. Sharman and A. S. Clough, Phys. Lett. 83B{\bf , }55
(1979).

\bibitem{}  M. Izycki, G. Backenstoss, L. Tauscher, P. Blum, R. Guigas, N.
Hassler, H. Koch, H. Poth, K. Fransson, A.Nilsson, P. Pavlopoulos and K.
Zioutas, {\it \ Z}. Phys. A297{\bf , }11 (1980).

\bibitem{}  P. M. Bird, A. S. Clough and K. R. Parker, Nucl. Phys. A404, 482
(1983).

\bibitem{}  M. Iwasaki, et al., Phys. Rev. Lett. 78{\bf , }3067 (1997).

\bibitem{}  A. D. Martin, Nucl. Phys. B179{\bf , }33 (1981).

\bibitem{}  B. R. Martin and M. Sakit, Phys. Rev. 183{\bf ,} 1352 (1969).

\bibitem{}  T. Waas, N. Kaiser and W. Weise, Phys. Lett. B365, 12 (1996).

\bibitem{}  J. C. Su, Commun. Theor. Phys. 18, 327 (1992).
\end{references}
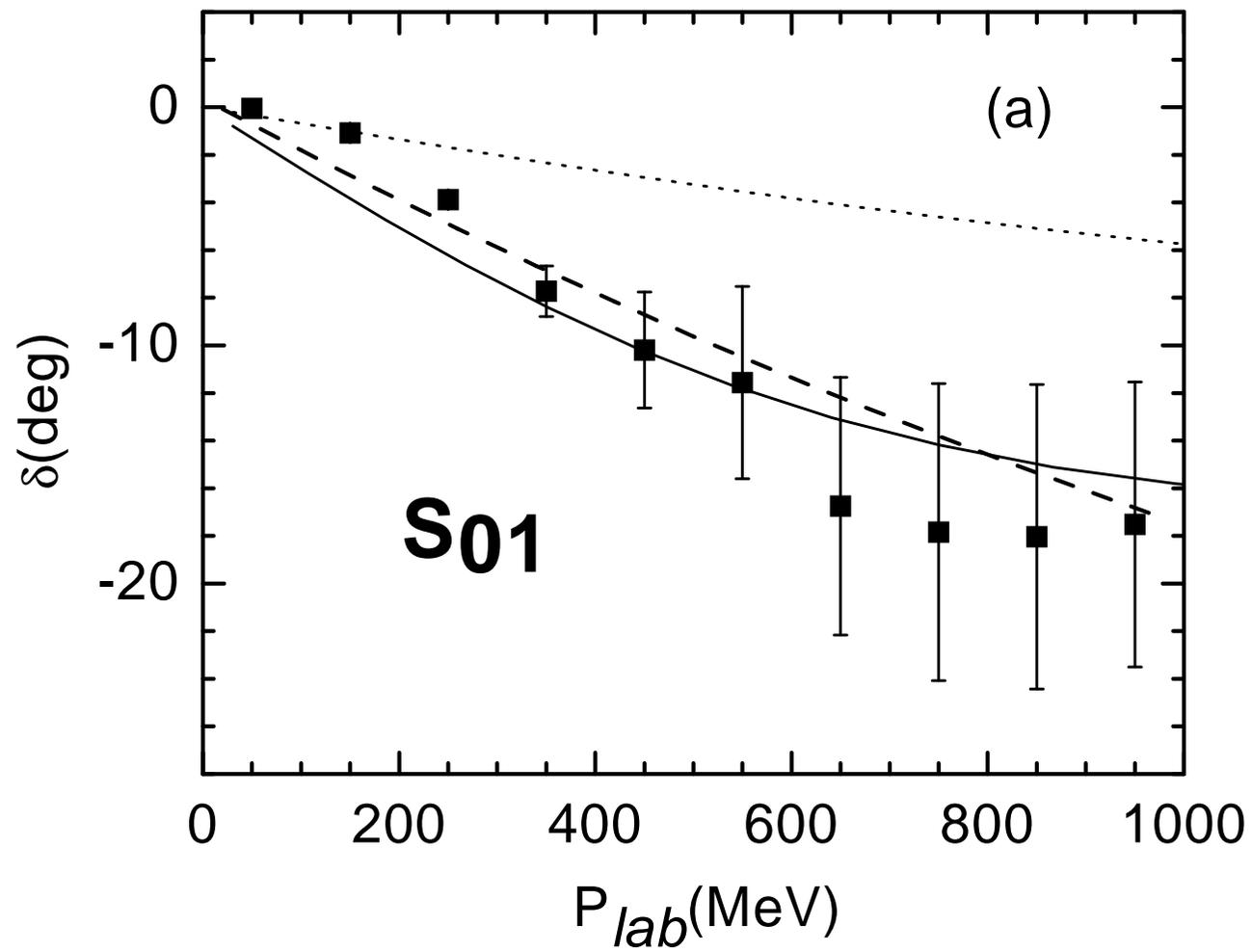
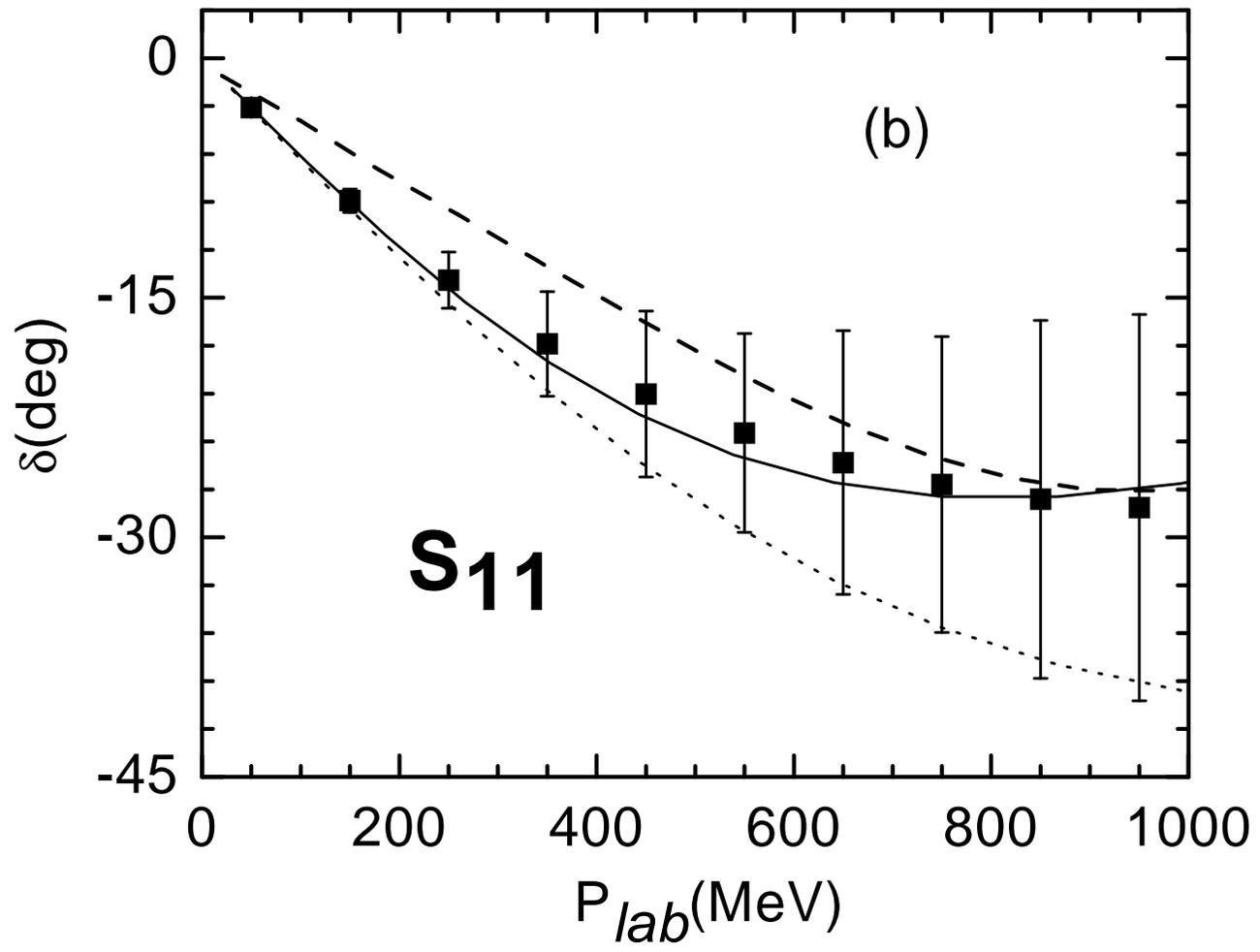
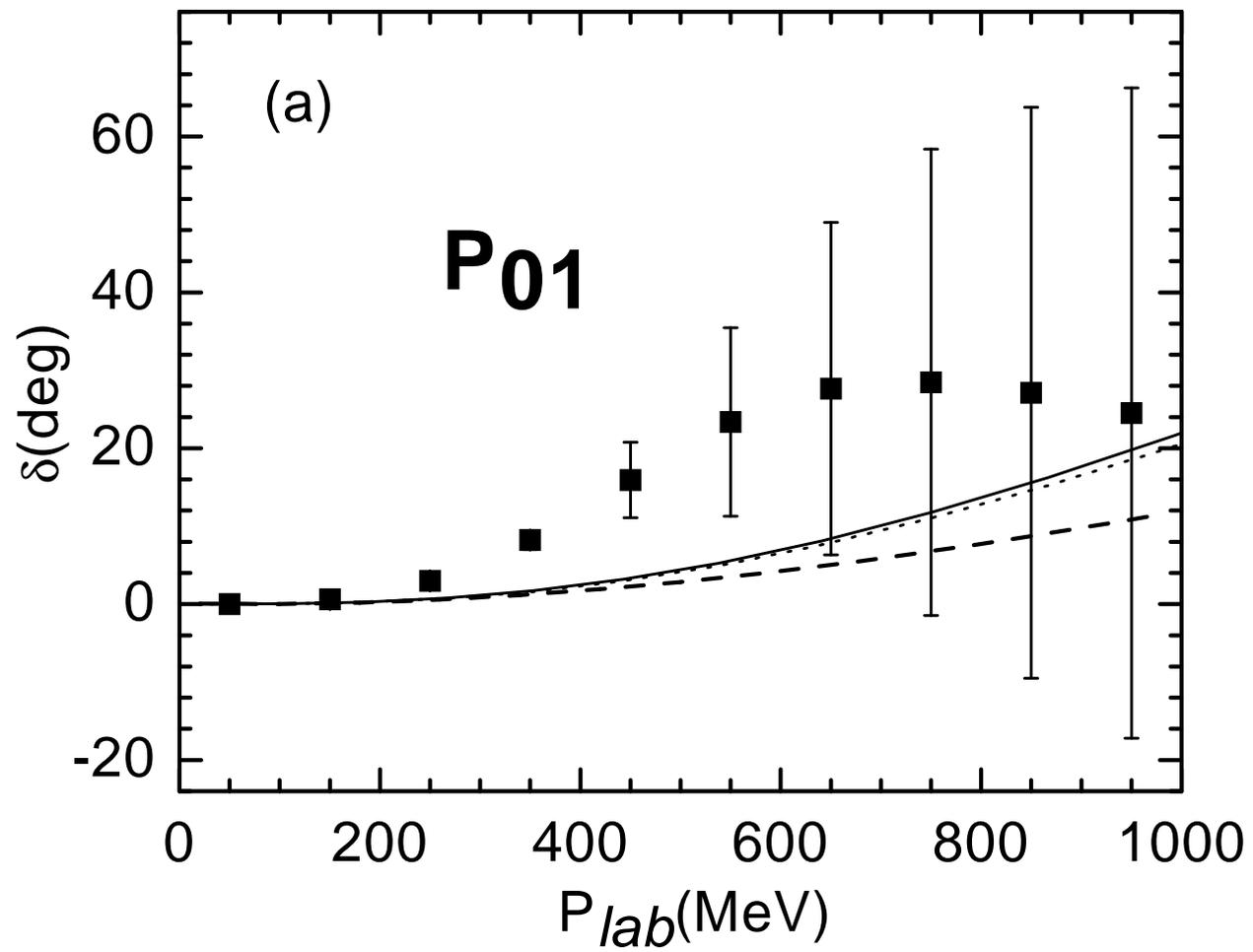
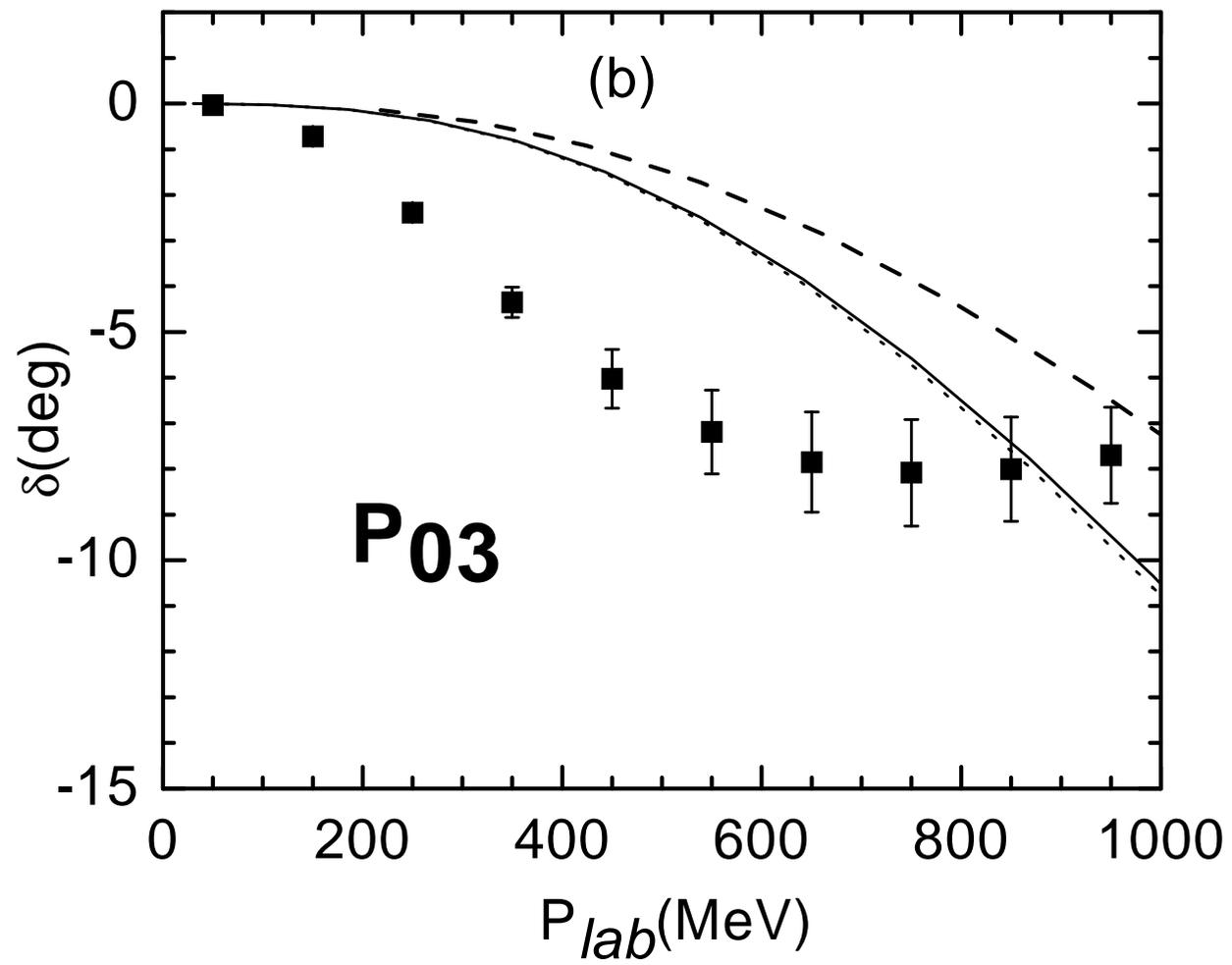
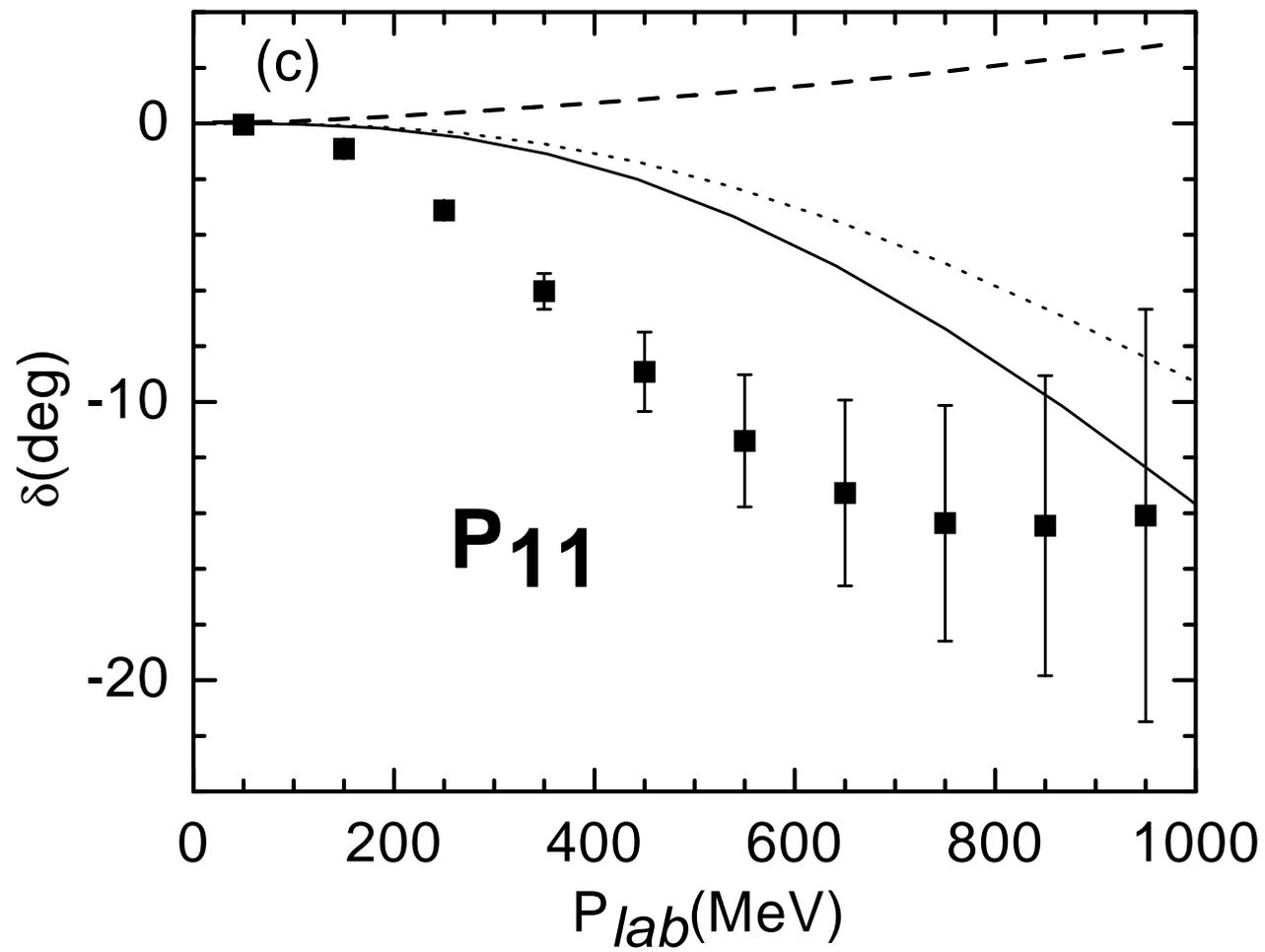
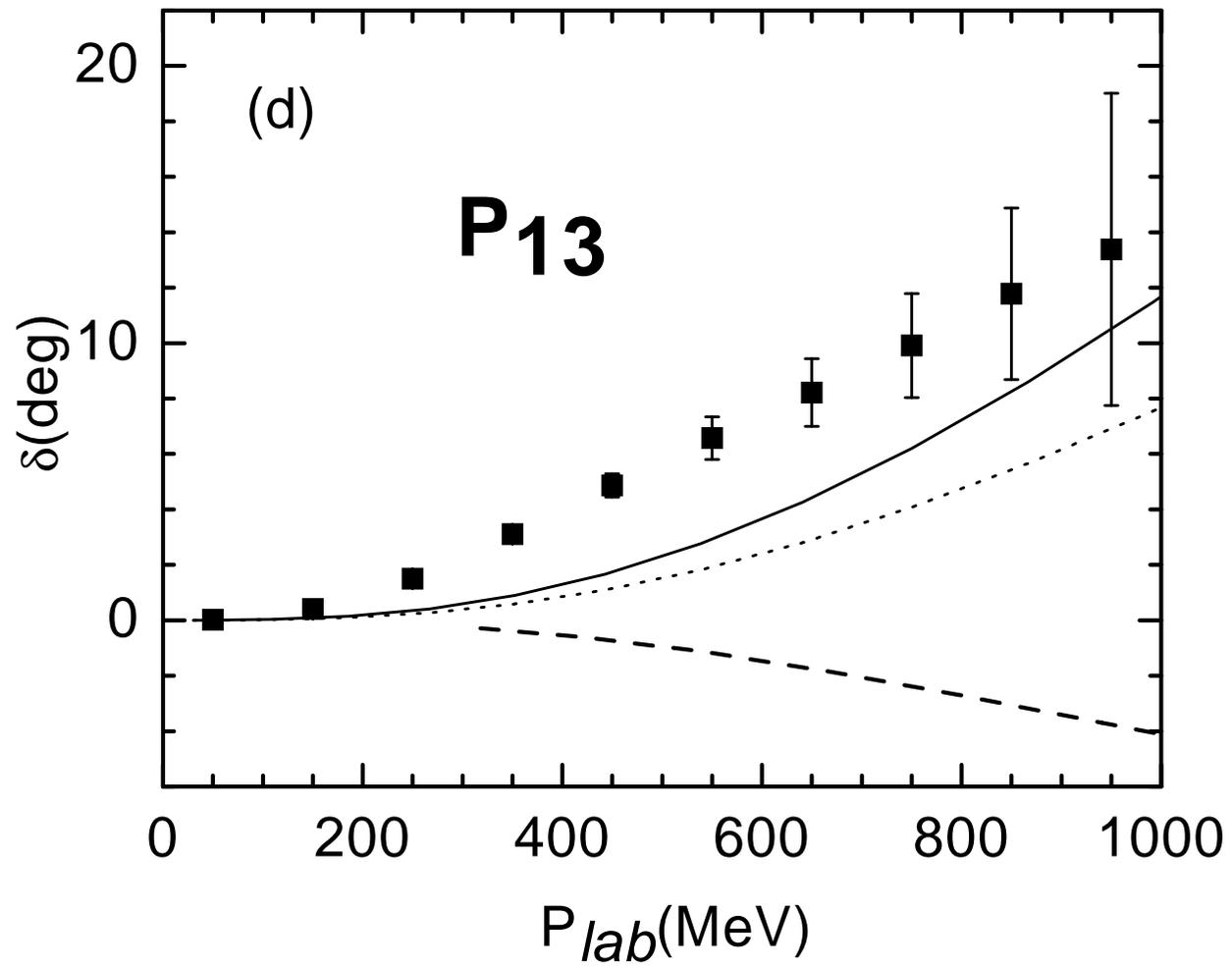
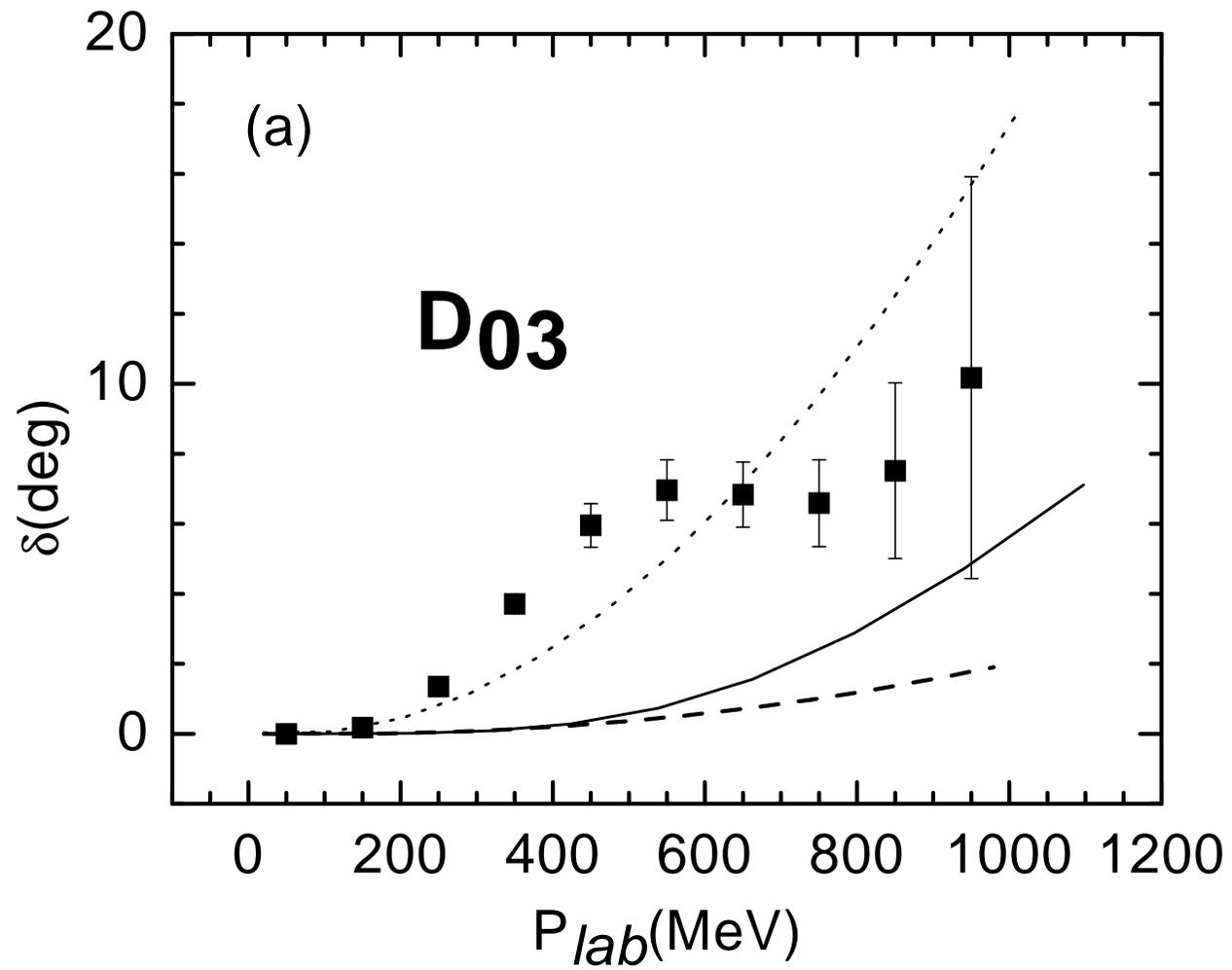
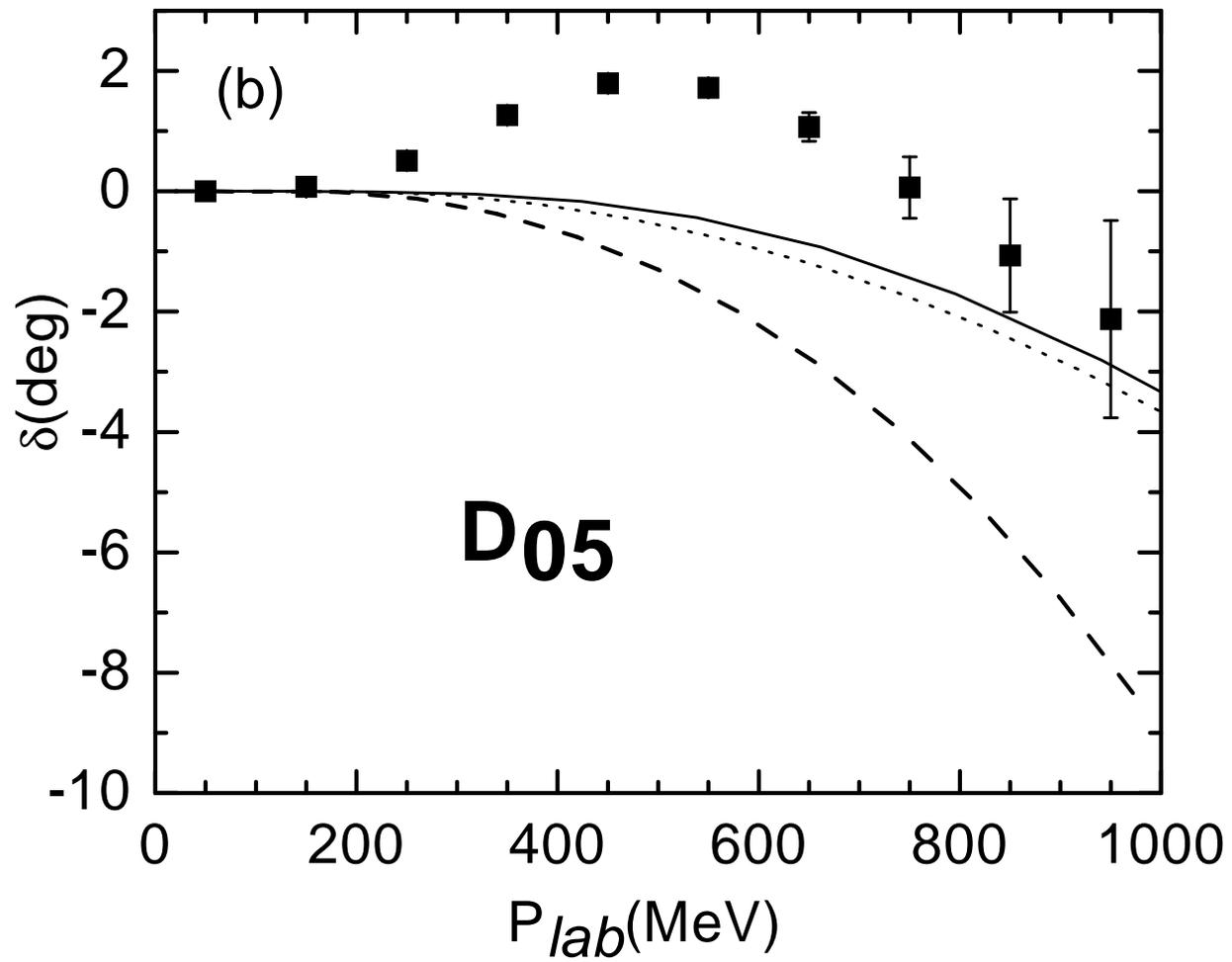
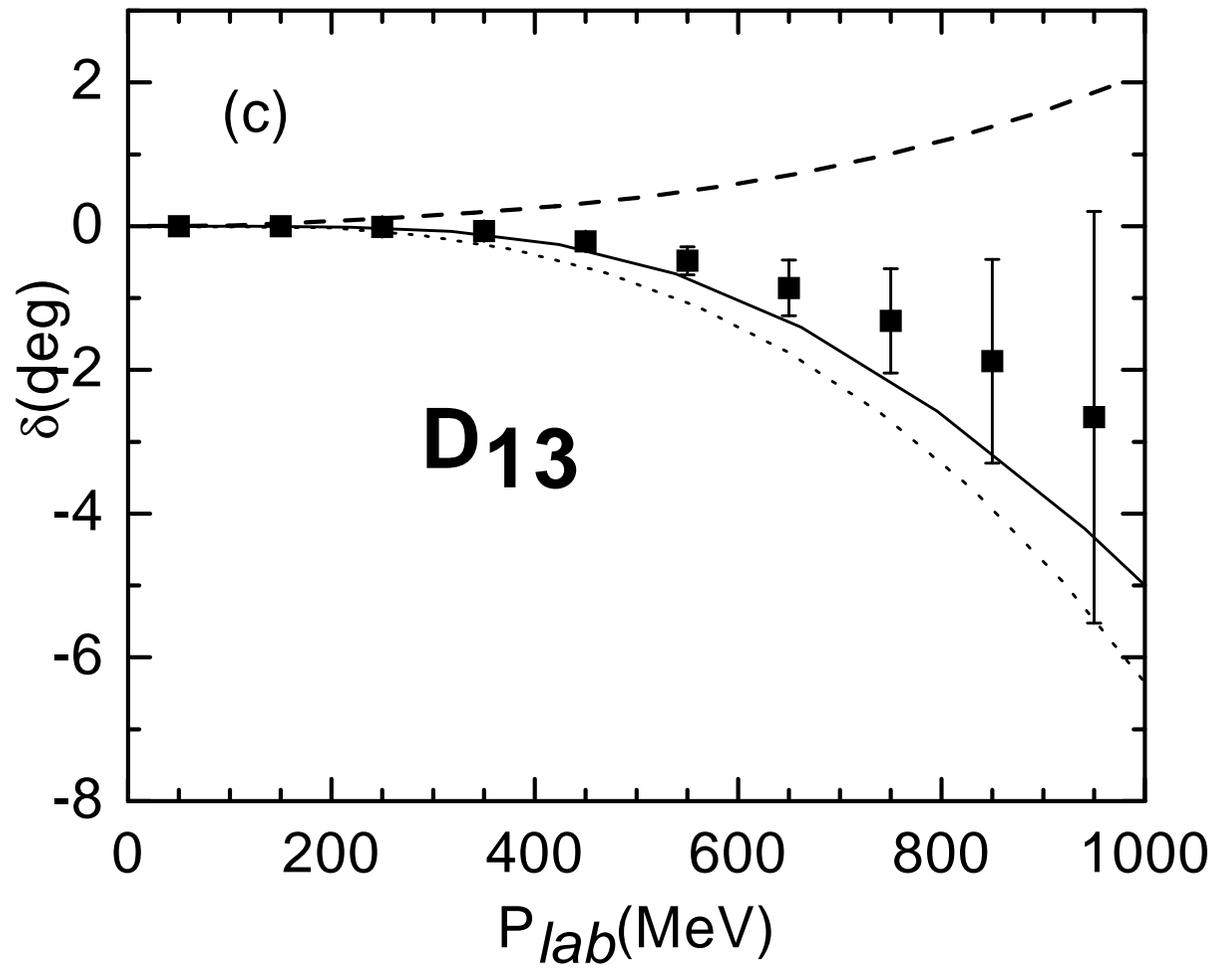
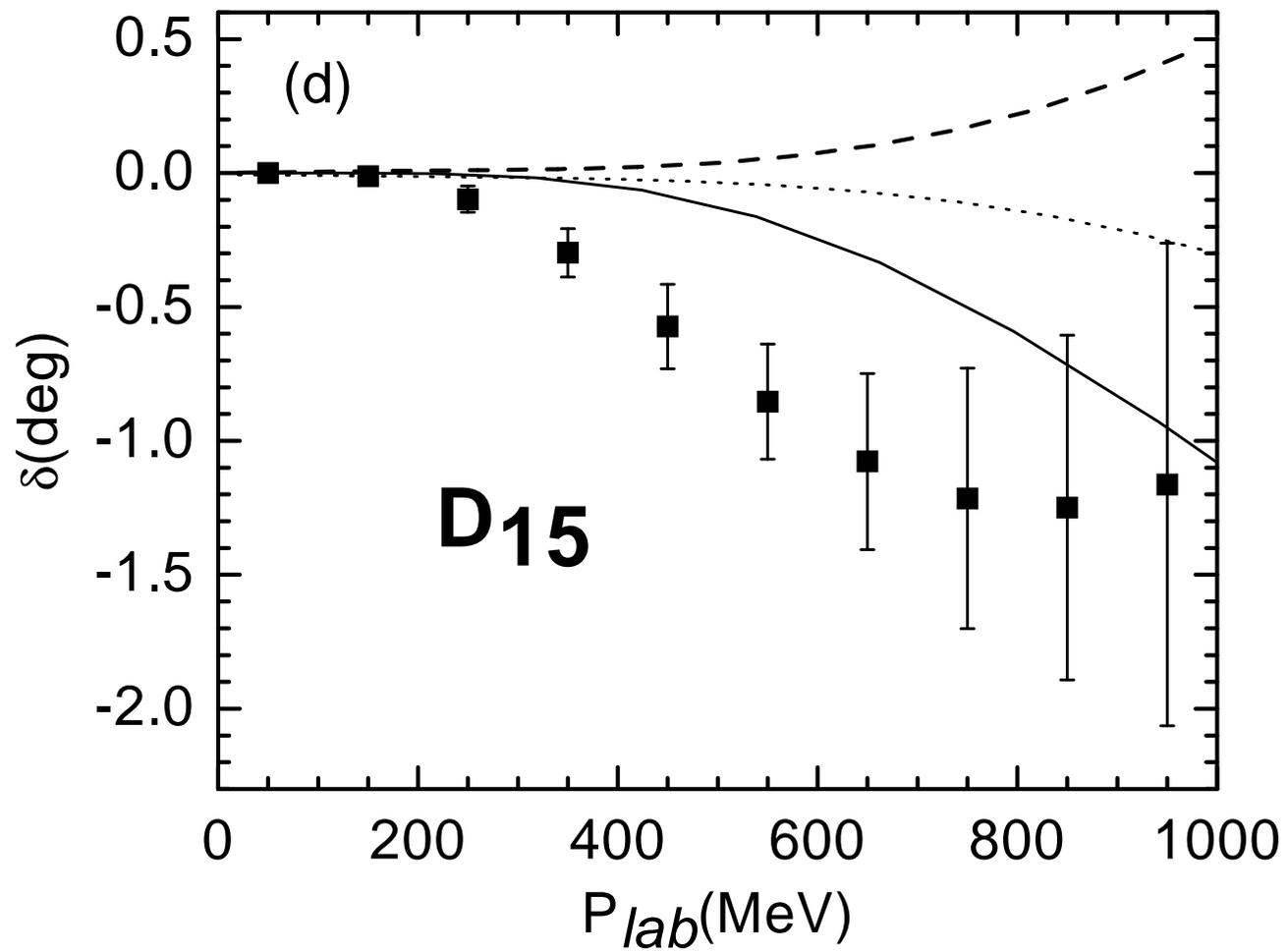
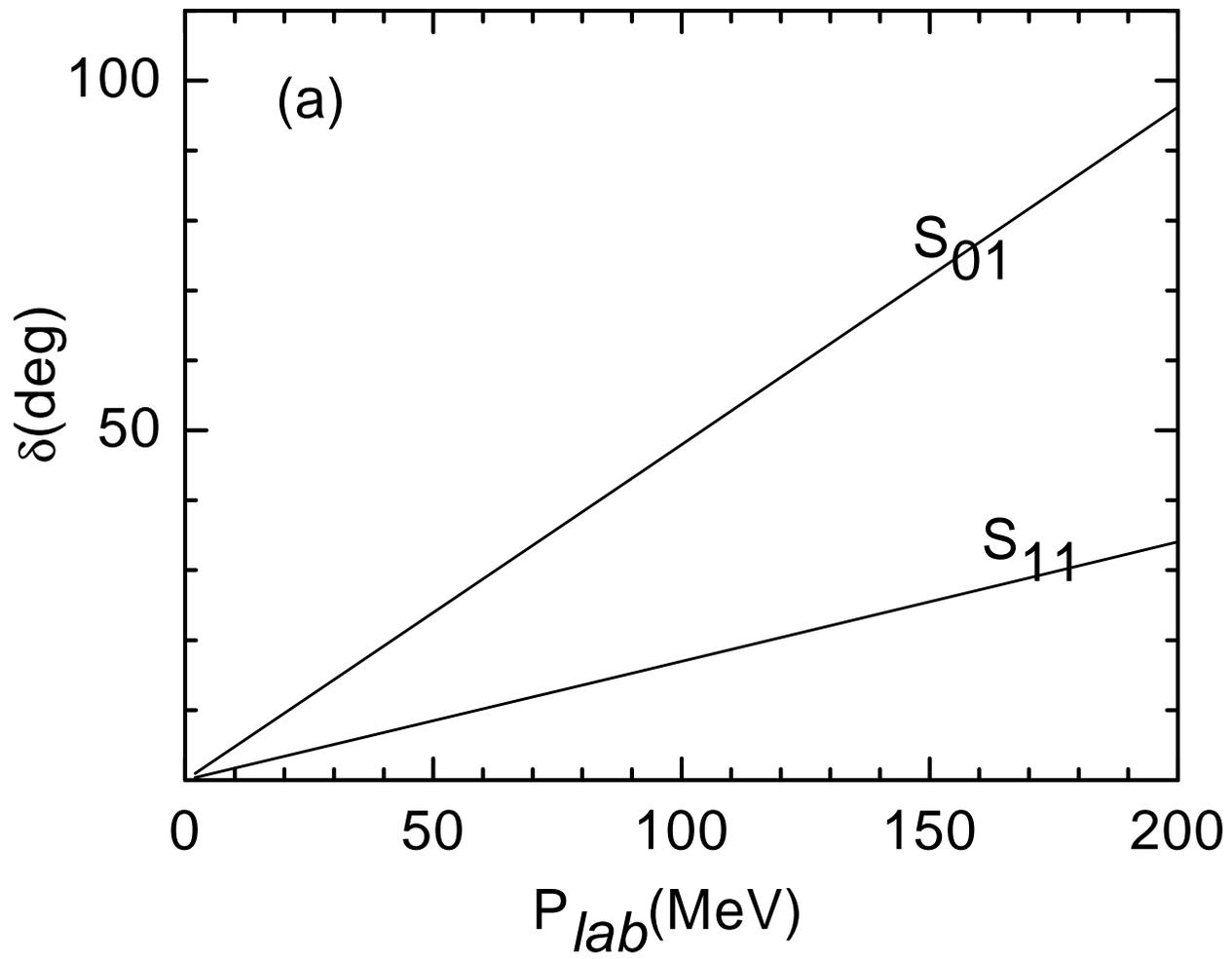
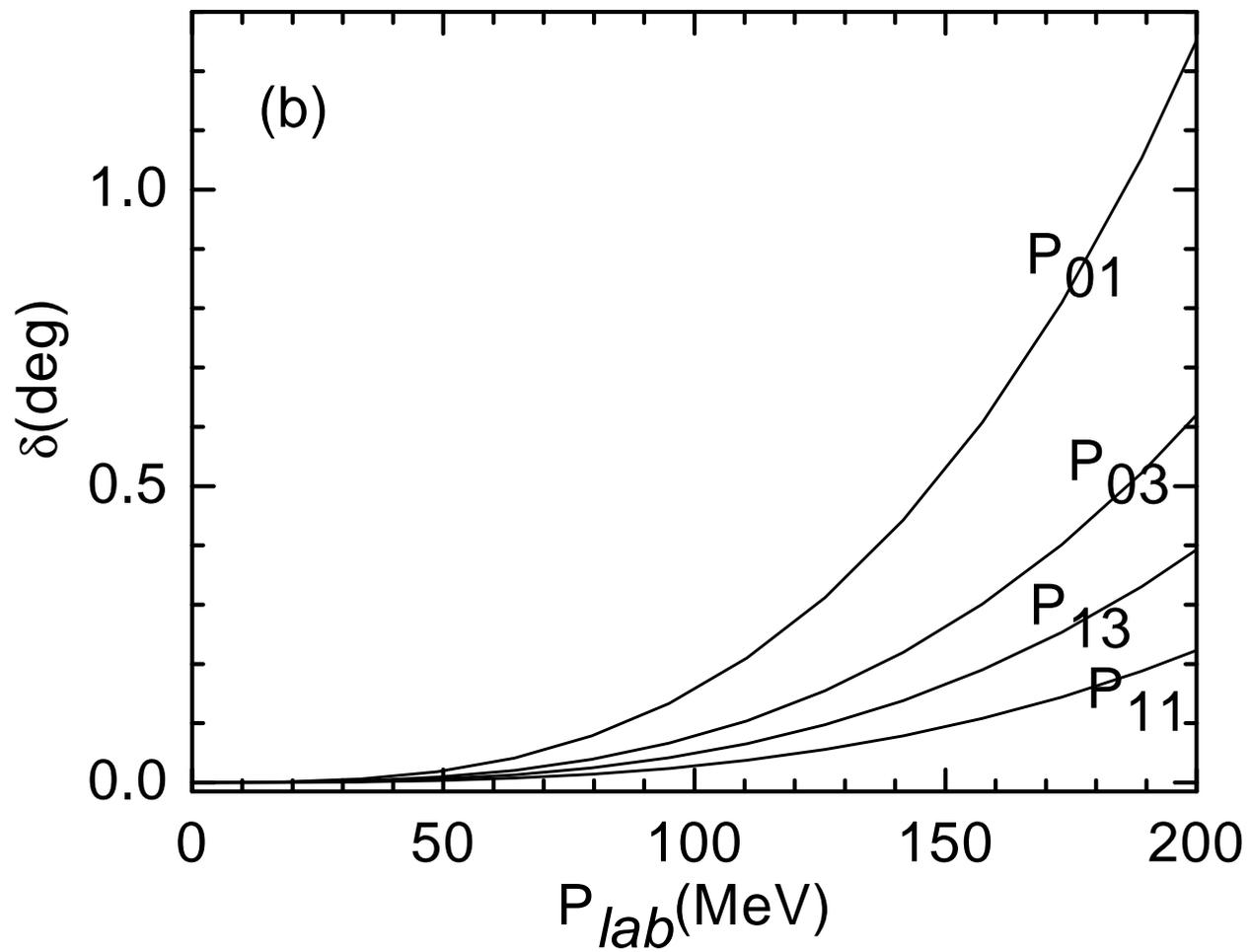
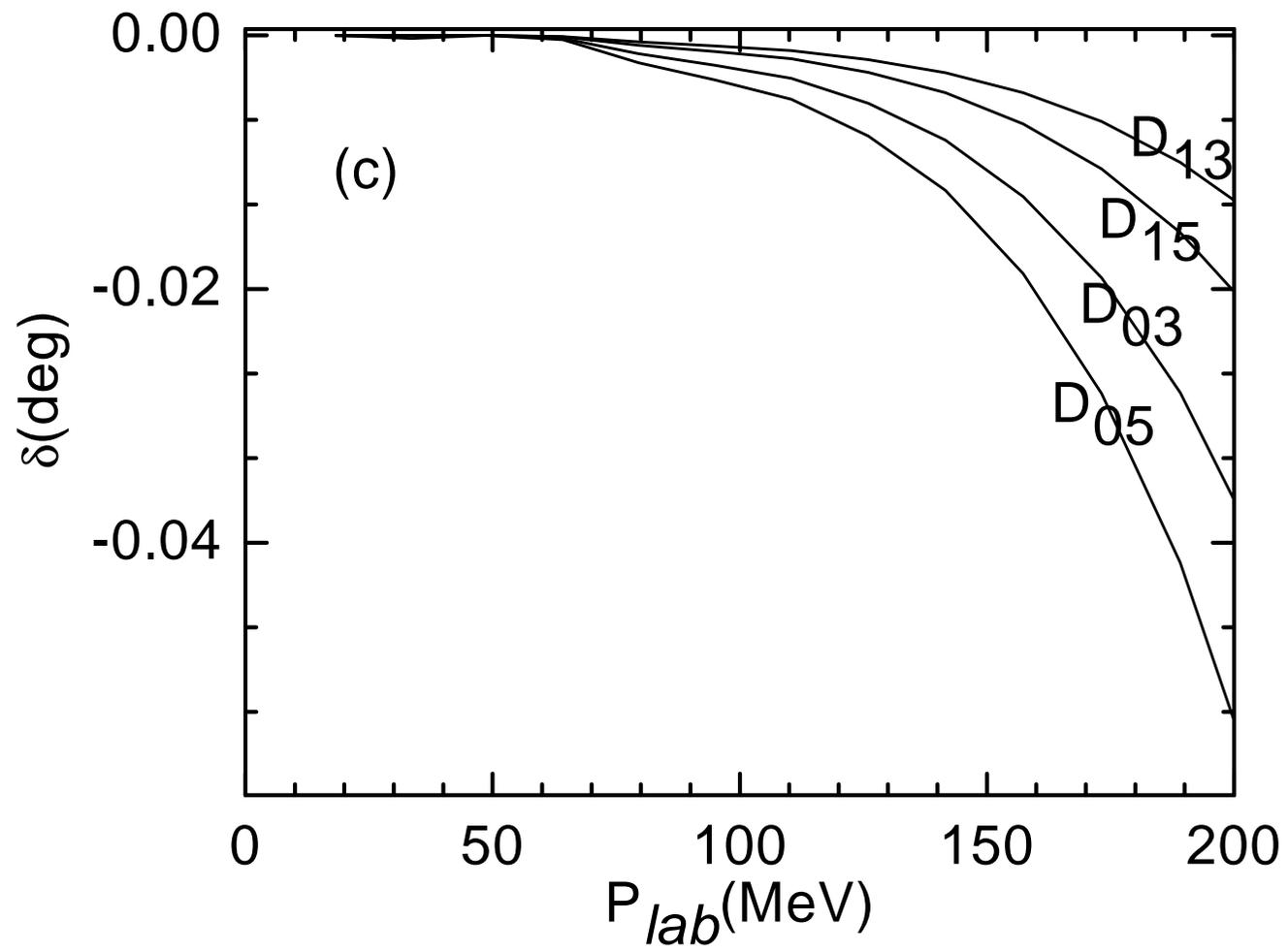
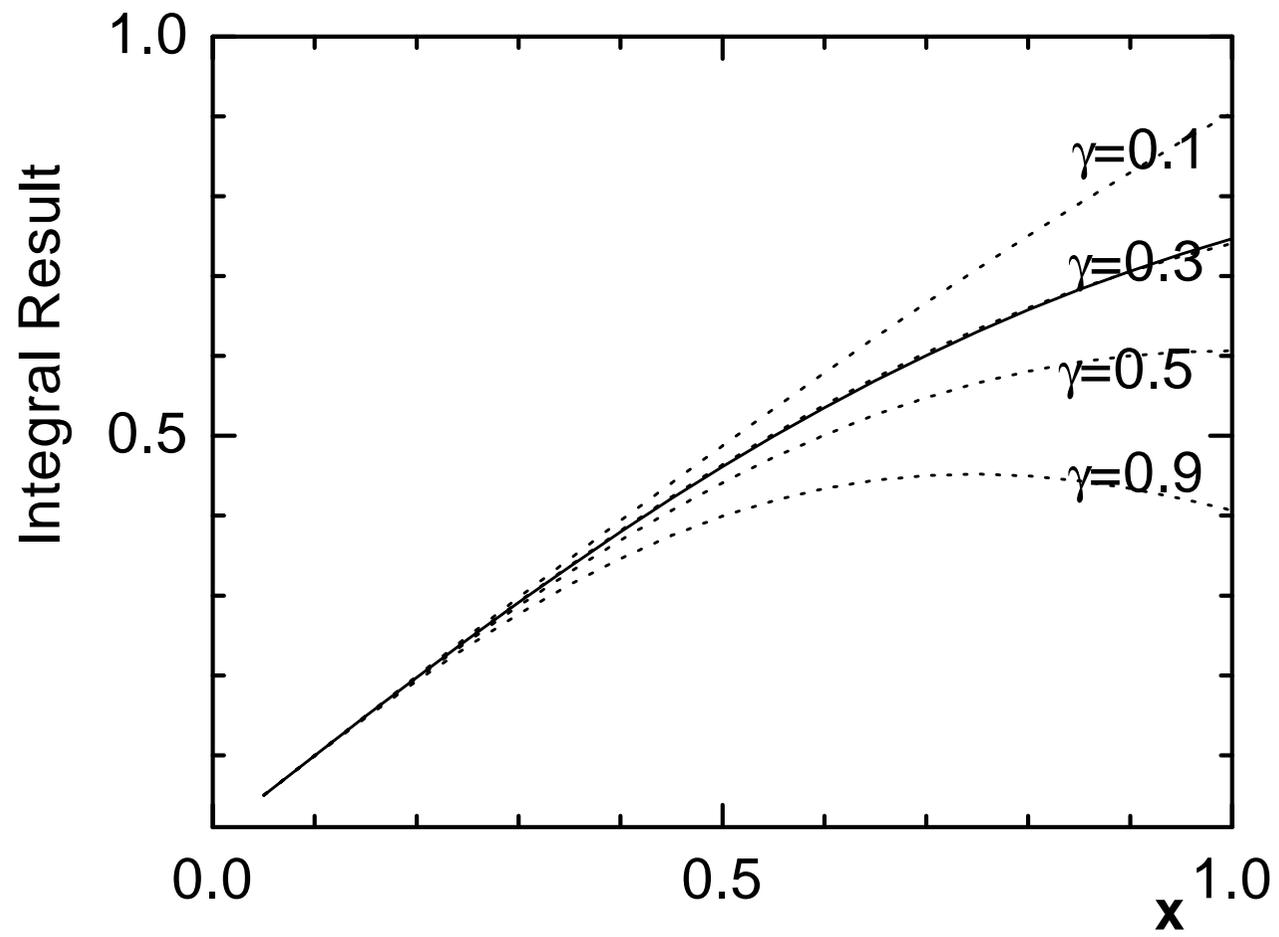
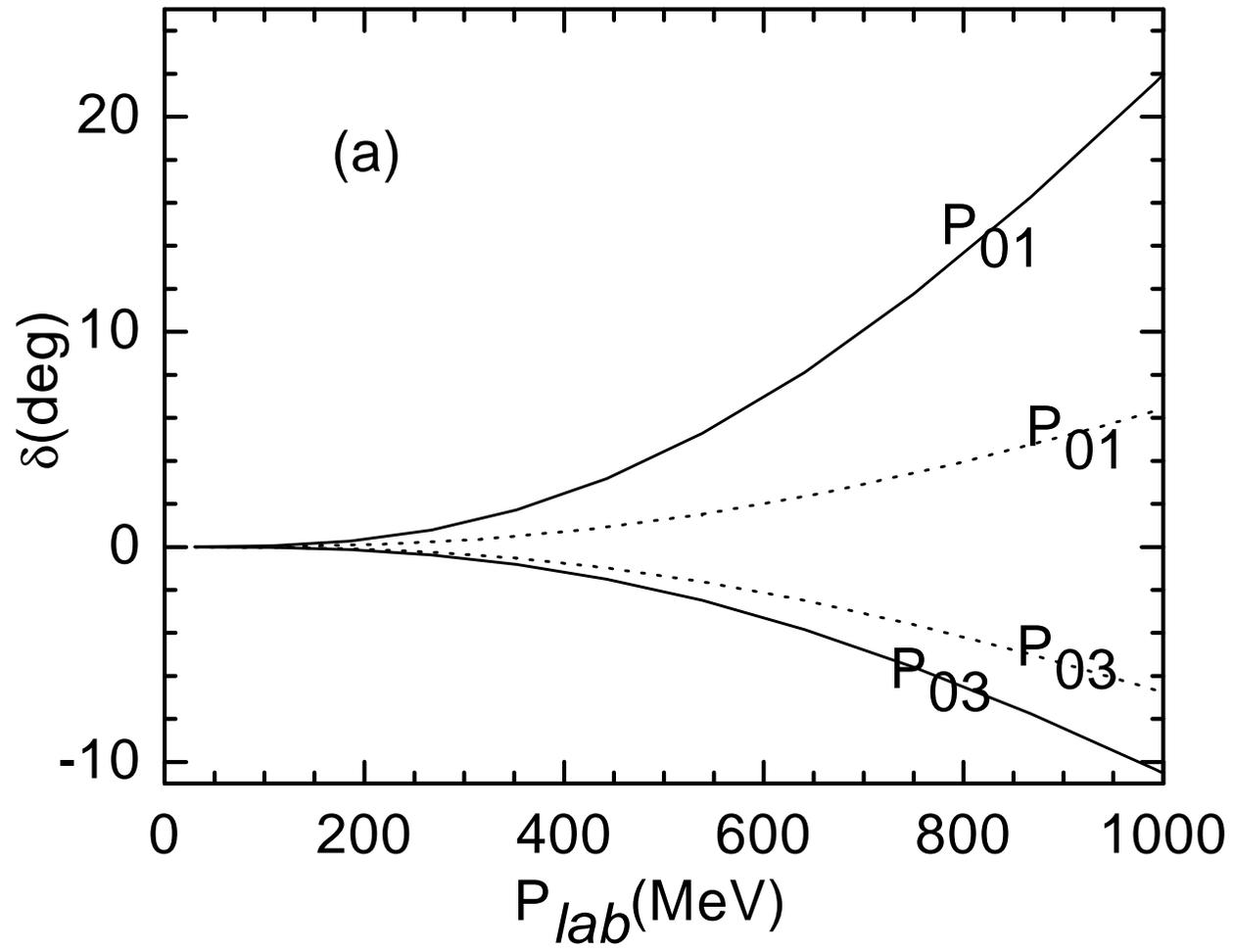
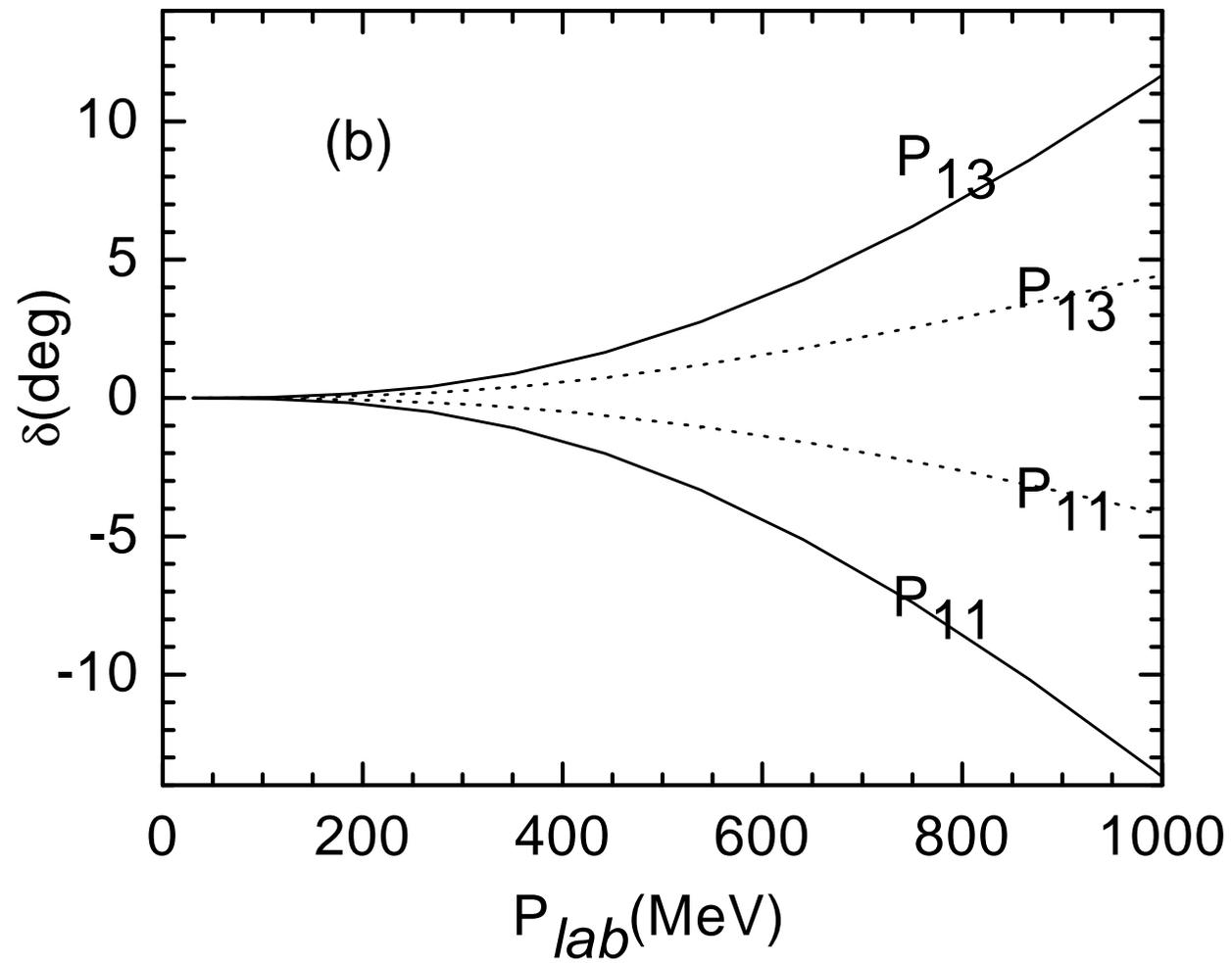
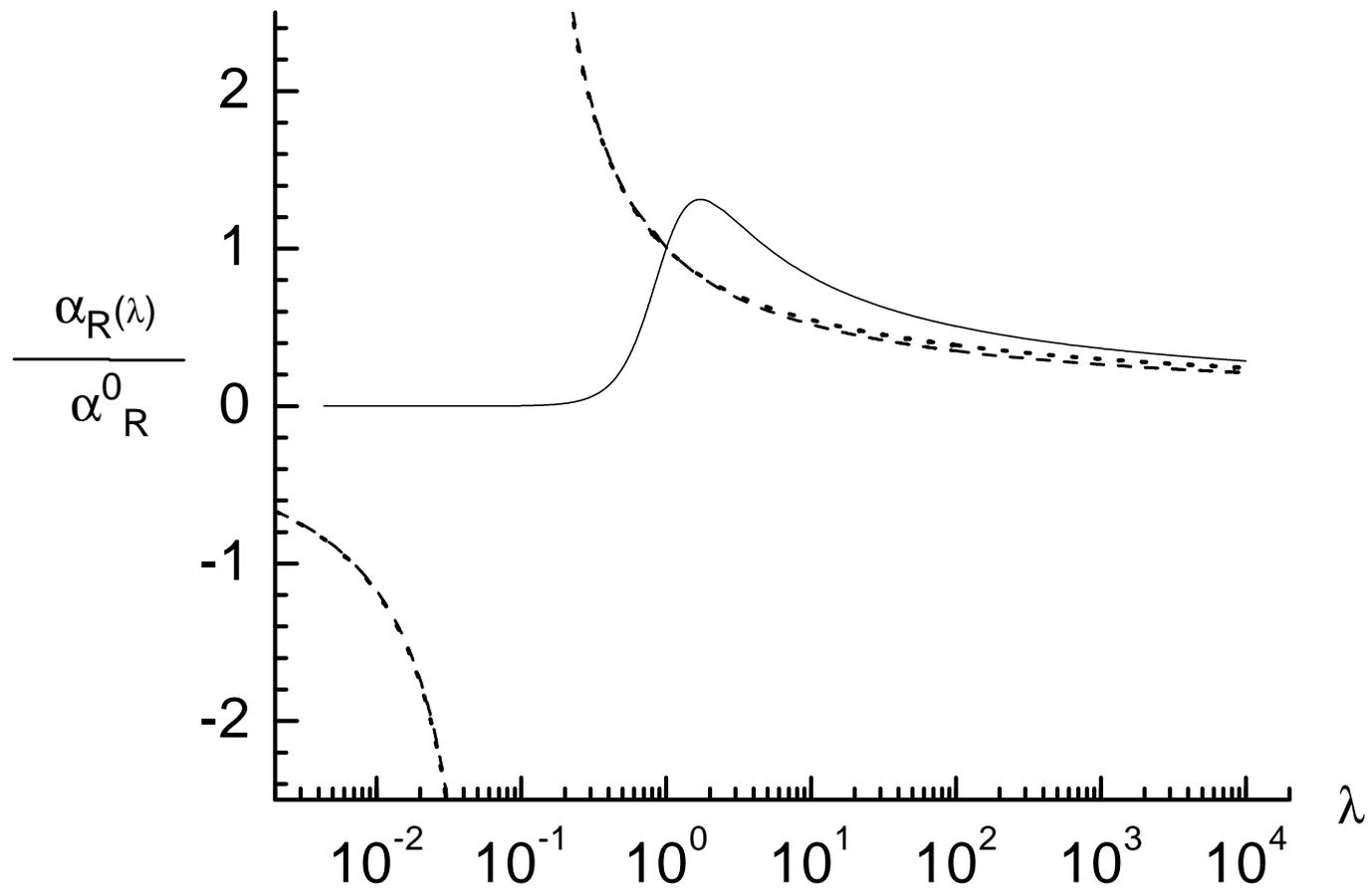
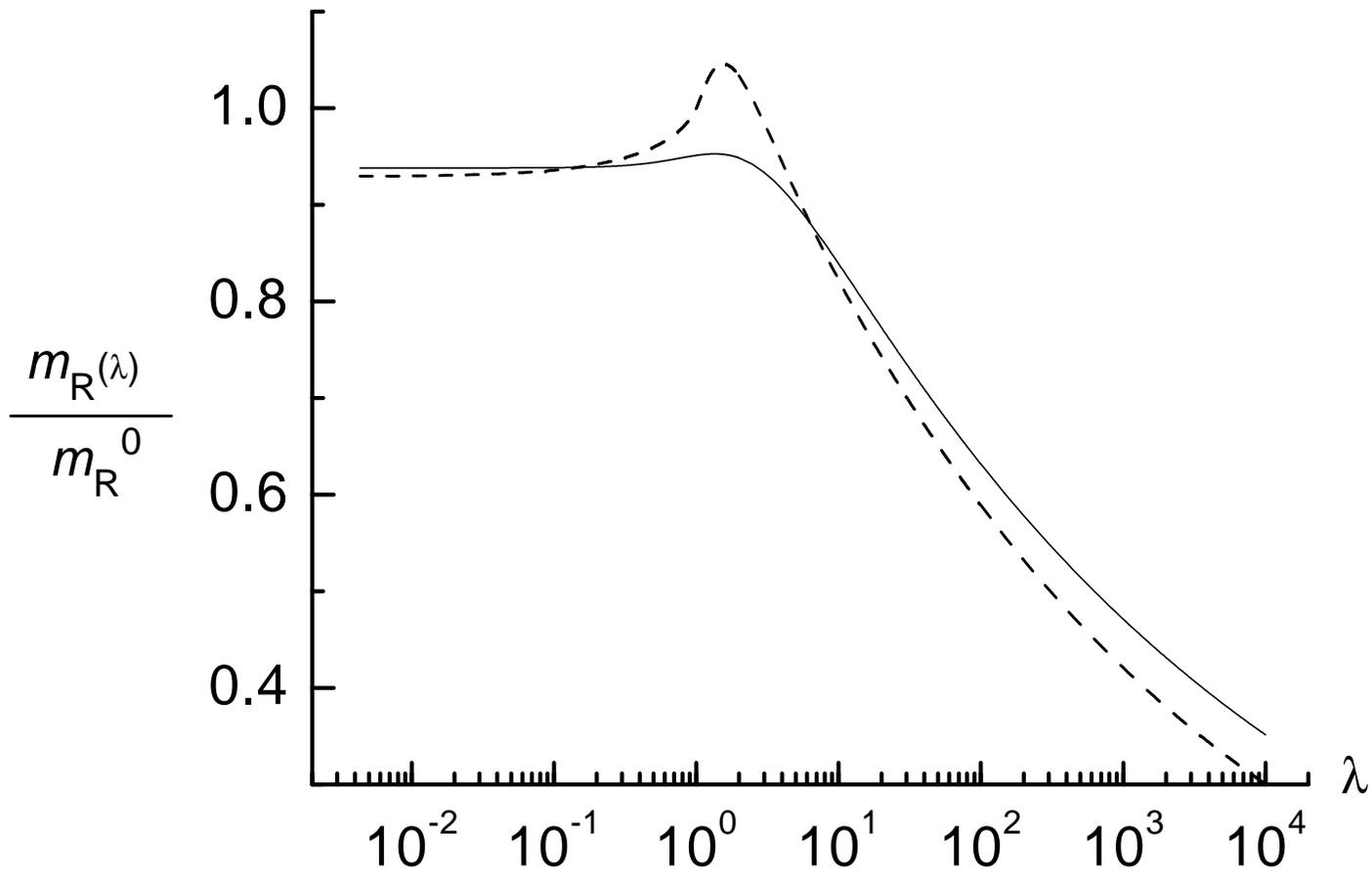
\end{document}